\documentclass[12pt]{article}
\pdfoutput=1

\usepackage{textcomp}
\usepackage{color}

\usepackage{putex}
\usepackage{autobreak}
\usepackage{graphicx}
\graphicspath{{plots/}}
\usepackage{tabularx}
\usepackage[font=small]{caption}
\usepackage{amsmath}
\usepackage{array}
\usepackage{subcaption}
\usepackage{epstopdf}
\usepackage{enumerate}
\usepackage{cite}
\usepackage{youngtab}
\usepackage{tensor}
\usepackage{slashed}
\usepackage[aligntableaux=center]{ytableau}
\usepackage[utf8]{inputenc}
\usepackage{rotating}
\usepackage{multirow}
\usepackage[
      colorlinks=true,
      linkcolor=blue,
      urlcolor=blue,
      filecolor=black,
      citecolor=red,
      bookmarks=false
      ]{hyperref}
\usepackage{cleveref}
\usepackage{bm}
\usepackage{tikz}
\usepackage{tikz-cd}
\usepackage{bbm}
\usepackage{newunicodechar}
\newunicodechar{Δ}{$\Delta$}

\newcommand{\grp}[1]{\mathrm{#1}}

\newcommand{\grU}{\grp{U}}
\newcommand{\grSU}{\grp{SU}}

\newcommand{\HH}{\mathbb{H}}

\newcommand{\abs}[1]{\left\lvert #1 \right\rvert}

\newcommand {\be} {\begin {equation}}
\newcommand {\ee} {\end {equation}}

\newcommand {\bes} {\begin {equation*}}
\newcommand {\ees} {\end {equation*}}

\newcommand{\es}[2] {\begin{equation} \label{#1} \begin{split} #2 \end{split} \end{equation}}

\newcommand{\R}{\mathbb{R}}

\newcommand{\cN}{{\mathcal N}}

\def\arccosh{\mop{arccosh}}

\newcommand{\beq}{\begin{equation}}
\newcommand{\eeq}{\end{equation}}

\def\ie{\begin{equation}\begin{aligned}}
\def\fe{\end{aligned}\end{equation}}

\newcommand{\la}{\langle}
\newcommand{\ra}{\rangle}

\newcommand{\m}{\mu}
\newcommand{\n}{\nu}

\newcommand{\ve}{{\varepsilon}}

\newcommand{\mZ}{{\mathbb Z}}
\newcommand{\mR}{{\mathbb R}}
\newcommand{\mH}{{\mathbb H}}

\newcommand{\mf}{\mathfrak }

\numberwithin{equation}{section}

\def\<{\langle}
\def\>{\rangle}

\def\SSP#1{}
\def\RD#1{}
\def\YW#1{}
\def\BO#1{}

\definecolor{mypurple}{rgb}{.5,0,.5}

\DeclareMathOperator{\SO}{SO}
\DeclareMathOperator{\SU}{SU}

\let\bar\relax
\def\bar{\overline}

\graphicspath{{plots/}}

\begin{document}

\preprint{PUPT-2652}

\institution{PU}{Joseph Henry Laboratories, Princeton University, Princeton, NJ 08544, USA}
\institution{PCTS}{Princeton Center for Theoretical Science, Princeton University, Princeton, NJ 08544, USA}
\institution{IAS}{Institute for Advanced Study, Princeton, NJ 08540, USA}
\institution{NYU}{Center for Cosmology and Particle Physics, New York University, New York, NY 10003, USA}

\title{Global Symmetry and Integral Constraint on Superconformal Lines in Four Dimensions}

\authors{Ross Dempsey,\worksat{\PU} Bendeguz Offertaler,\worksat{\PU} Silviu S. Pufu,\worksat{\PU,\PCTS,\IAS} and Yifan Wang\worksat{\NYU}}

\abstract{We study properties of point-like impurities preserving flavor symmetry and supersymmetry in four-dimensional ${\cal N} = 2$ field theories.  At large distances, such impurities are described by half-BPS superconformal line defects.  By working in the $\text{AdS}_2\times \text{S}^2$ conformal frame, we develop a novel and simpler way of deriving the superconformal Ward identities relating the various two-point functions of flavor current multiplet operators in the presence of the defect.  We use these relations to simplify a certain integrated two-point function of flavor current multiplet operators that, in Lagrangian theories, can be computed using supersymmetric localization.  The simplification gives an integral constraint on the two-point function of the flavor current multiplet superconformal primary with trivial integration measure in the $\text{AdS}_2\times \text{S}^2$ conformal frame.  We provide several consistency checks on our Ward identities.
}

\date{May 2024}

\maketitle

\tableofcontents

\section{Introduction}\label{sec:intro}

Line defects are point-like impurities in quantum systems. They participate actively in the many-body dynamics and 
provide order parameters for phase transitions. 
Canonical examples include the Kondo impurity in a metal \cite{Kondo:1964nea,Affleck:1995ge}, localized magnetic field lines in scalar field theories \cite{Parisen_Toldin_2017, Allais:2014fqa,Cuomo:2021kfm,Popov:2022nfq}, and Wilson-'t Hooft lines in gauge theories \cite{wilson1974confinement,t1978phase,Rey:1998ik,Maldacena:1998im,Kapustin:2005py}.
Of particular interest are conformal line defects, which arise as nontrivial fixed points from renormalization group flows of impurities coupled to a critical bulk system. Such a conformal line produces new structures in local observables even away from the defect, which in turn encode the dynamical signature of the impurity at long distances. 

Recently, modern tools in Conformal Field Theory (CFT) have been developed to analyze conformal defects and correlation functions of bulk insertions in their presence. In particular, the bootstrap axioms obeyed by correlators of local operators in CFT have been generalized to incorporate defects, and one can write down a bootstrap equation for the two-point function of bulk local-operators in the presence of the line defect \cite{Billo:2016cpy,Lauria:2018klo,Kobayashi:2018okw}.  This equation shares some similarities with the crossing equation for the four-point function of local operators without a defect (see \cite{Poland:2018epd,Chester:2019wfx,Poland:2022qrs} for recent reviews), such as the existence of multiple operator-product-expansion (OPE) channels. Unlike the case of the four-point function, however, the bootstrap equation in a CFT with a defect lacks positivity and is therefore insufficiently constraining.  Additional input is indispensable in this case. 

Extra dynamical input has also been useful in the case of the four-point function bootstrap in superconformal field theories (SCFTs), where a powerful technique known as supersymmetric localization (see \cite{Pestun:2016zxk} for a collection of reviews) provides exact results that have facilitated such bootstrap studies  \cite{Binder:2018yvd,Binder:2019jwn,Chester:2019jas,Chester:2020dja,Chester:2020vyz,Binder:2019mpb,Binder:2020ckj,Chester:2021gdw,Binder:2021cif,Alday:2021ymb,Chester:2021aun,Chester:2022sqb,Behan:2023fqq,Chester:2023ehi,Chester:2023qwo,Behan:2024vwg}. These exact results can generally be recast as certain linear combinations of correlation functions of operators in the same superconformal multiplet integrated over the conformal cross-ratios. Thus, they provide \textit{integral constraints} on the local correlators. In particular, together with either large $N$ or numerical bootstrap methods, they have been used efficiently to constrain a family of local correlators in the $d=4$ $\SU(N)$ $\cN=4$ super-Yang-Mills (SYM) theory at arbitrary coupling \cite{Binder:2019jwn,Chester:2019jas,Chester:2020dja,Chester:2020vyz,Chester:2021aun,Chester:2023ehi}. 
More recently, analogous exact results for local correlators in the presence of half-BPS Wilson-'t Hooft line defects in ${\cal N} = 4$ SYM were also derived in \cite{Pufu:2023vwo}. One of the main goals of this work is to elucidate the precise form of the corresponding integrated correlator from \cite{Pufu:2023vwo}.

Another motivation of our work concerns the interplay between symmetries and impurities.
When a CFT has a continuous global symmetry $G_F$ with conserved currents $j^I_\m$ (where $I$ is the adjoint index for $G_F$), a natural question to ask is how impurities immersed in the system affect this symmetry and vice versa.  At long distances, this question can be addressed by studying correlation functions of $j^I_\m$ in the presence of a conformal line defect $\mathbb{L}$. The residual conformal symmetry preserved by $\mathbb{L}$ implies that the one-point function $\la j^I_\m(x)\ra_\mathbb{L}$ vanishes, and the first non-trivial correlator is the two-point function $\la j^I_\m(x) j^J_\n(y)\ra_\mathbb{L}$ \cite{Billo:2016cpy}. The form of this correlator depends on whether the symmetry is preserved by the defect. For reasons that will become clear shortly, we focus on the case where the defect $\mathbb{L}$ preserves the $G_F$ symmetry. 
Even so, this correlator is much richer than in the absence of the defect.  There are multiple conformal structures consistent with current conservation, each being multiplied by functions of two conformally-invariant cross-ratios (see \cite{Herzog:2020bqw} for examples).\footnote{These defect structures encode how the symmetry is realized by the combined system and how the conserved currents arise from a dynamical mixture of degrees of freedom in the bulk and localized at the impurity. Such a phenomenon already arises in the context of the Kondo impurity, where the relevant symmetry is the $\grSU(2)$ spin symmetry of the electrons. While the Kondo impurity appears to break the bulk spin symmetry at short distance, at long distance the spin symmetry is recovered due to dynamical screening and reorganization between impurity and bulk degrees of freedom (see \cite{Affleck:1995ge} for a review).}

In this work, we study superconformal line defects in general 4d $\cN=2$ SCFTs with continuous flavor symmetry group $G_F$.\footnote{In the supersymmetric context, a flavor symmetry is a global symmetry that commutes with the supercharges.}   As explained in \cite{Agmon:2020pde},
unitarity, superconformal symmetry, and locality imply that such a superconformal line has to be half-BPS and it can never break $G_F$!\footnote{Here we are interested in \textit{local} line defects for which correlation functions of bulk local operators are single-valued. In particular the defect preserves the rotational symmetry in the transverse space.} The line preserves the $\mf{osp}(4^*|2)$ subalgebra of the $\cN=2$ superconformal algebra $\mf{su}(2,2|2)$.
Well-known examples include half-BPS Wilson-'t Hooft lines in $\cN=2$ superconformal gauge theories with flavor symmetry, $\cN=4$ SYM theory being a special case with $G_F=\SU(2)$ \cite{Alday:2009fs,Drukker:2009id,Drukker:2010jp}.  Superconformal symmetry implies that the conserved currents $j_\m^I$ belong to a short superconformal multiplet known as the flavor current multiplet, which also contains scalar operators $J_{ij}^I$, $K^I$, and ${\bar K}^I$ of scaling dimension $2$, $3$, and $3$, respectively, as well as fermionic operators of dimension $5/2$ \cite{Dolan:2002zh}.  The operators $J_{ij}^I$, also known as moment map operators, transform as triplets of the $\mf{su}(2)_R$ R-symmetry (with $i,j=1,2$) and are the superconformal primary operators of the multiplet.   

We are specifically interested in the two-point functions of the various flavor current multiplet operators in the presence of the superconformal line $\mathbb{L}$\@. 
In this case, the correlators are expected to be determined in terms of the two-point function of $J_{ij}^I$ via the superconformal Ward identities. Furthermore, the flavor  symmetry structure is fixed by the Killing form up to an overall constant for simple $G_F$. Thus it suffices to focus on a Cartan generator (i.e.,~a $\grU(1)$ subgroup of $G_F$), and we will henceforth omit the adjoint index $I$ on the current multiplet operators.

As will be explained in detail in the next two sections, the exact supersymmetric localization results  of interest here come from placing the SCFT on ${\rm S}^4$, deforming it by an ${\cal N} = 2$-preserving mass $m$, and taking derivatives with respect to $m$.  Evaluating these derivatives at $m=0$ gives integrated correlators of $J_{ij}$, $K$, and $\bar K$ in the SCFT\@.  In particular, after the Weyl mapping from $\R^4$ to ${\rm S}^4$, a line defect on $\R^4$ becomes supported on a great circle of ${\rm S}^4$.  If $\langle \mathbb{L} \rangle(m)$ is the expectation value of the defect  in the mass-deformed theory on ${\rm S}^4$, the quantity $\partial_m^2 \log \la \mathbb{L}\ra|_{m=0}$ is given by a linear combination of integrals of two-point functions of $J_{ij}$, $K$, and $\bar K$ in the presence of the defect.  This integral constraint can be simplified using the superconformal Ward identities relating these two-point functions.

To determine these Ward identities, we find that it is much more efficient to work in the ${\rm AdS}_2\times {\rm S}^2$ conformal frame, with the asymptotic boundary condition specified by the superconformal line $\mathbb{L}$, as was considered in \cite{Kapustin:2005py}. Intuitively, the asymptotic boundary $\mR\times {\rm S}^2$ of ${\rm AdS}_2\times {\rm S}^2$ is related by a Weyl transformation to the boundary of the tubular neighborhood of the straight line defect, and inserting the line defect is equivalent to specifying a set of boundary conditions there. The ${\rm AdS}_2\times {\rm S}^2$ conformal frame makes manifest the $\mf{osp}(4^*|2)$ residual superconformal symmetry in terms of (super)isometries of the spacetime. This turns out to simplify dramatically the derivation of the $\mf{osp}(4^*|2)$ Ward identities.\footnote{Intuitively, this is because the (super-)Poincar\'e algebra (suitably generalized to curved spacetime) is much simpler than the (super-)conformal algebra.} Indeed, we find simple expressions for $\la KK\ra_\mathbb{L},\la K\bar K\ra_\mathbb{L},\la \bar K\bar K\ra_\mathbb{L}$ in terms of the ${\rm AdS}_2\times {\rm S}^2$ Laplacian acting on an expression proportional to $\la J J\ra_\mathbb{L}$. This is reminiscent of the supersymmetry Ward identity that relates such two-point functions in the absence of the defect (i.e.,~a trivial defect), where $K,\bar K$ are related to $J$ by acting with the Poincar\'e supercharges twice. Our results indicate that such relations persist in the presence of a line defect provided that we work in the ${\rm AdS}_2\times {\rm S}^2$ conformal frame.

Putting together these superconformal Ward identities and the combinations of current-multiplet two-point functions that appear in the integrated correlator $\partial_m^2 \log \la \mathbb{L}\ra|_{m=0}$, we arrive at an extremely simple relation (see  \eqref{TwoPointAdS}, \eqref{FreeL}, and \eqref{SecondDerSimp4}),
\ie 
\partial_m^2\log \la \mathbb{L}\ra |_{m=0}=-4\pi^2\int_{\mH^2\times {\rm S}^2} d^4 x\,  \sqrt{g} \,  \left\la J_{11}(x) J_{22}(y) \right\ra_{\mathbb{L}, \text{conn}}  \,,
\label{integralconstraint}
\fe
which takes the form of an integral of the \textit{connected }two-point function of $J_{ij}$ over $\mH^2\times {\rm S}^2$ (which is the Euclidean continuation of ${\rm AdS}_2\times {\rm S}^2$). This result, when combined with analytical and numerical bootstrap methods, paves the way to solving the un-integrated defect correlator as outlined in \cite{Pufu:2023vwo}. As a preliminary step, we verify explicitly that \eqref{integralconstraint} is compatible with existing results for the half-BPS fundamental Wilson loop $\mathbb{L}=\mathbb{W}$ in $\cN=4$ $\grSU(N)$ SYM at large $N$ from localization \cite{Pufu:2023vwo} and from holography \cite{Barrat:2021yvp}. We emphasize that this check only makes use of a small portion of the defect data determined in \cite{Pufu:2023vwo}. More generally, the result from \cite{Pufu:2023vwo} constrains the two-point function via \eqref{integralconstraint} at arbitrary Yang-Mills coupling (and $\theta$-angle), and deserves to be analyzed further.

The two-point functions of operators from the flavor current multiplet and the mass derivative on the left-hand side of \eqref{integralconstraint} were also considered in \cite{Billo:2023ncz}. In \cite{Billo:2023ncz}, Ward identities are derived in the $\mathbb{R}^4$ conformal frame, whereas we use the $\mathbb{H}^2 \times {\rm S}^2$ conformal frame. Our result \eqref{integralconstraint} differs from that given in \cite{Billo:2023ncz}; it would be interesting to find the source of the discrepancy.\footnote{We were informed by the authors of \cite{Billo:2023ncz} that they have recently reconsidered the derivation presented in \cite{Billo:2023ncz}, and that they have independently obtained the integral constraint \eqref{integralconstraint}.  The updated analysis will appear in a forthcoming publication \cite{BilloToAppear}.}  

The rest of the paper is organized as follows. In Section~\ref{SETUP}, we start by reviewing $\cN=2$ SCFTs with continuous global symmetry, the associated current multiplet, and the corresponding supersymmetric mass deformation on ${\rm S}^4$. We introduce the half-BPS superconformal line defect in Section~\ref{sec:line}, explain the kinematic structure for correlators of the current multiplet in the presence of the line defect, and discuss how they simplify in the ${\rm AdS}_2\times {\rm S}^2$ conformal frame. In Section~\ref{sec:massivedefect}, we discuss the supersymmetric mass deformation with the line defect and express the integrated correlator obtained from two mass derivatives of the partition function in terms of the two-point functions of the current multiplet. We derive simple superconformal Ward identities relating such two-point functions in Section~\ref{WARD} by working directly in the ${\rm AdS}_2\times {\rm S}^2$ conformal frame.  In Section~\ref{sec:integralconstraint}, we derive the integral constraint.  In Section~\ref{sec:sugra}, we apply our general result to the case of half-BPS Wilson line in the $\cN=4$ SYM theory, establishing the precise connection between the integrated correlator defined in \cite{Pufu:2023vwo} and the two-point function of the current multiplet for the $\grSU(2)_F$ flavor symmetry.  Using the leading order result of \cite{Barrat:2021yvp}, we provide a check on our integral constraint.  Several technical details are relegated to the Appendices.

\section{\texorpdfstring{${\cal N} = 2$ SCFT with $\grU(1)$ flavor symmetry}{N = 2 SCFT with U(1) flavor symmetry}}
\label{SETUP}

As mentioned in the Introduction, we are interested in what constraints the mass-deformed ${\rm S}^4$ partition function with a circular defect insertion imposes on correlation functions of the SCFT\@.  In this section, we begin by defining the setup for part of this question, namely ${\cal N} = 2$ SCFTs with $\grU(1)$ flavor symmetry and the corresponding mass deformation on ${\rm S}^4$, first without the defect.  We will include the defect in Sections~\ref{sec:line} and~\ref{sec:massivedefect}.
 
Throughout this paper, we will adopt the conventions of \cite{Freedman:2012zz,Lauria:2020rhc}. 
In particular, since we will discuss the SCFT on curved manifolds related by Weyl transformations from the flat spacetime, we adopt the notation of \cite{Lauria:2020rhc} that the frame indices are denoted by $a,b=0,\dots,3$ and the tangent indices are denoted by $\m,\n=0,\dots,3$.  We will use four-component fermions with their spinor indices suppressed.  The $\mathfrak{su}(2)_R$ R-symmetry indices are denoted by $i, j, \ldots$, and for the fermions, the placement of the R-symmetry indices will be correlated with the chirality of the fermions in a way that will be specified in each case.  In Lorentzian signature, charge conjugation is implemented by complex conjugation combined with the raising / lowering of all $\mathfrak{su}(2)_R$ indices of the fermions and of the spinor parameters.

\subsection{\texorpdfstring{${\cal N} = 2$ SCFTs and Weyl rescalings}{N = 2 SCFTs and Weyl rescalings}}
\label{sec:N2setup}

${\cal N} = 2$ SCFTs are invariant under the $\mf{su}(2, 2|2)$ superalgebra, which is a real form of $\mf{su}(4|2)$ that contains as a bosonic subalgebra $\mf{so}(4, 2) \times \mf{su}(2)_R \times \mf{u}(1)_R$. Tthe first factor is the conformal algebra and the latter two factors denote the R-symmetries.  The conformal group is generated by translations $P_a$, rotations $M_{ab}$, special conformal transformations $K_a$, and dilatation $D$.  We denote the $\mf{su}(2)_R$ and $\mf{u}(1)_R$ R-symmetry generators by $U_i{}^j$, with $i, j = 1, 2$, and $T$, respectively.  The fermionic generators of $\mathfrak{su}(2, 2|2)$ are the left-handed $Q_i$ and $S^i$ and the right-handed $Q^i$ and $S_i$.  See Table~\ref{SCGeneratorTable} for a list of generators of the $\mathfrak{su}(2, 2|2)$ algebra.

\begingroup
\renewcommand{\arraystretch}{1.3}
\begin{table}[htp]
	\begin{center}
		\begin{tabular}{c|c|c|c|c}
			generator & dimension & Lorentz rep & $\mf{su}(2)_R$ irrep & $\mf{u}(1)_R$ charge \\
			\hline
			$P_a$ & $1$ & $(\frac 12, \frac 12)$ & ${\bf 1}$ & $0$  \\
			$K_a$ & $-1$ & $(\frac 12, \frac 12)$ & ${\bf 1}$ & $0$  \\
			$D$ & $0$ & $(0, 0)$ & ${\bf 1}$ & 0  \\
			$M_{ab}$ & $0$ & $(1, 0) + (0, 1)$ & ${\bf 1}$ & 0  \\
			\hline
			$Q_i$ & $\frac 12$ & $(\frac 12, 0)$ & ${\bf 2}$ & $-\frac 12$  \\
			$Q^i$ & $\frac 12$ & $(0, \frac 12)$ & ${\bf 2}$ & $\frac 12$ \\
			$S_i$ & $\frac 12$ & $(0, \frac 12)$ & ${\bf 2}$ & $-\frac 12$ \\
			$S^i$ & $\frac 12$ & $(\frac 12, 0)$ & ${\bf 2}$ & $\frac 12$ \\
			\hline
			$U_i{}^j$ & $0$ & $(0, 0)$ & ${\bf 3}$ & $0$ \\
			$T$ & $0$ & $(0, 0)$ & ${\bf 1}$ & $0$ 
		\end{tabular}
	\end{center}
	\caption{Generators of the superconformal algebra and their quantum numbers.  The Lorentz irreps $(\frac 12, 0)$ and $(0, \frac 12)$ denote left-handed and right-handed spinors, respectively.}\label{SCGeneratorTable}
\end{table}
\endgroup

CFTs (and in particular SCFTs) can be studied on flat space $\R^{1, 3}$, as is most often done, where the metric is $ds^2=-dt^2 + d\vec{x}^2$ and the standard frame is $e^a = dx^a$, with $x^a = (t, \vec{x})$.  But they can also be studied more generally on any conformally flat space in a canonical way, where the metric and frame are rescaled by appropriate powers of the Weyl factor $\Omega(t, \vec{x})$:
\es{ConfFlat}{
	ds^2 = \Omega(t, \vec{x})^{-2} (-dt^2 + d\vec{x}^2) \,, \qquad
	e^a = \Omega(t, \vec{x})^{-1} dx^a  \,.
}
On the rescaled space, a scalar primary operator $\phi(t, \vec{x})$ of dimension $\Delta$ is related to the corresponding flat space operator $\phi_\text{flat}(t, \vec{x})$ also by a rescaling,
\es{phiRelation}{
	\phi(t, \vec{x}) = \Omega(t, \vec{x})^{\Delta} \phi_\text{flat}(t, \vec{x})\,.
}
Consequently, going from flat space to the space \eqref{ConfFlat} is achieved by multiplying all the correlation functions by a factor of $\Omega(t_i, \vec{x}_i)^{\Delta_i}$ for every scalar primary operator $\phi_i$ of dimension $\Delta_i$ (and similarly for spinning primaries in the frame basis).  In this paper we will be primarily concerned with two particular examples of conformally flat spaces: ${\rm S}^4$ and ${\rm AdS}_2 \times {\rm S}^2$ (and also the Euclidean continuation of the latter, $\mH^2\times {\rm S}^2$).

For an SCFT, local operators belonging to the same supermultiplet are linked together by supersymmetry.   In analogy with infinitesimal rotations acting on the wavefunction of a  particle with spin in non-relativistic quantum mechanics, the supersymmetry transformations have ``orbital parts'' related to the fact that the spacetime point where the operator is inserted undergoes an infinitesimal change, and ``intrinsic parts'' which cannot be attributed to such a change.  The orbital parts can be absorbed into the coefficients of (in general, some other) intrinsic supersymmetry transformations.  (In the quantum mechanics example, the orbital part of an infinitesimal rotation is an infinitesimal translation with a position-dependent coefficient that depends on the parameters of the rotation.)  In flat space, Poincar\'e supersymmetry transformations have only an intrinsic part.  Superconformal transformations, on the other hand, have both an intrinsic part and an orbital part, the latter being a Poincar\'e supersymmetry transformation with position-dependent coefficients.\footnote{In radial quantization, the (super)conformal generators acting on operators at the origin have only an intrinsic part, and the orbital part arises when the generators act on operators away from the origin as a consequence of the (super)conformal algebra. }  We will see examples below.

Thus, as in \cite{Freedman:2012zz,Lauria:2020rhc}, we denote by $(Q_i, Q^i, S_i, S^i)$ the action of only the intrinsic part of the supersymmetry transformations, and define
\es{deltaDef}{
	\delta \equiv \bar \epsilon^i Q_i + \bar \epsilon_i Q^i + \bar \eta^i S_i + \bar \eta_i S^i \,,
}
with anti-commuting spinor parameters $(\epsilon^i, \epsilon_i, \eta^i, \eta_i)$ of the same chirality as the supercharges they multiply.  
Note that $\bar\lambda\equiv \lambda^T C$ denotes the Majorana conjugate as in \cite{Freedman:2012zz} with charge conjugation matrix $C$;  $\bar \lambda$ should not be confused with the Dirac conjugate.
 Because intrinsic transformations must be accompanied by their corresponding orbital parts, the parameters $(\epsilon^i, \epsilon_i, \eta^i, \eta_i)$ cannot be arbitrary. 
 Instead, they are constrained by the requirement that $\delta$ in \eqref{deltaDef} should be a symmetry of the SCFT\@.

To deduce what such a requirement implies (on the conformally flat space \eqref{ConfFlat}), we can couple the SCFT to a flat conformal supergravity background with vanishing background fields except for the metric and frame, which are assumed to take on the values in \eqref{ConfFlat},\footnote{Conformal supergravity also has composite gauge fields, two of which are also non-zero in this case.  The first is the spin connection, which is identified with the spin connection of the frame \eqref{ConfFlat}.  The second is the gauge field $f_\mu{}^a$ for special conformal transformations.  We will not need the expression for the latter in this work.} and require that the variations of all fields in the background supergravity multiplet vanish.  One can see from Eq.~(20.69) of \cite{Freedman:2012zz} that the only non-trivial variations are those of the gravitini, and they read $\delta \psi_\mu^i = D_\mu \epsilon^i - \gamma_\mu \eta^i$ and $\delta \psi_{\mu\, i} = D_\mu \epsilon_i - \gamma_\mu \eta_i$, where $D_\mu$ is a covariant derivative.   The vanishing of these expressions gives the conformal Killing spinor equations
\es{epsDelta}{
	D_\mu \epsilon^i = \gamma_\mu \eta^i \,, \qquad D_\mu \epsilon_i = \gamma_\mu \eta_i \,.
}

In flat space ($\Omega = 1$), the solution of these equations is
\es{epsetaSolnFlat}{
	\epsilon_\text{flat}^i &= \alpha^i + x^a \gamma_a \beta^i \,, \qquad \eta_\text{flat}^i = \beta^i \,, \\
	\epsilon_{\text{flat}\, i} &= \alpha_i + x^a \gamma_a \beta_i \,, \qquad \eta_{\text{flat}\, i} = \beta_i \,,
}
where now $(\alpha^i, \alpha_i, \beta^i, \beta_i)$ are arbitrary constant complex parameters subject to chirality constraints (see Table~\ref{ChiralityParamTable}) and reality conditions $\alpha_i^* = B \alpha^i$, $\beta_i^* = B \beta^i$, where $B$ is a matrix that obeys $B^{-1} \gamma_a^* B = \gamma_a$ for all $a$ (see Chapter 5 of \cite{Freedman:2012zz}).   The parameters $(\alpha^i, \alpha_i, \beta^i, \beta_i)$ parameterize the infinitesimal $\mf{su}(2,2|2)$ superconformal symmetry transformations.  Thus, we can write
\es{deltaAgain}{
	\delta_\text{flat} = \bar \alpha^i Q_i + \bar \alpha_i Q^i
	+ \bar \beta^i \left( S_i - x^a \gamma_a Q_i \right) 
	+ \bar \beta_i \left( S^i - x^a \gamma_a Q^i \right)  \,.
}
This equation can be interpreted as saying that $(\alpha^i, \alpha_i)$ parameterize the full Poincar\'e supersymmetry transformations, which have only intrinsic parts.  By contrast, $(\beta^i, \beta_i)$ parameterize the full superconformal  transformations, which have both intrinsic and orbital parts.  Note that at $x^a = 0$, which is the only point on $\R^{1, 3}$ that is fixed under dilatations, the orbital parts vanish and we are only left with the intrinsic transformations.

\begingroup
\renewcommand{\arraystretch}{1.3}
\begin{table}[htp]
	\begin{center}
		\begin{tabular}{c|c|c}
			parameter & Lorentz rep & $\mf{u}(1)_R$  \\
			\hline
			$\epsilon^i$ & $(\frac 12, 0)$ & $\frac 12$ \\
			$\epsilon_i$ & $(0, \frac 12)$ & $-\frac 12$ \\
			$\eta^i$ & $(0, \frac 12)$ & $\frac 12$ \\
			$\eta_i$ & $(\frac 12, 0)$ & $-\frac 12$
		\end{tabular} \hspace{1in}
		\begin{tabular}{c|c|c}
			parameter & Lorentz rep & $\mf{u}(1)_R$  \\
			\hline
			$\alpha^i$ & $(\frac 12, 0)$ & $\frac 12$ \\
			$\alpha_i$ & $(0, \frac 12)$ & $-\frac 12$ \\
			$\beta^i$ & $(0, \frac 12)$ & $\frac 12$ \\
			$\beta_i$ & $(\frac 12, 0)$ & $-\frac 12$
		\end{tabular}
	\end{center}
	
	\caption{Assignments of Lorentz irreps and $\mf{u}(1)_R$ charges of the various supersymmetry parameters ensuring that the variation $\delta$ is a $\mf{u}(1)_R$ neutral (and $\mf{su}(2)_R$ neutral) Lorentz scalar. }\label{ChiralityParamTable}
\end{table}%
\endgroup

On the conformally flat space \eqref{ConfFlat}, the solution of \eqref{epsDelta} can be obtained from transforming the flat space solution as follows:\footnote{In deriving the transformation properties of $(\eta^i, \eta_i)$ it is useful to note that the spin connection associated with the frame \eqref{ConfFlat} is $\omega_{ab} = (\partial_a \log \Omega) dx_b - (\partial_b \log \Omega) dx_a$, where $dx_a = \eta_{ab} dx^b$, $\eta_{ab}$ being the flat space metric, and that the spinor covariant derivative is defined as $D_\mu \equiv  \partial_\mu + \frac 14 \omega_{\mu ab} \gamma^{ab}$.}
\es{epsRescaling}{
	\epsilon^i &= \Omega^{-1/2} \epsilon^i_\text{flat} \,, \qquad \
	\eta^i = \Omega^{1/2} \left( \eta^i_\text{flat} - \frac 12 \partial_a (\log \Omega) \gamma^a \epsilon^i_\text{flat} \right) \,, \\
	\epsilon_i &= \Omega^{-1/2} \epsilon_{i \, \text{flat}}  \,, \qquad
	\eta_i = \Omega^{1/2} \left( \eta_{i\, \text{flat}} - \frac 12 \partial_a (\log \Omega) \gamma^a \epsilon_{i\, \text{flat}} \right) \,,
}
and more explicitly given by
\es{epsConfFlat}{
	\epsilon^i &= \Omega^{-1/2} \Bigl(  \alpha^i + x^a \gamma_a \beta^i \Bigr)   \,, \qquad
	\eta^i = \Omega^{1/2} \left( \beta^i  - \frac {x^a \beta^i + \gamma^a \alpha^i + \gamma^a{}_b x^b \beta^i}{2} \partial_a (\log \Omega)  \right) \,, \\
	\epsilon_i &= \Omega^{-1/2} \Bigl(  \alpha_i + x^a \gamma_a \beta_i \Bigr)   
	\,, \qquad
	\eta_i = \Omega^{1/2} \left( \beta_i  - \frac {x^a \beta_i + \gamma^a \alpha_i + \gamma^a{}_b x^b \beta_i}{2} \partial_a (\log \Omega)  \right) \,.
}
Plugging this expression in \eqref{deltaDef} we see that, in general, on a conformally flat space both the Poincar\'e and superconformal transformations have orbital and intrinsic parts.  These transformations are still parameterized by $(\alpha^i, \alpha_i, \beta^i, \beta_i)$, which are space-time independent constants.

\subsection{Flavor current multiplet}
\label{FLAVOR}

We are interested in ${\cal N} =2$ SCFTs with a $\grU(1)$ flavor symmetry (e.g., an abelian subgroup of the full flavor symmetry).    The $\grU(1)$ flavor current $j_a$ belongs to a short $\mf{su}(2,2|2)$ multiplet referred to as a flavor current multiplet.  In addition to $j_a$, this multiplet contains an $\mf{su}(2)_R$ triplet of scalars $J_{ij}$ of dimension $2$, fermionic operators $\xi^i$ and $\xi_i$  of dimension $5/2$,  complex conjugate scalar operators $K$ and $\bar K$ of dimension $3$ that have opposite $\mf{u}(1)_R$ charges, as well as the conformal descendants of all these operators (see Table~\ref{ConservedCurrentTable}).  As written, the scalar operators $J_{ij}$ form the components of a rank-two symmetric tensor, and they satisfy the reality condition $(J_{ij})^* = J^{ij} = \varepsilon^{ik} \varepsilon^{jl} J_{kl}$.\footnote{We use the convention $\ve^{12}=\ve_{12}=1$ for the epsilon tensors that raise and lower the $\mf{su}(2)_R$ indices using the NW-SE convention \cite{Freedman:2012zz, Lauria:2020rhc}.}

\begingroup
\renewcommand{\arraystretch}{1.3}
\begin{table}[htp]
	\begin{center}
		\begin{tabular}{c|c|c|c|c}
			operator & $\Delta$ & Lorentz rep & $\mf{su}(2)_R$ irrep & $\mf{u}(1)_R$ charge \\
			\hline
			$J_{ij}$ & $2$ & $(0, 0)$ & ${\bf 3}$ & 0 \\
			$\xi^i$ & $\frac 52$ & $(\frac 12, 0)$ & ${\bf 2}$ & $-\frac 12$ \\
			$\xi_i$ & $\frac 52$ & $(0, \frac 12)$ & ${\bf 2}$ & $\frac 12$ \\
			$K$ & $3$ & $(0, 0)$ & ${\bf 1}$ & $-1$ \\
			$\bar K$ & $3$ & $(0, 0)$ & ${\bf 1}$ & $1$ \\
			$j_a$ & $3$ & $(\frac 12, \frac 12)$ & ${\bf 1}$ & $0$ 
		\end{tabular}
	\end{center}
	\caption{Conformal primary operators in the flavor current multiplet and their quantum numbers.  As in Table~\ref{SCGeneratorTable}, the Lorentz irreps $(\frac 12, 0)$ and $(0, \frac 12)$ denote left-handed and right-handed spinors, respectively.}\label{ConservedCurrentTable}
\end{table}%
\endgroup

For the conformal primaries of the flavor current multiplet, we can infer the transformation rules from (3.103) of \cite{Lauria:2020rhc}:\footnote{In (3.103) in \cite{Lauria:2020rhc}, the transformation rules of a tensor multiplet are given.  The tensor multiplet fields are a triplet of scalars $L_{ij}$, fermions $\varphi^i$, $\varphi_i$, complex scalars $G$ and $\bar G$, and an anti-symmetric tensor field $E_{\mu\nu}$.  This multiplet is the same as a flavor current multiplet for which one has solved the conservation condition for the current in terms of the field strength of an anti-symmetric tensor:  $j^\mu = \frac{1}{e} \varepsilon^{\mu\nu\rho\sigma} \partial_\nu E_{\rho \sigma}$.  Thus, we have the additional identifications $J_{ij} = L_{ij}$, $\xi_i = \varphi_i$, $\xi^i = \varphi^i$, $K = G$, $\bar K = \bar G$.}
\es{SUSYVars}{
	\delta J_{ij} &= \bar \epsilon_{(i} \xi_{j)}  + \varepsilon_{ik} \varepsilon_{jl} \bar \epsilon^{(k} \xi^{l)} \,, \\
	\delta \xi^i &= \frac 12 (\slashed{\partial} J^{ij})\, \epsilon_j
	+ \frac 12 \varepsilon^{ij} \slashed{j} \epsilon_j - \frac 12 K \epsilon^i + 2 J^{ij} \eta_j   \,, \\
	\delta \xi_i &= \frac 12 (\slashed{\partial} J_{ij})\, \epsilon^j 
	+ \frac 12 \varepsilon_{ij} \slashed{j} \epsilon^j - \frac 12 \bar K \epsilon_i + 2 J_{ij} \eta^j  \,, \\ 
	\delta K &= - \bar \epsilon_i \slashed{D} \xi^i 
	+ 2 \bar \eta_i \xi^i \,, \\
	\delta \bar K &= - \bar \epsilon^i \slashed{D} \xi_i 
	+ 2 \bar \eta^i \xi_i \,, \\
	\delta j^a &=   \frac 12 \bar \epsilon^i \gamma^{ab} (D_b \xi^j) \varepsilon_{ij}  
	+  \frac 12 \bar \epsilon_i \gamma^{ab} (D_b \xi_j) \varepsilon^{ij}  
	+  \frac 32 \bar \eta^i  \gamma^a \xi^j \varepsilon_{ij} 
	+  \frac 32 \bar \eta_i  \gamma^a \xi_j \varepsilon^{ij}  \,,
}
with $D_\mu$ being a covariant derivative and $D_a \equiv  e_a{}^\mu D_\mu$.  These transformations should be used only after plugging in the solutions of the conformal Killing spinor equations in \eqref{epsConfFlat}.

\subsection{Half-BPS SCFT deformations using flavor current operators}
\label{COUPLING}

Let us consider supersymmetry-preserving deformations of an SCFT by operators from the flavor current multiplet.  As shown in \cite{Cordova:2016xhm}, in flat space one can construct such a deformation that is half-BPS and in theories with Lagrangians such a deformation can be seen as introducing a (complex) mass parameter.  On the round sphere (after Euclidean continuation), one can also construct an analogous half-BPS deformation, but the supersymmetries being preserved will be different from those preserved in flat space.

An efficient way of studying deformations involving the flavor current multiplet is by coupling it to a {\em background} off-shell vector multiplet.  In ${\cal N} = 2$ superconformal theories, the off-shell vector multiplet consists of a gauge field $A_\mu$, as well as complex scalars $X$ and $\bar X$, left-handed and right-handed fermions $\Omega_i$ and $\Omega^i$, respectively, and a triplet of real scalars $Y_{ij}$ obeying $Y_{ij} = Y_{ji}$ and $(Y_{ij})^* = Y^{ij} = \varepsilon^{ik} \varepsilon^{jl} Y_{kl}$  (see Table~\ref{VectorMultipletTable}).  
\begingroup
\renewcommand{\arraystretch}{1.3}
\begin{table}[htp]
	\begin{center}
		\begin{tabular}{c|c|c|c|c}
			field & $\Delta$ & Lorentz rep & $\mf{su}(2)_R$ irrep & $\mf{u}(1)_R$ charge \\
			\hline
			$A_\mu$ & $1$ & $(\frac 12, \frac 12)$ & ${\bf 1}$ & $0$  \\
			$X$ & $1$ & $(0, 0)$ & ${\bf 1}$ & $1$ \\
			$\bar X$ & $1$ & $(0, 0)$ & ${\bf 1}$ & $-1$ \\
			$\Omega^i$ & $\frac 32$ & $(0, \frac 12)$ & ${\bf 2}$ & $-\frac 12$ \\
			$\Omega_i$ & $\frac 32$ & $(\frac 12, 0)$ & ${\bf 2}$ & $\frac 12$ \\
			$Y_{ij}$ & $2$ & $(0, 0)$ & ${\bf 3}$ & 0 \\
		\end{tabular}
	\end{center}
	\caption{Field content of the vector multiplet.}\label{VectorMultipletTable}
\end{table}%
\endgroup
The supersymmetry variations specialized to the case of a conformally flat space where the conformal Killing spinor equations \eqref{epsDelta} hold are:
\es{SUSYVarsVector}{
	\delta X &= \frac 12 \bar \epsilon^i \Omega_i \,, \\
	\delta \bar X &= \frac 12 \bar \epsilon_i \Omega^i \,, \\
	\delta \Omega_i &= \slashed{\partial} X \epsilon_i + \frac{1}{4} \gamma^{ab} F_{ab}^-  \varepsilon_{ij} \epsilon^j + Y_{ij} \epsilon^j + 2 X \eta_i \,, \\
	\delta \Omega^i &= \slashed{\partial} \bar X \epsilon^i + \frac{1}{4} \gamma^{ab} F_{ab}^+ \varepsilon^{ij} \epsilon_j + Y^{ij} \epsilon_j + 2 \bar X \eta^i \,, \\
	\delta A_\mu &= \frac 12 \varepsilon^{ij} \bar \epsilon_i \gamma_\mu \Omega_j 
	+ \frac 12 \varepsilon_{ij} \bar \epsilon^i \gamma_\mu \Omega^j \,, \\
	\delta Y_{ij} &= \frac 12 \bar \epsilon_{(i} \slashed{D} \Omega_{j)}
	+ \frac 12 \varepsilon_{ik} \varepsilon_{jl} \bar \epsilon^{(k} \slashed{D} \Omega^{l)} \,,
}
where $F_{\mu\nu} = \partial_\mu A_\nu - \partial_\nu A_\mu$ is the field strength, and $F^+_{ab}$ and $F^-_{ab}$ are its self-dual and anti-self-dual parts. 

One can check that the coupling (see~(3.34) of \cite{Binder:2021euo})
\es{SAJ}{
	S_{A-J} = \int d^4 x\, \sqrt{-g} \left( A_\mu j^\mu - X K - \bar X\, \bar K + J_{ij} Y^{ij} - \bar \xi^i \Omega_i  - \bar \xi_i \Omega^i \right)  
}
is a superconformal invariant, provided that both the fields of the vector multiplet and the operators of the flavor current multiplet are transformed appropriately.  (In particular, one can check that the supersymmetry variation of the integrand can be written as a total derivative.)  However, since we treat the vector multiplet as a background by giving its fields expectation values, superconformal symmetry will be broken by these expectation values.  The preserved supersymmetries can be found by setting to zero all the variations in \eqref{SUSYVarsVector} and solving for the parameters $(\alpha^i, \alpha_i, \beta^i, \beta_i)$ in the solution \eqref{epsConfFlat} to the Killing spinor equations.

As an example, the mass deformation in flat space is obtained by setting $X_\text{flat} = \mathfrak{m}/2$, $\bar X_\text{flat} = \bar{\mathfrak{m}}/2$, with the position-independent parameters $(\mathfrak{m}, \bar{\mathfrak{m}})$ being complex conjugates of each other, and setting to zero all other fields $A_{\text{flat}\, \mu} = Y_{\text{flat}\, ij} = \Omega_{\text{flat}\, i} = \Omega_\text{flat}^i = 0$.  The supersymmetry variations of the bosonic fields in \eqref{SUSYVarsVector} vanish automatically, while those of the fermions vanish provided that $\beta^i = \beta_i = 0$.  Thus, the flat space deformation
\es{FlatSpaceMass}{
	S_{\mathfrak{m}, \bar{\mathfrak{m}}, \text{flat}} = - \frac 12 \int d^4 x\,  \left(  \mathfrak{m} K_\text{flat} + \bar{\mathfrak{m}} \bar K_\text{flat}  \right)
} 
preserves half of the supersymmetries, namely the Poincar\'e supersymmetries generated by $(\alpha^i, \alpha_i)$ and breaks the superconformal symmetries.  It also breaks the dilatations, the special conformal transformations, and the $\mf{u}(1)_R$ symmetry (because $X$ and $\bar X$ carry non-zero $\mf{u}(1)_R$ charges), but it preserves translations, Lorentz transformations, and $\mf{su}(2)_R$.

The normalization of $(\mathfrak{m}, \bar{\mathfrak{m}})$ above is such that these are the usual mass parameters for a free hypermultiplet, provided that we take the $\mathfrak{u}(1)$ flavor symmetry to be that under which the two complex scalar fields comprising the hypermultiplet have charges $\pm 1$.  See Appendix~\ref{FREE}.

\subsection{\texorpdfstring{Half-BPS mass deformation on round ${\rm S}^4$}{Half-BPS mass deformation on round S4}}
\label{MASSS4}

Lastly, let us discuss the half-BPS mass deformation on a round ${\rm S}^4$, for which the partition function can be computed exactly using supersymmetric localization in Lagrangian theories.  ${\rm S}^4$ is a Euclidean manifold, so we should first perform a Euclidean continuation.  This is achieved by sending $t \to -i \tau$ (or $x^0 \to -i x^4$), $\gamma^0 \to -i \gamma^4$, $\gamma_0 \to i \gamma_4$, $A^0 \to -i A^0$, $A_0 \to i A_0$, etc.~in all the formulas.  The supersymmetry variations are always derived in Lorentzian signature first, and then continued to Euclidean signature using these rules.

The round ${\rm S}^4$ of radius $r$ has conformal factor 
\es{ConfFactorS4}{
	\Omega_\text{sphere}(x) = \frac{1 + x^2/r^2}{2} \,,
}
where $x = (\vec{x}, \tau)$ is a 4-component vector and its norm squared, $x^2$, is computed with the standard inner product on $\R^4$.  Thus, the expressions \eqref{epsConfFlat} for the conformal  Killing spinors become
\es{epsS4}{
	\epsilon_\text{sphere}^i &= \sqrt{2} \frac{ \alpha^i + x^a \gamma_a \beta^i }{\sqrt{1 + \frac{x^2}{r^2}}}   \,, \qquad
	\eta_\text{sphere}^i = \frac{1}{\sqrt{2}} \left( \frac {\beta^i - \frac{1}{r^2} x^a \gamma_a \alpha^i }{\sqrt{1 + \frac{x^2}{r^2}}}  \right) \,, \\
	\epsilon_{\text{sphere}\, i} &= \sqrt{2} \frac{ \alpha_i + x^a \gamma_a \beta_i }{\sqrt{1 + \frac{x^2}{r^2}}}   \,, \qquad
	\eta_{\text{sphere}\, i} = \frac{1}{\sqrt{2}} \left( \frac {\beta_i - \frac{1}{r^2} x^a \gamma_a \alpha_i }{\sqrt{1 + \frac{x^2}{r^2}}}  \right) \,.
}

The analog of the mass deformation \eqref{FlatSpaceMass} on ${\rm S}^4$ involves giving rotationally-invariant expectation values to the background vector multiplet fields on the sphere.  This implies that $A_{\text{sphere}\, \mu} = \Omega^i_\text{sphere} = \Omega_{\text{sphere}\, i} =  0$, since any expectation values for these fields would necessarily break rotational symmetry.  Unlike their flat space analogs, the sphere parameters $(\eta_\text{sphere}^i, \eta_{\text{sphere}\, i})$ depend non-trivially on position. Consequently if we were to give constant expectation values to $X_\text{sphere}$ and $\bar X_\text{sphere}$ and set $Y_{\text{sphere}\, ij}$ to zero, we would be breaking all supersymmetries since $\delta \Omega_\text{sphere}^i$ and $\delta \Omega_{\text{sphere}\, i}$ would not vanish for any choice of the $(\alpha^i, \alpha_i, \beta^i, \beta_i)$ parameters.  We are thus forced to give a non-zero constant expectation value to the field $Y_{\text{sphere}\, ij}$.  Without loss of generality we can take this expectation value to be in the second direction in $\mathfrak{su}(2)_R$, 
\es{ExpValues}{
	X_\text{sphere} = \frac{\mathfrak{m}}{2} \,, \qquad \bar X_\text{sphere} = \frac{\bar{\mathfrak{m}}}{2} \,, \qquad
	Y_{\text{sphere}\, ij}  = \frac{1}{2} y (\tau_2)_{ij} \,,
}
for some constant $y$ that we will determine in terms of $(\mathfrak{m}, \bar{\mathfrak{m}})$.  Note that in the conventions of \cite{Freedman:2012zz}, we have $(\tau_2)_{ij} = \delta_{ij}$ and $(\tau_2)^{ij} = -\delta^{ij}$.  With these choices, the supersymmetry variations of the fermions become
\es{SUSYVarsVectorS4}{
	\delta \Omega_{\text{sphere}\, i} &=   \frac{1}{\sqrt{2}} \frac{ -y  \alpha^i + \mathfrak{m} \beta_i + x^a \gamma_a (-y  \beta^i - \frac{1}{r^2}  \mathfrak{m} \alpha_i ) }{\sqrt{1 + \frac{x^2}{r^2}}} \,, \\
	\delta \Omega_\text{sphere}^i   &=  \frac{1}{\sqrt{2}} \frac{ -y \alpha_i + \bar{\mathfrak{m}} \beta^i + x^a \gamma_a (-y  \beta_i - \frac{1}{r^2}  \bar{\mathfrak{m}} \alpha^i ) }{\sqrt{1 + \frac{x^2}{r^2}}} \,,
}
where the placement of the $\mf{su}(2)_R$ indices does not match between the various terms because $\mf{su}(2)_R$ is broken by the non-zero value of $Y_{\text{sphere}\, ij}$.  The vanishing of these variations gives four sets of equations that close within each set, namely
\es{alphabetaeqs}{
	\begin{cases}
		y  \alpha^i = \mathfrak{m} \beta_i  \,, \\
		y  \beta_i = - \frac{1}{r^2}  \bar{\mathfrak{m}} \alpha^i  \,, 
	\end{cases} \qquad
	\begin{cases}
		y \alpha_i = \bar{\mathfrak{m}} \beta^i  \,, \\
		y  \beta^i = - \frac{1}{r^2}  \mathfrak{m} \alpha_i \,,
	\end{cases}
}
for $i=1,2$.  All these equations have non-trivial solutions provided that $y^2 =  - \mathfrak{m} \bar{\mathfrak{m}} / r^2$, or $y = \pm i \sqrt{\mathfrak{m} \bar{\mathfrak{m}}} / r$.

Thus, we have two choices for our supersymmetric vector multiplet background given by
\es{ExpValuesSolved}{
	X_\text{sphere} = \frac{1}{2} \mathfrak{m} \,, \qquad \bar X_\text{sphere} = \frac{1}{2} \bar{\mathfrak{m}} \,, \qquad
	Y_{\text{sphere}\, ij}  = \pm i (\tau_2)_{ij}  \frac{\sqrt{\mathfrak{m} \bar{\mathfrak{m}}}}{2r} \,.
}
Plugging these values in the Euclidean version of \eqref{SAJ}, we obtain the mass deformation on the sphere:
\es{SmSphere}{
	S_{\mathfrak{m}, \bar{\mathfrak{m}}, \text{sphere}} =  \frac 12 \int d^4 x\, \sqrt{g_\text{sphere}} \left(   \mathfrak{m} K_\text{sphere} + \bar{\mathfrak{m}} \bar K_\text{sphere} \pm i \frac{\sqrt{\mathfrak{m} \bar{\mathfrak{m}}}}{r} (J_{\text{sphere}\, 11} + J_{\text{sphere}\, 22})  \right)  \,,
}
where the overall sign difference with \eqref{FlatSpaceMass} comes from the fact that the Euclidean action equals minus the analytically continued Lorentzian action.

The deformation \eqref{SmSphere} of the SCFT Lagrangian on ${\rm S}^4$ preserves the $\mf{so}(5)$ rotational symmetry of the sphere, the $\mf{u}(1)$ subalgebra of $\mf{su}(2)_R$ that is generated by $\tau_2$, as well as half of the supersymmetries of the superconformal field theory because Eqs.~\eqref{alphabetaeqs} have an eight-parameter family of solutions.  Together, these symmetries generate the superalgebra $\mf{osp}(2|4)$.  On the other hand, the deformation \eqref{SmSphere} breaks all other conformal symmetries, it breaks $\mf{su}(2)_R$ to a $\mf{u}(1)$ subalgebra as we just mentioned, and it also breaks $\mf{u}(1)_R$.

\section{\texorpdfstring{Line defect in ${\cal N} = 2$ SCFT with flavor symmetry}{Line defect in N = 2 SCFT with flavor symmetry}}\label{sec:line}

In the previous section we introduced ${\cal N} = 2$ SCFTs and their mass deformations, both in flat space and on a round ${\rm S}^4$.  Let us now turn to discussing half-BPS line defects in ${\cal N} = 2$ SCFTs, starting in flat space.  In Section~\ref{CONFORMAL}, we will start with a review of conformal line defects and correlation functions of local operators in the presence of a defect in a general non-supersymmetric CFT\@.  Then, in Section~\ref{SUSYDEFECT}, we will discuss the case of a half-BPS defect in ${\cal N} =2$ SCFTs.   We will continue in Section~\ref{CORRELDEFECT} with the correlation functions of flavor current multiplet operators in the presence of such a defect.  Lastly, we will end in Section~\ref{ADS2} with how a Weyl rescaling to ${\rm AdS}_2 \times {\rm S}^2$ simplifies these correlation functions.

Throughout this section, we will work in a Lorentzian signature, with a time-like defect located at a fixed point in space.  We will eventually be interested in continuing our results to Euclidean signature, and for this reason we will restrict our attention to local operator insertions in the presence of the defect that are space-like separated from each other (but not from the defect).

\subsection{Time-like line defect in 4d CFT}
\label{CONFORMAL}

Let us start with the general non-supersymmetric case.  Let us denote the coordinate along the defect by $x^0 = t$ and the transverse coordinates by $\vec{x} = (x^1, x^2, x^3)$, and assume the defect is located at $\vec{x} = 0$.  Such a defect breaks the conformal algebra according to
\es{BreakingConformal}{
	\mf{so}(4, 2) \quad \longrightarrow \quad \mf{so}(1,2) \times \mf{so}(3) \,,
}
where the $\mf{so}(1,2)$ is generated by the dilatation and by the translation and special conformal transformation in the time direction, and the $\mf{so}(3)$ acts by rotating $\vec{x}$ as a fundamental vector.  In terms of the conformal generators introduced at the beginning of Section~\ref{sec:N2setup} (see also Table~\ref{SCGeneratorTable}), the breaking pattern \eqref{BreakingConformal} is
\es{BreakingConformalAlgebra}{
	\{ P_a, K_a, D, M_{ab} \} \quad \longrightarrow \quad \{ D, P_0, K_0\} \times \{M_{12}, M_{23}, M_{31} \} \,.
}

In the presence of the defect, we denote expectation values by $\langle \cdots \rangle_\mathbb{L}$, where $\mathbb{L}$ denotes the line defect.  If we want to refer to expectation values without the defect, we will omit the subscript $\mathbb{L}$.  The expectation values are normalized such that $\langle 1 \rangle_\mathbb{L} = 1$.  Correlators of local operators are restricted by the $\mf{so}(1,2) \times \mf{so}(3)$ symmetry.  For instance, the expectation value (one-point function) of a scalar primary operator $\phi_\text{flat}(t, \vec{x})$ of dimension $\Delta$ is
\es{OnePt}{
	\langle \phi_\text{flat}(t, \vec{x}) \rangle_\mathbb{L} = \frac{a_{\phi, \mathbb{L}}}{\abs{\vec{x}}^\Delta} \,,
}
for some numerical coefficient  $a_{\phi, \mathbb{L}}$, and the expectation values of operators of odd or half-integer spin vanish.  The two-point functions are expressed in terms of functions of two conformally-invariant cross ratios.  For instance, the two-point function of scalar primary operators $\phi_{\text{flat}\, i}(t_i, \vec{x}_i)$ with dimensions $\Delta_i$ is
\es{TwoPt}{
	\langle \phi_{\text{flat}\, 1}(t_1, \vec{x}_1) \phi_{\text{flat}\, 2}(t_2, \vec{x}_2) \rangle_\mathbb{L} = \frac{F_{\phi_1 \phi_2, \mathbb{L}} (\xi, \eta)}{\abs{\vec{x}_{1}}^{\Delta_1} \abs{\vec{x}_{2}}^{\Delta_2}} \,,
}
where $F_{\phi_1 \phi_2, \mathbb{L}}$ is an arbitrary function of $\mf{so}(1,2) \times \mf{so}(3)$ invariants $\xi$ and $\eta$ that can be constructed from the coordinates $(t_i, \vec{x}_i)$.  A convenient choice for our purposes will be
\es{xietaDef}{
	\xi = \frac{ -(t_1 - t_2)^2 + \abs{\vec{x}_{1}}^2 + \abs{\vec{x}_{2}}^2}{ 2\abs{\vec{x}_{1}}  \abs{\vec{x}_{2}}} \,, \qquad
	\eta = \frac{\vec{x}_{1} \cdot \vec{x}_{2}}{ \abs{\vec{x}_{1}}  \abs{\vec{x}_{2}}} \,,
}
whose interpretation will become clearer below.

\subsection{\texorpdfstring{Half-BPS line defect in ${\cal N} = 2$ SCFT}{Half-BPS line defect in N = 2 SCFT}}
\label{SUSYDEFECT}

In an SCFT, the defect breaks the superconformal algebra down to a subalgebra that can potentially be larger than $\mf{so}(1,2) \times \mf{so}(3)$, depending on whether or not the defect preserves some supersymmetry.  In an ${\cal N} = 2$ SCFT, the largest subalgebra of the superconformal algebra $\mf{su}(2, 2|2)$ that a time-like line defect can preserve is $\mf{osp}(4^*|2)$, 
which is a half-BPS subalgebra containing 8 real supercharges, and this is the unique superconformal algebra that contains the conformal subalgebra in \eqref{BreakingConformal}
(see \cite{Agmon:2020pde} for details).\footnote{The fact that this is the largest amount of supersymmetry preserved can also be seen after mapping to ${\rm AdS}_2 \times {\rm S}^2$, where $\mf{osp}(4^*|2)$ is the maximal ${\cal N} = 2$ SUSY algebra.}  As far as the bosonic symmetry is concerned, the defect breaks the conformal and R-symmetry according to
\es{Breaking}{
	\mf{so}(4, 2) \times \mf{su}(2)_R \times \mf{u}(1)_R \quad \longrightarrow \quad  \mf{so}(1,2) \times \mf{so}(3) \times \mf{su}(2)_R \,.
} 
In particular, the defect preserves the $\mf{su}(2)_R$ R-symmetry of the ${\cal N} =2$ SCFT but completely breaks $\mf{u}(1)_R$.  This breaking is parameterized by an angle $\vartheta$, and the preserved supercharges depend on this angle.

To get some guidance for what a half-BPS defect looks like, let us consider a free ${\cal N} = 2$ Abelian gauge theory with dynamical vector multiplet fields $(A_\mu, X, \bar X, \Omega_i, \Omega^i)$, which obey the same supersymmetry algebra as the background vector multiplet considered in Section~\ref{COUPLING}.  In this theory, which is an ${\cal N} = 2$ SCFT, we can consider a time-like Wilson line defect
\es{WilsonLine}{
	W = \exp \left[ i q \int dt \left( A_{\text{flat}\, 0} + \lambda X_\text{flat} + \bar \lambda\, {\bar X}_\text{flat} \right) \right] \,,
}
where in addition to the gauge field we also have terms in the exponent proportional to the vector multiplet scalars $X_\text{flat}$ and ${\bar X}_\text{flat}$ with coefficients $\lambda$ and $\bar \lambda$, respectively, that are each other's complex conjugates.   The usual Wilson line is that with $\lambda = \bar \lambda = 0$.  From \eqref{SUSYVarsVector}, the SUSY variation of the integrand in the exponent is
\es{deltaExponent}{
	\delta (A_{\text{flat}\, 0} + \lambda X_\text{flat} + \bar \lambda \,{\bar X}_\text{flat}) =  \frac 12 \bar \Omega_{\text{flat}\, i} \left( \lambda  \epsilon_\text{flat}^i  -  \varepsilon^{ij}  \gamma^0 \epsilon_{\text{flat}\, j}  \right) 
	+ \frac 12 \bar \Omega_\text{flat}^i \left( \lambda  \epsilon_{\text{flat}\, i}  -  \varepsilon_{ij} \gamma^0 \epsilon_\text{flat}^j  \right)  \,.
}
This quantity vanishes provided that  $\epsilon_\text{flat}^i = \lambda^{-1} \varepsilon^{ij}  \gamma^0 \epsilon_{\text{flat}\, j}$ and $\epsilon_{\text{flat}\, i} = \bar \lambda^{-1} \varepsilon_{ij}  \gamma^0 \epsilon_\text{flat}^j$.  If $\abs{\lambda} \neq 1$, these equations have no non-trivial solutions, and in this case the Wilson line \eqref{WilsonLine} is non-supersymmetric.  On the other hand, if $\lambda = e^{i \vartheta}$ for some angle $\vartheta$, \eqref{epsetaSolnFlat} implies
\es{OSpParams}{
	\alpha^i &= e^{-i \vartheta} \varepsilon^{ij} \gamma^0 \alpha_j \,, \qquad
	\beta^i = e^{-i \vartheta} \varepsilon^{ij} \gamma^0 \beta_j \,.
}
These are $8$ equations for $16$ variables, and they have an $8$-parameter family of solutions.  In this case, the Wilson line \eqref{WilsonLine} is half-BPS\@.  

From \eqref{deltaAgain}, we see that the supercharges preserved by the time-like defect on $\R^{1,3}$ are 
\es{SuperchargesPreserved}{
	Q^i - e^{-i \vartheta} \varepsilon^{ij} \gamma^0 Q_j \,, \qquad
	S^i - e^{-i \vartheta} \varepsilon^{ij} \gamma^0 S_j \,, \qquad
	i = 1, 2\,.
}
These eight real super(conformal)charges together with the generators in \eqref{BreakingConformalAlgebra} and the $\mf{su}(2)_R$ generators $U_i{}^j$, generate $\mf{osp}(4^*|2)$.  While \eqref{deltaAgain} with arbitrary $(\alpha^i, \alpha_i, \beta^i, \beta_i)$ correspond to the fermionic generators of $\mf{su}(2,2|2)$, the restriction to $\mf{osp}(4^*|2)$ is imposed by \eqref{OSpParams}.

While we derived \eqref{OSpParams} and \eqref{SuperchargesPreserved} for the Wilson line \eqref{WilsonLine} in the free Maxwell theory, \eqref{OSpParams} and the preserved supercharges in \eqref{SuperchargesPreserved} are more general properties of half-BPS defects that extend in the time direction.  We may consider them to be the definitions of the $\mf{osp}(4^*|2)$-preserving defects we study in this paper.

\subsection{Correlators of flavor current operators in presence of defect}
\label{CORRELDEFECT}

In an SCFT, correlation functions of local operators in the presence of a line defect take the same form as in the non-supersymmetric case discussed in Section~\ref{CONFORMAL}, with the only difference being that certain correlation functions may be related to one another via superconformal Ward identities.  Let us focus on the one- and some of the two-point functions of flavor current multiplet operators introduced in Section~\ref{FLAVOR} in the presence of a half-BPS line defect.  The only conformal primary operators from the flavor current multiplet that can have non-zero expectation values are the scalars $K_\text{flat}$ and $\bar K_\text{flat}$:
\es{OnePoint}{
	\langle K_\text{flat}(t, \vec{x}) \rangle_\mathbb{L} &= \frac{a_{\mathbb{L}}}{\abs{\vec{x}}^3}  \,, \qquad \langle  \bar K_\text{flat}(t, \vec{x}) \rangle_\mathbb{L} = \frac{\bar a_{\mathbb{L}}}{{\abs{\vec{x}}^3}} \,, \\
	\langle  J_{\text{flat}\, ij} \rangle_\mathbb{L} &= \langle  \xi_\text{flat}^i \rangle_\mathbb{L} 
	= \langle \xi_{\text{flat}\, i} \rangle_\mathbb{L} = \langle j_\text{flat}^a \rangle_\mathbb{L} = 0 \,,
}  
where $a_\mathbb{L}$ and $\bar a_\mathbb{L}$ are numerical coefficients and the vanishing of the expectation values of $ J_{\text{flat}\, ij}$, $\xi_\text{flat}^i$, and $\xi_{\text{flat}\, i}$ is due to $\mf{su}(2)_R$ symmetry, while the expectation value of $ j_\text{flat}^a$ vanishes due to the $\mf{so}(1,2) \times \mf{so}(3)$ symmetry.  As we will see in Section~\ref{WARD}, supersymmetry Ward identities further imply that $a_\mathbb{L} = \bar a_\mathbb{L} = 0$.

The non-zero two-point functions of the scalar operators $J_{\text{flat}\, ij}$, $K_\text{flat}$, and $\bar K_\text{flat}$ take the form
\es{TwoPoint}{
	\langle  J_{\text{flat}\, ij}(t_1, \vec{x}_1)  J_{\text{flat}\, kl}(t_2, \vec{x}_2) \rangle_\mathbb{L}
	&= \frac{\varepsilon_{ik} \varepsilon_{jl} + \varepsilon_{il} \varepsilon_{jk}}{2} \frac{A_\mathbb{L}(\xi, \eta)}{\abs{\vec{x}_1}^2 \abs{\vec{x}_2}^2}   \,, \\
	\langle K_\text{flat}(t_1, \vec{x}_1)  K_\text{flat}(t_2, \vec{x}_2) \rangle_\mathbb{L} &= \frac{B_\mathbb{L}(\xi, \eta)}{\abs{\vec{x}_1}^3 \abs{\vec{x}_2}^3} \,, \\
	\langle {\bar K}_\text{flat}(t_1, \vec{x}_1) {\bar K}_\text{flat}(t_2, \vec{x}_2) \rangle_\mathbb{L} &= \frac{{\bar B}_\mathbb{L}(\xi, \eta)}{\abs{\vec{x}_1}^3 \abs{\vec{x}_2}^3} \,, \\
	\langle K_\text{flat}(t_1, \vec{x}_1) {\bar K}_\text{flat}(t_2, \vec{x}_2) \rangle_\mathbb{L} &= \frac{C_\mathbb{L}(\xi, \eta)}{\abs{\vec{x}_1}^3 \abs{\vec{x}_2}^3} \,,
}
for some functions $A_\mathbb{L}$, $B_\mathbb{L}$, ${\bar B}_\mathbb{L}$, and $C_\mathbb{L}$ of the invariants $\xi$ and $\eta$ in \eqref{xietaDef}.  For a defect that preserves charge conjugation symmetry, the functions $A_\mathbb{L}$ and $C_\mathbb{L}$ are real and ${\bar B}_\mathbb{L}$ is the complex conjugate of $B_\mathbb{L}$.  If the defect does not preserve charge conjugation symmetry, as will be the case of the Wilson line we study in Section~\ref{sec:sugra}, then, in general,  $A_\mathbb{L}$ and $C_\mathbb{L}$ are complex and ${\bar B}_\mathbb{L}$ is not the complex conjugate of $B_\mathbb{L}$.    We will determine the relations between these functions in Section~\ref{WARD}.

\subsection{\texorpdfstring{Weyl transformation to ${\rm AdS}_2 \times {\rm S}^2$}{Weyl transformation to AdS2 x S2}}
\label{ADS2}

It was pointed out starting with the work of \cite{Kapustin:2005py} that a natural conformally flat space for studying a time-like line defect is ${\rm AdS}_2 \times {\rm S}^2$, obtained by a Weyl rescaling that sends the line defect to the boundary of ${\rm AdS}_2$.  In our case, this is\footnote{Since quantities that are defined on ${\rm AdS}_2 \times {\rm S}^2$ (or its Euclidean continuation $\HH^2 \times {\rm S}^2$) appear a significant number of times below, we will not indicate this fact with a subscript as we did above for quantities defined on flat space or the round sphere.}
\es{WeylFactor}{
	\Omega(t, \vec{x}) = \abs{\vec{x}} \,.
}
Indeed, in this case the metric \eqref{ConfFlat} becomes
\es{AdS2S2Metric}{
	ds^2 = \frac{-dt^2 + d\vec{x}^2}{\abs{\vec{x}}^2} 
	= \frac{-dt^2 + dz^2}{z^2} 
	+ d\theta^2 + \sin^2 \theta \, d\phi^2 \,, 
}
where the first term on the RHS is the ${\rm AdS}_2$ metric in Poincar\'e coordinates, while the second term is the ${\rm S}^2$ metric in the standard spherical coordinates.  The ``radial'' direction $z$ of ${\rm AdS}_2$ is identified with the transverse distance $z = \abs{\vec{x}}$ to the defect, and the conformal boundary is located at $z=0$. The transverse ${\rm S}^2$ is embedded in the $\R^3$  parametrized by $\vec x$ via the unit vector $ \hat n \equiv {\vec{x} \over \abs{\vec{x}}} = (\sin \theta \cos \phi, \sin \theta \sin \phi, \cos \theta)$.  The coordinates used to parameterize ${\rm AdS}_2 \times {\rm S}^2$ are of course up to us, and we could equally well choose $(t, \vec{x})$ or $(t, z, \theta, \phi)$.

The advantage of ${\rm AdS}_2 \times {\rm S}^2$ over $\R^{1,3}$ is that the $\mf{so}(1,2) \times \mf{so}(3)$ residual conformal symmetry is realized as isometries of ${\rm AdS}_2 \times {\rm S}^2$, a fact that will become very useful in our derivation of the supersymmetric Ward identities in the next section.  A related simplification on ${\rm AdS}_2 \times {\rm S}^2$ is that correlation functions in the presence of the defect become simpler.   Indeed, after the rescaling \eqref{phiRelation}, we see from \eqref{OnePt} that the one-point functions on ${\rm AdS}_2 \times {\rm S}^2$ become completely independent of position:
\es{OnePtAdS2S2}{
	\langle \phi(t, \vec{x}) \rangle_\mathbb{L} = a_{\phi, \mathbb{L}} \,.
}
Similarly, the two-point functions of scalar operators are just functions of the $\mf{so}(1,2) \times \mf{so}(3)$ invariants $\xi$ and $\eta$  defined in \eqref{xietaDef}:
\es{TwoPtAdS2S2}{
	\langle \phi_1(t_1, \vec{x}_1) \phi_2(t_2, \vec{x}_2) \rangle_\mathbb{L} = F_{\phi_1 \phi_2, \mathbb{L}} (\xi, \eta) \,.
}
In fact, the invariants $\xi$ and $\eta$ have simple interpretations in ${\rm AdS}_2 \times {\rm S}^2$.  The invariant $\eta$  is the inner product of the two unit vectors in $\R^3$ that correspond to the points on ${\rm S}^2$, $\eta  = \hat n_1 \cdot \hat n_2$.  The invariant $\xi$ has a similar interpretation for the ${\rm AdS}_2$ factor.  If we define the embedding coordinates of ${\rm AdS}_2$ in $\R^{2, 1}$ as $X = \left( \frac{t}{z}, \frac{1 + t^2 - z^2 }{2 z}, \frac{1 - t^2 + z^2 }{2 z} \right)$, with inner product in the embedding space taken using the flat metric with signature $(-, +, -)$, then one can check that $\xi = - X_1 \cdot X_2$. 

For the flavor current multiplet on ${\rm AdS}_2 \times {\rm S}^2$, we thus have the one-point functions
\es{OnePointAdS}{
	\langle K \rangle_\mathbb{L} &= a_\mathbb{L} \,, \qquad \langle  \bar K \rangle_\mathbb{L} = {\bar a}_\mathbb{L} \,, \qquad
	\langle  J_{ij} \rangle_\mathbb{L} = \langle  \xi^i \rangle_\mathbb{L} 
	= \langle \xi_{i} \rangle_\mathbb{L} = \langle j^a \rangle_\mathbb{L} = 0 \,,
} 
and the two-point functions 
\es{TwoPointAdS}{
	\langle  J_{ij}(t_1, \vec{x}_1)  J_{kl}(t_2, \vec{x}_2) \rangle_\mathbb{L}
	&= \frac{\varepsilon_{ik} \varepsilon_{jl} + \varepsilon_{il} \varepsilon_{jk}}{2} A_\mathbb{L}(\xi, \eta)  \,, \qquad \langle K(t_1, \vec{x}_1) {\bar K}(t_2, \vec{x}_2) \rangle_\mathbb{L} = C_\mathbb{L}(\xi, \eta) \,, \\
	\langle K(t_1, \vec{x}_1)  K(t_2, \vec{x}_2) \rangle_\mathbb{L} &= B_\mathbb{L}(\xi, \eta) \,, \qquad \qquad \qquad \qquad
	\langle {\bar K}(t_1, \vec{x}_1) {\bar K}(t_2, \vec{x}_2) \rangle_\mathbb{L} = {\bar B}_\mathbb{L}(\xi, \eta) 
	\,.
}

Note that from \eqref{epsConfFlat}, we can identify the conformal Killing spinors on ${\rm AdS}_2 \times {\rm S}^2$:
\es{epsAdS}{
	\epsilon^i &= \frac{ \alpha^i + x^a \gamma_a \beta^i}{\sqrt{\abs{\vec{x}}}}  \,, \qquad
	\epsilon_i = \frac{ \alpha_i + x^a \gamma_a \beta_i}{\sqrt{\abs{\vec{x}}}}  \,.
}  
The expressions for $(\eta^i, \eta_i)$ can be calculated from the equations $\slashed{D} \epsilon^i = 4 \eta^i$ and $\slashed{D} \epsilon_i = 4 \eta_i$.  While arbitrary $(\alpha^i, \alpha_i, \beta^i, \beta_i)$ parameterize the supersymmetries of $\mf{su}(2, 2|2)$ preserved by the defect, in order to restrict to the supersymmetries of $\mf{osp}(4^*|2)$ we should choose these parameters so that they obey \eqref{OSpParams}.

\section{Supersymmetric mass deformation with defect}
\label{sec:massivedefect}

We will now combine the two ingredients introduced thus far, namely the mass deformation of ${\cal N} = 2$ SCFTs on ${\rm S}^4$ and the half-BPS defect of ${\cal N} = 2$ SCFTs.  In the previous section we introduced the half-BPS defect in Lorentzian signature where it was extended along the time direction, first on $\R^{1, 3}$ and then on ${\rm AdS}_2 \times {\rm S}^2$.  One can of course analytically continue this setup to Euclidean signature by sending $x^0 \to -i x^4$, and one can perform a Weyl transformation to any other conformally flat space.  In particular, on ${\rm S}^4$, if we take the Weyl factor to be that in \eqref{ConfFactorS4}, the defect becomes extended along the great circle parameterized by $x^4$ at $x^1=x^2=x^3=0$.

The particular case of a circular Wilson loop was studied in \cite{Pestun:2007rz} and in many other follow-ups in general ${\cal N} = 2$ SCFTs that have Lagrangian gauge theory descriptions in terms of vector multiplets and hypermultiplets.  In these theories, Ref.~\cite{Pestun:2007rz} showed that the expectation value of the circular Wilson loop can be computed exactly using the technique of supersymmetric localization.  It was also noticed in \cite{Pestun:2007rz} that it is possible to introduce a mass parameter $m$ for the hypermultiplets while preserving the supercharge used for localization, and the circular Wilson loop is still invariant under this supercharge.  This fact suggests that, for a general half-BPS defect, it may be possible for the mass deformation on ${\rm S}^4$ and the defect to share at least one supercharge in common.  Moreover, as we will see, derivatives of the mass deformed ${\rm S}^4$ partition function with the defect insertion are related to integrated connected correlators of flavor multiplet operators in the presence of the defect.

In Section~\ref{EXPECTATION} we start by clarifying what we mean by expectation values and define the notion of connected correlators when a defect is present.  Then, in Section~\ref{SPHEREMASS} we relate the mass parameter $m$ mentioned above to the mass parameters $\mathfrak{m}$ and $\bar{\mathfrak{m}}$ introduced in Section~\ref{MASSS4}.  In Section~\ref{DERIVATIVES} we will relate more concretely derivatives of the mass-deformed ${\rm S}^4$ partition function to integrated correlators of flavor multiplet operators.  Lastly, in Section~\ref{SIMPLIFICATION} we will provide a first simplification of these integrated correlators.

\subsection{Expectation values and correlation functions}
\label{EXPECTATION}

First, let us discuss the definitions of expectation values and correlation functions used in this work. We start with correlation functions of local point-like operators.  Usually, when a QFT is defined via the path integral, one can compute correlation functions by explicitly inserting operators in the path integral.  For instance, in a free theory, one can compute correlation functions by performing Wick contractions.  In general, such a procedure must be supplemented by a regularization and renormalization prescription.  A more abstract way of defining correlators is to couple local operators to position-dependent sources and take functional derivatives of the partition function with respect to these sources.  The aforementioned choices of the regularization and renormalization prescription can be rephrased in terms of local terms in the sources that can be added to the generating functional.  If we want the correlation functions to preserve a given symmetry of the theory, we should ensure that the partition function, seen as a functional of the sources, has the corresponding symmetry.  Equivalently, one has to choose the contact terms in the correlation functions appropriately in order for the symmetry Ward identities to hold.

This is a sufficiently important point to warrant considering a non-supersymmetric example for illustration. 
Consider the theory of a free massless complex scalar $\phi$ in Euclidean flat space.  This theory has a $\grU(1)$ symmetry under which $\phi$ gets multiplied by a phase.  We can introduce a background gauge field $A_\mu$ for this $\grU(1)$ global symmetry, as well as a source $\lambda$ for the operator $\abs{\phi}^2$.  A choice of the partition function that is invariant under background gauge transformations of $A_\mu$ is
 \es{ActionFreeScalar}{
  Z[A_\mu, \lambda] = \int D\phi \,  e^{ - \int d^4 x\, \bigl[ \abs{\partial_\mu \phi  - i A_\mu \phi}^2 + \lambda \abs{\phi}^2 \bigr]}
   =  \int D\phi \,   e^{- \int d^4x \, \bigl[ \abs{\partial_\mu \phi}^2 + A^\mu j_\mu + A_\mu A^\mu \abs{\phi}^2 
     + \lambda \abs{\phi}^2 \bigr]} \,,
 }
where the $\grU(1)$ current is $j_\mu = i (\phi^* \partial_\mu \phi - \phi \partial_\mu \phi^*)$, and where the operators appearing in the exponents are assumed to be normal ordered (i.e.,~Wick self-contractions are excluded when computing correlators).    We can now consider the three-point function $I^{\mu\nu}(x, y, z) = \langle j^\mu(x) j^\nu(y) \abs{\phi}^2(z) \rangle$, first computed using Wick contractions, and then using functional differentiation from \eqref{ActionFreeScalar}.  

Performing Wick contractions with the scalar propagator $G(x, y) = \frac{1}{4 \pi^2 (x - y)^2}$, we obtain
 \es{IOpIns}{
  I_\text{Wick}^{\mu\nu}(x, y, z) &= - 2 \frac{\partial G(x, y)}{\partial x^\mu} \frac{\partial G(y, z)}{\partial y^\nu} G(x, z)
   -2 \frac{\partial G(x, z)}{\partial x^\mu} \frac{\partial G(x, y)}{\partial y^\nu} G(y, z)\\
    &{}+2\frac{\partial^2 G(x, y)}{\partial x^\mu \partial y^\nu} G(x, z) G(y, z) + 2 \frac{\partial G(x, z)}{\partial x^\mu} \frac{\partial G(y, z)}{\partial y^\nu} G(x, y) \,. 
 }
If we assume that this expression holds everywhere (which is a choice of regularization and renormalization prescription), one can check that the current conservation Ward identity is not obeyed.  Indeed, taking the divergence of this expression at $x$ and using $\nabla^2 G(x, y) = - \delta^{(4)}(x - y)$, we can easily check that
 \es{NotObeyed}{
  \frac{\partial}{\partial x^\mu} I^{\mu\nu}_\text{Wick}(x, y, z) &= 2  G(y, z)^2 \frac{\partial }{\partial x_\nu} \delta^{(4)}(x - y) \,,
 }
which does not vanish.   
On the other hand, from the gauge-invariant partition function \eqref{ActionFreeScalar}, it is clear that in order for the Ward identity to be obeyed, one has to supplement $\langle j^\mu(x) j^\nu(y) \abs{\phi}^2(z) \rangle$ with the contact term obtained by taking functional derivatives of the quadratic term in $A_\mu$ in \eqref{ActionFreeScalar} (also known as a ``seagull'' term).  Defining 
 \es{IFuncDiffDef}{
  I^{\mu\nu}(x, y, z) &= - \frac{1}{Z} \frac{\delta^3 Z}{ \delta A_\mu(x) \delta A_\nu(y) \delta \lambda(z)} \biggr|_{A_\mu = \lambda = 0} \\
   &= I_\text{Wick}^{\mu\nu}(x, y, z)
    - 2 \delta^{\mu\nu} \delta^{(4)}(x-y) \langle \abs{\phi}^2(y) \abs{\phi}^2(z) \rangle_\text{Wick} \,,
 }
and performing the Wick contractions for the additional term, we find 
 \es{IFuncDiff}{
  I^{\mu\nu}(x, y, z) 
   &= I_\text{Wick}^{\mu\nu}(x, y, z)
    - 2 \delta^{\mu\nu} \delta^{(4)}(x-y) G(y, z)^2 \,.
 }
With \eqref{NotObeyed}, it is now straightforward to see that the Ward identity is obeyed:
 \es{CurrentCons}{
   \frac{\partial}{\partial x^\mu} I^{\mu\nu}(x, y, z) &= 0 \,.
 }
The lesson is that current conservation holds generally if the correlation function is obtained by functional differentiation starting from a partition function that is invariant under background gauge transformations.  It may not hold otherwise.

Ending our digression, let us now assume that the correlators used in this work come from functional differentiation of a generating functional that is invariant under background conformal supergravity gauge transformations.  With this choice, the supersymmetry Ward identities will hold everywhere, including at coincident points.  If we want to relate the correlators used here to correlators in some other regularization and renormalization scheme, the correlators here may include several explicit coincident-point contributions coming from lower-point functions in that other scheme.\footnote{The importance of such coincident-point contributions for Ward identities of conformal symmetry has been emphasized in \cite{Osborn:1993cr,Erdmenger:1996yc}. For constraints from supersymmetry on contact terms, see for example\cite{Closset:2012vg,Closset:2012vp,Gomis:2015yaa,Gomis:2016sab,Papadimitriou:2017kzw,Schwimmer:2018hdl,Papadimitriou:2019gel,Katsianis:2019hhg,Katsianis:2020hzd,Bzowski:2020tue}.}

Note that in a specific regularization and renormalization scheme, the coupling \eqref{SAJ} is not the actual term we add to the action;  in general, we may need additional higher order terms in the background fields in order to ensure invariance under supersymmetry and background gauge transformations.  Instead, the coupling \eqref{SAJ} should be interpreted as a definition of the correlators involving flavor current multiplet operators.  In particular, (in Euclidean signature) we define the functional derivatives of $Z$ w.r.t.~$A_\mu$, $X$, $\bar X$, $Y^{ij}$, etc.~to correspond to the correlators of $-j^\mu$, $K$, $\bar K$, $-J_{ij}$, etc.  The supersymmetry Ward identities for the correlators defined this way follow from the fact that the partition function $Z[A_\mu, X, \bar X, Y^{ij}, \Omega_i, \Omega^i]$ is constructed so that it is invariant under the supersymmetry transformation rules \eqref{SUSYVarsVector} of the background vector multiplet.

Another point that we should clarify before we proceed is the definition of expectation values and correlation functions in the presence of the defect. In Lorentzian signature, the presence of the time-like defect changes the Hilbert space of the theory, so the expectation value $\langle \cdots \rangle_\mathbb{L}$ with the defect and $\langle \cdots \rangle$ without the defect are a priori independent.  In Euclidean signature, however, we can think of the defect as an operator $\mathbb{L}$ that we can insert in the path integral, and which can have a nontrivial expectation value $\langle \mathbb{L} \rangle$.  Thus, we can define
 \es{ExpDefect}{
  \langle \cdots \rangle_\mathbb{L}  = \frac{\langle \cdots \mathbb{L} \rangle}{\langle \mathbb{L} \rangle} \,,
 }
where the ellipses stand for whatever other operator insertions we might have.  This definition is consistent with the normalization $ \langle 1 \rangle_\mathbb{L} = 1$ used so far.  When we define ``connected correlators'' below, we will include the defect as an operator that is on the same footing as the other operators.  For example, for a one-point function, we have
 \es{OnePtConn}{
  \langle {\cal O}_1 \rangle_{\mathbb{L}, \text{conn}} \equiv 
   \frac{\langle {\cal O}_1 \mathbb{L} \rangle_\text{conn}}{\langle \mathbb{L} \rangle}
    = \frac{\langle {\cal O}_1 \mathbb{L} \rangle - \langle {\cal O}_1 \rangle \, \langle \mathbb{L} \rangle}{\langle \mathbb{L} \rangle} = \langle {\cal O}_1 \rangle_{\mathbb{L}} - \langle {\cal O}_1 \rangle \,,
 } 
and for a two-point function a similar calculation gives
 \es{TwoPtConn}{
  \langle {\cal O}_1 {\cal O}_2 \rangle_{\mathbb{L}, \text{conn}} 
   = \langle {\cal O}_1 {\cal O}_2 \rangle_{\mathbb{L}} - \langle {\cal O}_1 {\cal O}_2 \rangle
    - \langle {\cal O}_1 \rangle_{\mathbb{L}}\, \langle {\cal O}_2 \rangle
     - \langle {\cal O}_2 \rangle_{\mathbb{L}}\, \langle {\cal O}_1 \rangle 
      + 2 \langle {\cal O}_1 \rangle \, \langle {\cal O}_2 \rangle \,.
 } 
Note that if the operators ${\cal O}_i$ have no overlap with the identity operator, then their expectation values in the CFT vacuum state vanish and the expressions \eqref{OnePtConn} and \eqref{TwoPtConn} simplify to $\langle {\cal O}_1 \rangle_{\mathbb{L}, \text{conn}} = \langle {\cal O}_1 \rangle_{\mathbb{L}}$ and $\langle {\cal O}_1 {\cal O}_2 \rangle_{\mathbb{L}, \text{conn}} 
   = \langle {\cal O}_1 {\cal O}_2 \rangle_{\mathbb{L}} - \langle {\cal O}_1 {\cal O}_2 \rangle$, respectively.  As usual, the generating functional for connected correlators is the logarithm of the partition function.  It can be shown that the generating functional for connected correlators in the presence of the defect (with the above definition of a connected correlator) is the logarithm of the ratio between the partition function with a defect insertion and the partition function without one (see around \eqref{FreeL}).

\subsection{Preserving supersymmetry}
\label{SPHEREMASS}

The next question to answer is how the mass parameter $m$ mentioned at the beginning of this section is related to the complex mass parameters $\mathfrak{m}$ and $\bar{\mathfrak{m}}$ introduced in Section~\ref{MASSS4}.  As we will see, in general, the mass deformation \eqref{SmSphere} and the half-BPS defect do not share any supersymmetries in common, but for a specific choice of $\mathfrak{m}$ and $\bar{\mathfrak{m}}$ they are both invariant under four supercharges.

Indeed, after continuing to Euclidean signature, \eqref{OSpParams} shows that the defect preserves the supersymmetry transformations where the parameters $(\alpha^i, \alpha_i, \beta^i, \beta_i)$ obey 
\es{OSpParamsEuc}{
	\alpha^i &= -i e^{-i \vartheta} \varepsilon^{ij} \gamma^4 \alpha_j \,, \qquad
	\beta^i = -i e^{-i \vartheta} \varepsilon^{ij} \gamma^4 \beta_j \,. 
}
On the other hand, the mass deformation preserves the supersymmetries that obey \eqref{alphabetaeqs}, with $y = \pm i \sqrt{\mathfrak{m} \bar{\mathfrak{m}}} / r$.  For general $\vartheta$, $\mathfrak{m}$, and $\bar{\mathfrak{m}}$, the equations \eqref{alphabetaeqs} and \eqref{OSpParamsEuc} have no non-trivial solutions, so the defect breaks all the supersymmetries of the mass-deformed theory.

However, we can identify a relation between $\vartheta$, $\mathfrak{m}$, and $\bar{\mathfrak{m}}$ for which \eqref{alphabetaeqs} and \eqref{OSpParamsEuc} do have non-trivial solutions simultaneously.  Plugging in the expressions $\beta^1 = \frac{y}{\bar{\mathfrak{m}}} \alpha_1$ and $\beta_2 = \frac{y}{\mathfrak{m}} \alpha^2$ into the equation  $\beta^1 = -i e^{-i \vartheta}  \gamma^4 \beta_2$, one obtains $ \alpha_1 = -i \frac{\bar{\mathfrak{m}}}{\mathfrak{m}} e^{-i \vartheta} \gamma^4  \alpha^2$.  This expression is consistent with the equation $ \alpha_1 = -i e^{i \vartheta} \gamma^4  \alpha^2$ in \eqref{OSpParamsEuc} provided that $\bar{\mathfrak{m}} / \mathfrak{m} = e^{2 i \vartheta}$.  A similar analysis applied to the equations involving $\beta_1$, $\beta^2$, $\alpha^1$, and $\alpha_2$ gives the same condition.

Thus, in order to preserve some supersymmetry, we need to choose 
\es{mdeltaChoice}{
	\mathfrak{m} = m e^{- i \vartheta} \,, \qquad
	\bar{\mathfrak{m}} = m e^{i \vartheta} \,, \qquad y = \pm i \frac{m}{r} \,,
}
with real $m$, and  the mass deformation becomes
\es{SmSphere2}{
	S_{m} = \frac{m}{2} \int d^4 x\, \sqrt{g} \left(  e^{-i\vartheta} K_\text{sphere} + e^{i \vartheta} \bar K_\text{sphere} \pm \frac{i}{r} (J_{\text{sphere}\, 11} + J_{\text{sphere}\, 22})  \right)  \,.
}
For these choices, Eqs.~\eqref{alphabetaeqs} and \eqref{OSpParamsEuc} have a four-parameter family of solutions, so the mass deformation and the defect preserve simultaneously four supercharges. 

On ${\rm S}^4$, the defect extends along a great circle, so both the mass deformation and the defect also preserve an $\mf{so}(3) \times \mf{so}(2
)$ subalgebra of the $\mf{so}(5)$ isometry group of ${\rm S}^4$.  They also preserve the $\mf{u}(1)$ subalgebra of $\mf{su}(2)_R$ that is generated by $\tau_2$ in our conventions.  Thus, the symmetry that is simultaneously preserved by the defect and the mass deformation is $\mf{so}(3) \times \mf{u}(1)^2$ along with four supercharges.  These combine to form an $\mathfrak{su}(2|1) \times \mathfrak{u}(1)$ superalgebra.

Let us point out that the deformation \eqref{SmSphere2} can also be written in terms of the operators of the SCFT on any conformally flat space.  For an operator $\phi$ of dimension $\Delta$ on a space with conformal factor $\Omega(\tau, \vec{x})$, we have 
\es{phiSphereTophi}{
	\phi_\text{sphere}(\tau, \vec{x})  = \left( \frac{2}{1 + x^2/r^2} \right)^\Delta \phi_\text{flat}(\tau, \vec{x})
	= \left( \frac{2 \Omega(\tau, \vec{x})}{1 + x^2/r^2} \right)^\Delta \phi(\tau, \vec{x}) \,,
}
where $x^2 = \tau^2 + \abs{\vec{x}}^2$.   Using this relation and the fact that $\sqrt{g_\text{sphere}} = 16/(1 + x^2/r^2)^4$, we can write \eqref{SmSphere2} as
\es{SmSphereGeneral}{
	S_{m} = \frac{m}{2} \int \frac{d\tau\, d^3\vec{x}}{\Omega^4} \,  \left[ \frac{2 \Omega }{1 + x^2/r^2}  \left(  e^{-i\vartheta} K + e^{i \vartheta} \bar K \right)  \pm \frac{i}{r}  \left( \frac{2 \Omega}{1 + x^2/r^2}  \right)^2 (J_{11} + J_{22})  \right]  \,.
}
Note that $d\tau\, d^3\vec{x} / \Omega^4$ is the right measure factor on the conformally flat space \eqref{ConfFlat}, and that the power  $4 - \Delta$  with which $2 \Omega/(1 + x^2/r^2)$ appears in front of an operator of dimension $\Delta$, which is expected by implementing the Weyl transformation on the background fields.

\subsection{\texorpdfstring{Derivatives of ${\cal I}_\mathbb{L}(m)$ at $m=0$}{Derivatives of ZL(m) at m = 0}}
\label{DERIVATIVES}

Let us denote by $Z_\mathbb{L}(m)$ and $Z(m)$ the partition functions of the mass deformed theory on ${\rm S}^4$ with and without the defect, respectively, and define
 \es{FreeL}{
  {\cal I}_\mathbb{L}(m) \equiv  \log \frac{Z_\mathbb{L}(m)}{Z(m)} = \log \langle \mathbb{L} \rangle(m)\,.
 }
As per the second equality, we may interpret ${\cal I}_\mathbb{L}(m)$ as $\log \langle \mathbb{L} \rangle(m)$ in the mass deformed theory.  Since, as we reviewed in Section~\ref{MASSS4}, the mass deformation can be written in terms of the operators of a flavor current multiplet, then the derivatives of ${\cal I}_\mathbb{L}(m)$ evaluated at $m=0$ should be related to integrated connected correlators of operators in the flavor current multiplet of the SCFT in the presence of the defect.

For simplicity and without loss of generality, let us set the radius of the sphere to $r=1$, since it can always be restored using dimensional analysis.  Since $m$ appears linearly in \eqref{SmSphereGeneral}, taking derivatives with respect to $m$ produces correlation functions of $-S_{m}/m$:\footnote{As mentioned in Section~\ref{EXPECTATION}, we define our correlators by the functional differentiation from a supersymmetric partition function, seen as a functional of the background vector multiplet fields.  In this definition,  the chain rule implies that the $n$th derivative of ${\cal I}_\mathbb{L}$ w.r.t.~$m$ produces only an integrated $n$-point correlator and no additional lower-point correlators.}
\es{nthDer}{
	\frac{d^n {\cal I}_\mathbb{L}}{dm^n} \bigg|_{m=0} = \left\langle 
	\left[ - \int \frac{d\tau\, d^3\vec{x}}{2\Omega^4} \,  \left[ \frac{2 \Omega }{1 + x^2}  \left(  e^{-i\vartheta} K + e^{i \vartheta} \bar K \right)  \pm i  \left( \frac{2 \Omega}{1 + x^2}  \right)^2 (J_{11} + J_{22})  \right]  \right]^n \right\rangle_{\mathbb{L}, \text{conn}}  \,.
}
It does not matter on which conformally flat space we evaluate this expression because the powers of $\Omega$ in the correlation functions are designed to precisely cancel the explicit powers of $\Omega$ in \eqref{nthDer}.  Note that this is a connected correlator in the presence of the defect, defined around \eqref{OnePtConn}--\eqref{TwoPtConn}.

Thus, let us specialize to the case of $\HH^2 \times {\rm S}^2$, where $\Omega(\tau, \vec{x}) = \abs{\vec{x}}$, and where correlation functions are simplest.  For the first derivative of ${\cal I}_\mathbb{L}$, we use the expectation values in \eqref{OnePointAdS} and obtain
\es{FirstDer}{
	{\cal I}_\mathbb{L}'(0) = -\frac 12 \int \frac{d\tau\, d^3 \vec{x}}{\abs{\vec{x}}^4} \, \left( \frac{2 \abs{\vec{x}}}{1 + x^2}  \right)
	\left( e^{-i \vartheta} a_\mathbb{L} + e^{i \vartheta} \bar a_\mathbb{L} \right) \,.
}
As already mentioned, this expression vanishes because $a_\mathbb{L} = \bar a_\mathbb{L} = 0$, as we will see in Section~\ref{WARD}.  Using \eqref{TwoPointAdS}, we can write the second derivative of ${\cal I}_\mathbb{L}$ at $m=0$ as
\es{SecondDerSimp}{
	{\cal I}_\mathbb{L}''(0) =  -\frac 12 I_2\left[A_{\mathbb{L}, \text{conn}}  \right] + \frac 14 I_3 \left[ e^{- 2 i\vartheta} B_{\mathbb{L}, \text{conn}} + 2 C_{\mathbb{L}, \text{conn}} + e^{2 i \vartheta} \bar B_{\mathbb{L}, \text{conn}} \right] \,,
}
where the ``conn'' index means that the corresponding quantities are the analogs of those defined in \eqref{TwoPointAdS} for the connected correlators, and where we have introduced the notation
\es{IDeltaF}{
	I_\Delta[F] \equiv \int \frac{d\tau_1\, d^3 \vec{x}_1}{\abs{\vec{x}_1}^4} \, \frac{d\tau_2 \, d^3 \vec{x}_2}{\abs{\vec{x}_2}^4} \,  \left( \frac{2 \abs{\vec{x}_1}}{1 + x_1^2}  \right)^{4 - \Delta} \left( \frac{2 \abs{\vec{x}_2}}{1 + x_2^2}  \right)^{4- \Delta} F(\xi, \eta) \,,
}
with $\xi$ and $\eta$ being the Euclidean analogs of \eqref{xietaDef}, namely
\es{xietaEucDef}{
	\xi = \frac{ (\tau_1 - \tau_2)^2 + \abs{\vec{x}_{1}}^2 + \abs{\vec{x}_{2}}^2}{ 2\abs{\vec{x}_{1}}  \abs{\vec{x}_{2}}} \,, \qquad
	\eta = \frac{\vec{x}_{1} \cdot \vec{x}_{2}}{ \abs{\vec{x}_{1}}  \abs{\vec{x}_{2}}} \,.
}

We can simplify the expression \eqref{SecondDerSimp} in two ways.  The first is to explicitly perform six out of the eight integrals in \eqref{IDeltaF}.  The second, which we will undertake in the next section, is to relate $B_{\mathbb{L}, \text{conn}}$, $\bar B_{\mathbb{L}, \text{conn}}$, and $C_{\mathbb{L}, \text{conn}}$ to $A_{\mathbb{L}, \text{conn}}$.  In the end, the entire expression in \eqref{SecondDerSimp} will involve only two integrals of the function $A_{\mathbb{L}, \text{conn}}$.

\subsection{Simplification of integrated correlators}
\label{SIMPLIFICATION}

It was shown in \cite{Billo:2023ncz} that one can perform six of the eight integrals in \eqref{IDeltaF}, obtaining an expression involving only two integrals.  Here, we present a different method that yields the same answer.  One advantage of the method we present below is that, at an intermediate stage of the calculation, we will obtain an expression for \eqref{IDeltaF} involving an integral over $\HH^2 \times {\rm S}^2$ that will be very useful when it is combined with the Ward identity derived in the next section.

Since the factor $d\mu \equiv d\tau \, d^3\vec{x} / \abs{\vec{x}}^4$ is the integration measure on $\HH^2 \times {\rm S}^2$, this measure can thus be written in a factorized form
\es{Measure}{
	d\mu  = d\mu^{\HH} d\mu^{\rm S} \,, \qquad
	d\mu^{\HH} = \frac{d\tau \, dz}{z^2} \,, \qquad
	d\mu^{\rm S} = \sin \theta\, d\theta \, d\phi \,,
}
where $d\mu^{\HH}$ is the measure on $\HH^2$ and $d\mu^{\rm S}$ is the measure on ${\rm S}^2$. We have used the notation $z = \abs{\vec{x}}$ and the parametrization of a point on ${\rm S}^2$ in terms of a unit vector $\hat n = \frac{\vec{x}}{\abs{\vec{x}}} =  (\sin \theta \cos \phi, \sin \theta \sin \phi, \cos \theta)$ embedded in $\R^3$.   
In the following discussion, in addition to the embedding coordinates $\hat n$ for the points of ${\rm S}^2$, it will also be useful to introduce the $\R^{2, 1}$ embedding coordinates for $\HH^2$ defined by
\es{embedding}{
	X = \left(
	\frac{\tau}{z},  \frac{1 - \tau^2 - z^2}{2z} ,  
	\frac{1 + \tau^2 + z^2}{2z}
	\right)  \,.
}
With the standard metric of signature $(+, +, -)$ on $\R^{2, 1}$, $X$ obeys $X \cdot X = -1$.  In the following discussion, we will place a subscript $i$ on all the quantities defined between \eqref{Measure}--\eqref{embedding} when they refer to $(\vec{x}_i, \tau_i)$ instead of $(\vec{x}, \tau)$.

As mentioned below \eqref{TwoPtAdS2S2} in the Lorentzian context, in terms of the embedding coordinates we have $\xi = -X_1 \cdot X_2$ and $\eta = \hat n_1 \cdot \hat n_2$.   Thus, in embedding space, \eqref{IDeltaF} becomes 
 \es{IDeltaFAgain}{
  I_\Delta[F] = \int d\mu_1^{\HH} \, d\mu_1^S \, d\mu_2^{\HH} \, d\mu_2^S \,     \frac{F(-X_1 \cdot X_2, \hat n_1 \cdot \hat n_2)}{\left( -X_1 \cdot X_* \right)^{4-\Delta} \left( -X_2 \cdot X_*  \right)^{4-\Delta}} \,,
 }
where $X_*$ is the special point on $\HH^2$ given by $X_* = (0, 0, 1)$.  The expression \eqref{IDeltaFAgain} makes manifest a property that was not visible before: the integrand is invariant under simultaneous $\SO(2, 1)$ transformations on $X_1$, $X_2$, and $X_*$.  Let us consider the $\SO(2, 1)$ transformation that interchanges the points $X_2$ and $X_*$.  This transformation acts on point $Y$ in $\R^{2, 1}$ by
 \es{SO21Transf}{
  Y \quad \longmapsto \quad f(Y) = - Y - \frac{Y \cdot ( X_2 + X_*)}{ 1  -  X_2 \cdot X_*} (X_2 + X_*) \,.
 }
It is straightforward to check that, indeed, $f(X_2) = X_*$ and $f(X_*) = X_2$, and that $f$ is an $\SO(2, 1)$ linear transformation.  Applying \eqref{SO21Transf} just to the integrand in \eqref{IDeltaFAgain}, we obtain
 \es{IDeltaFAfterSO21}{
  I_\Delta[F] = \int d\mu_1^{\HH} \, d\mu_1^S \, d\mu_2^{\HH} \, d\mu_2^S \,     \frac{F(-f(X_1) \cdot X_*, \hat n_1 \cdot \hat n_2)}{\left( -f(X_1) \cdot X_2 \right)^{4-\Delta} \left( -X_2 \cdot X_*  \right)^{4-\Delta}} \,.
 }
We can then change variables in the integration over $X_1$ from $X_1$ to $X = f(X_1)$.  This change of variables has unit Jacobian since \eqref{SO21Transf} is an $\SO(2, 1)$ transformation.  After this change of variables, as well as the renaming $\hat n_1 \to \hat n$,  the expression \eqref{IDeltaFAfterSO21} is written equivalently as
\es{IDeltaFAgain3}{
	I_\Delta[F] = 4 \pi  \int  d\mu^{\HH} \, d\mu^{\rm S} \, \left[ \int \frac{d\mu_2^{\HH}}{ \left( -X \cdot X_2 \right)^{4-\Delta} \left( -X_* \cdot X_2  \right)^{4-\Delta}} \right]  F(-X \cdot X_*, \hat n \cdot \hat n_*) \,,
}
where we also notice that the integral over $\hat n_1$ in \eqref{IDeltaFAgain} does not depend on the value of $\hat n_2$, so we can fix $\hat n_2 = \hat n_* = (0, 0, 1)$ and replace the integration over $\hat n_2$ by $\Vol({\rm S}^2) = 4 \pi$.

Let us denote the quantity in the square brackets in \eqref{IDeltaFAgain3} by 
\es{DFunc}{
	D_\Delta(-X \cdot X_*) \equiv \int \frac{d\mu_2^{\HH}}{ \left(- X \cdot X_2 \right)^{4-\Delta}\left( -X_* \cdot X_2  \right)^{4-\Delta}} \,.
}
The RHS is a function of $X \cdot X_*$ because the integrand is $\SO(2, 1)$-invariant, provided that we transform simultaneously $X$, $X_*$, and $X_2$.  For $\Delta = 2, 3$, the integrals can be evaluated explicitly (see Appendix~\ref{DDELTA}):
\es{DExplicit2}{
	D_2 (\xi) &= - \frac{2\pi}{\xi^2 - 1} + \frac{2 \pi \xi \arccosh \xi}{(\xi^2 - 1)^{3/2}} \,, \\
	D_3 (\xi) &= \frac{2 \pi \arccosh \xi}{\sqrt{\xi^2 - 1}}   \,.
}

Going back to \eqref{IDeltaFAgain3} and using the notation $\xi = -X \cdot X_*$ and $\eta =  \hat{n}\cdot\hat{n}_*$, \eqref{IDeltaFAgain3} becomes
\es{IDeltaFAgain4}{
	I_\Delta[F] = 4 \pi  \int_{\HH^2 \times {\rm S}^2}  d^4 x \, \sqrt{g}  \, D_\Delta(\xi)  F(\xi, \eta) \,.
}
Going back further to \eqref{SecondDerSimp}, we can write the second derivative of ${\cal I}_\mathbb{L}$ explicitly as
\es{SecondDerSimp2}{
	{\cal I}_\mathbb{L}''(0) =  \pi  \int_{\HH^2 \times {\rm S}^2}  &d^4 x \, \sqrt{g}
	\biggl[-2 D_2(\xi) A_{\mathbb{L}, \text{conn}}(\xi, \eta) \\
	 &{}+ D_3(\xi)  \left( e^{- 2 i\vartheta} B_{\mathbb{L}, \text{conn}}(\xi, \eta) + 2 C_{\mathbb{L}, \text{conn}}(\xi, \eta) + e^{2 i \vartheta} \bar B_{\mathbb{L}, \text{conn}}(\xi, \eta) \right)\biggr] 
	\,.
}
This expression can be simplified further by performing two out of the four integrals, but we will do so only after relating $B_{\mathbb{L}, \text{conn}}$, $\bar B_{\mathbb{L}, \text{conn}}$, and $C_{\mathbb{L}, \text{conn}}$ to $A_{\mathbb{L}, \text{conn}}$.

\section{\texorpdfstring{Ward identity from ${\rm AdS}_2 \times {\rm S}^2$}{Ward identity from AdS2 x S2}} 
\label{WARD}

In this section we derive the Ward identities that relate the $\langle KK\rangle$, $\langle K \bar K \rangle$, and $\langle \bar K \bar K\rangle$ correlators to that of the superconformal primary $\langle J_{ij} J_{kl} \rangle$, and we will also explain why $\langle K \rangle = \langle \bar K \rangle = 0$.  We will work in Lorentzian signature, but the results can be trivially continued to Euclidean signature.  For readability, in this section we will drop the indices on the functions $A$, $B$, $\bar B$, $C$, as well as the indices on the expectation values.  The formulas we present hold equally well in the case of no defect, or for the full correlator in the presence of the defect $\mathbb{L}$, or for the connected correlator in the presence of the defect.

Let us focus on the two-point functions first and relate the functions $B$ and $C$ that appear in \eqref{TwoPointAdS} to the function $A$ appearing in the same equation. Without loss of generality, let us take the first operator to be at a generic position parameterized by $(t_1, \vec{x}_1) = (t, \vec{x})$, and the second operator to be at $(t_2, \vec{x}_2) = (0, \hat z)$, where $\hat z = (0, 0, 1)$.   (In standard ${\rm AdS}_2 \times {\rm S}^2$ coordinates, the second insertion is at $(t_2, z_2) = (0, 1)$ and at the North pole of the sphere, at $\theta_2 = 0$.)    For simplicity, we refer to the first insertion point as $x_1 = x$ and the second insertion point as $x_2 = 0$.   With this choice, the invariants $\xi$ and $\eta$ can be identified with
\es{xietaParticular2}{
	\xi = \frac{1 - t^2 + \abs{\vec{x}}^2}{2\abs{\vec{x}}}
	= \frac{1 - t^2 + z^2}{2z} \,, \qquad
	\eta = \frac{x^3}{\abs{\vec{x}}} = \cos \theta \,,
} 
depending on whether we use $(t, \vec{x})$ or $(t, z, \theta, \phi)$ as our coordinates.

The procedure for relating $B$ and $C$ to $A$ relies on the Ward identity
\es{WardId}{
	\langle K(x)   \delta \delta' J_{22}(0) \rangle 
	= \langle \delta' \delta K(x)  J_{22}(0) \rangle \,.
}
The SUSY variations $\delta$ and $\delta'$ have parameters $(\epsilon_i, \epsilon^i, \eta_i, \eta^i)$ and $(\epsilon_i', \epsilon^{\prime i}, \eta_i', \eta^{\prime i})$, respectively, that have to be chosen appropriately. 

Let us start with the LHS of \eqref{WardId}.  Note that invariance under $\mf{su}(2)_R$ implies that only the operators $K$, $\bar K$, and $j^a$ can have a non-zero two-point function with $K$,  so we only have to keep track of these operators when we compute  $\delta \delta' J_{22}(0)$.  From \eqref{SUSYVars}, we have
\es{ddpJ22}{
	\delta \delta' J_{22} &=  \frac 12 \left( \bar \epsilon^{\prime 1}   \gamma^a \epsilon_2
	-\bar \epsilon'_{2} \gamma^a  \epsilon^1 \right)  j_a
	- \frac 12  \bar \epsilon^{\prime 1}  \epsilon^1\, K
	-\frac 12  \bar \epsilon'_{2}    \epsilon_2 \,   \bar K   + \text{other ops} \,.
} 
If we want to ensure that in \eqref{WardId} we obtain only the correlators in \eqref{TwoPointAdS}, we should ensure that the coefficient of $j_a(0)$ vanishes by choosing the supersymmetry parameters to obey the four equations 
\es{jaVanish}{
	\bar \epsilon^{\prime 1}(0)   \gamma^a \epsilon_2(0)
	=\bar \epsilon'_{2}(0) \gamma^a  \epsilon^1(0) \,, \qquad a = 0, 1, 2, 3 \,.
} 
On the other hand, on the RHS of \eqref{WardId}, $\mf{su}(2)_R$ symmetry implies that only the operator $J_{11}$ and its derivatives can contribute, so when we compute $\delta' \delta K$ using \eqref{SUSYVars}, we keep track only of terms involving $J_{11}$ and its derivatives:
\es{deltadeltaK}{
	\delta' \delta K
	&= -\frac 12 \bar \epsilon_2 \slashed{D}   \slashed{D} (J_{11} \epsilon'_2 )   
	+  \bar \eta_2   \slashed{D} (J_{11} \epsilon'_2 ) + \text{other ops} \,.
}
Using $\slashed{D} \epsilon^i = 4 \eta^i$, $\slashed{D} \epsilon_i = 4 \eta_i$, and $\slashed{D} \eta^i  = \slashed{D} \eta_i = 0$, it is straightforward to show that \eqref{deltadeltaK} simplifies to the ${\rm AdS}_2 \times {\rm S}^2$ Laplacian of a product between $J_{11}$ and a bilinear in the Killing spinors:
\es{Id}{
	\delta' \delta K &= - \frac 12 \square \left(\bar \epsilon_2 \epsilon_2' J_{11} \right) + \text{other ops} \,.
}

Plugging this into \eqref{WardId} and using the two-point functions in \eqref{TwoPointAdS} for a choice of $\delta$ and $\delta'$ obeying \eqref{jaVanish}, we find 
\es{WardIdExplicit}{
	\bar \epsilon^{1}(0)  \epsilon^{\prime 1}(0)\, B(x)
	+ \bar \epsilon_{2}(0)    \epsilon'_2(0) \,   C(x) 
	=  \square \Bigl[\bar \epsilon_2(x) \epsilon_2'(x) A(x) \Bigr] \,.
}
We can make further simplifications using \eqref{epsAdS} and the relations \eqref{OSpParams} to restrict the Killing spinors to parameterize $\mf{osp}(4^*|2)$ transformations.  Note that after using these relations we can write both \eqref{jaVanish} and \eqref{WardIdExplicit} only in terms of $\alpha_2$, $\alpha_2'$, $\beta_2$, $\beta_2'$.  Each of these quantity is a chiral spinor so it has two free complex parameters, for a total of $8$ free complex parameters.  The equations \eqref{jaVanish} imply (see Appendix~\ref{WARDDETAILS} for more details)
\es{jaVanish3}{
	a = 0&: \qquad \bar \alpha_2'  \alpha_2
	+\bar \beta_2' \beta_2   =0 \,, \\
	a = 3&: \qquad   \bar \alpha_2' \gamma^0  \beta_2 - \bar \beta_2'  \gamma^0\alpha_2 
	=0 \,, \\
	a =1, 2&: \qquad    \bar \alpha_2' \gamma^{0a3} \beta_2 + \bar \beta_2'  \gamma^{0a3} \alpha_2 
	=0   \quad
	\Longrightarrow  \quad
	\bar \alpha_2' \gamma^{a} \beta_2 - \bar \beta_2'  \gamma^{a} \alpha_2  = 0 \,.
}

Using the conditions \eqref{jaVanish3}, we find (see also Appendix~\ref{WARDDETAILS} for a couple of intermediate steps)
\es{WardIdExplicit3}{
	e^{-2 i\vartheta} (X 
	- Y
	)   \, B(x) + (X  + Y )   \,   C(x) & =  \square \Bigl[\frac{ X (1 - t^2 + \abs{\vec{x}}^2 )  +  2 x^3 Y  }{2 \abs{\vec{x}}}   A(x) \Bigr] \,, \\
	\text{with } \quad X &\equiv \bar \alpha_{2}  \alpha'_2 \,, \qquad 
	Y \equiv \frac{\bar \alpha_{2}   \gamma^3 \beta'_2
		-\bar \beta_2  \gamma^3  \alpha'_2}{2} \,.
} 
As mentioned above, we have $8$ complex parameters in $\alpha_2$, $\alpha_2'$, $\beta_2$, $\beta_2'$.  We have four equations in \eqref{jaVanish3}, and both \eqref{jaVanish3} and \eqref{WardIdExplicit3} are invariant under the rescalings $(\alpha_2, \beta_2, \alpha_2', \beta_2') \to (\lambda \alpha_2, \lambda \beta_2, \lambda' \alpha_2', \lambda' \beta_2')$ with independent complex parameters $\lambda$ and $\lambda'$.  Thus, we have enough free parameters to be able to set the ratio $X/Y$ in \eqref{WardIdExplicit3} arbitrarily.  Taking $X/Y = -1$ and $X/Y = 1$, we obtain the final form of our Ward identities:\footnote{Because we adopted the functional differentiation definition of correlation functions, as described in Section~\ref{EXPECTATION}, these Ward identities hold everywhere, including at coincident points (at $\xi = \eta = 1$).  Depending on the form of $A(\xi, \eta)$, it is possible that $B$ and/or $C$ will have a distributional piece supported at $\xi = \eta = 1$.}
\es{WardFinal}{
	B(\xi, \eta) &= \frac{e^{2 i \vartheta}}{2} \square \biggl[ (\xi - \eta) A(\xi, \eta) \biggr] \,, \qquad
	C(\xi, \eta) = \frac{1}{2} \square \biggl[ (\xi + \eta) A(\xi, \eta) \biggr] \,,
}
where we have identified $\frac{1 - t^2 + \abs{\vec{x}}^2 }{2 \abs{\vec{x}}}$ and $x^3 /  \abs{\vec{x}}$ with $\xi$ and $\eta$, respectively, as per \eqref{xietaParticular2}.  A similar calculation shows that $\bar B$ is related to $A$ via an expression similar to the first equation in \eqref{WardFinal}:\footnote{Since the Ward identities do not depend on specifics of the defect, this expression can also be obtained from \eqref{WardFinal} by complex conjugation if the defect preserves charge conjugation symmetry (such as Wilson lines in ${\rm SU}(2)$ SYM), in which case $A$ is a real function.}
 \es{WardFinalExtra}{
  \bar B(\xi, \eta) &= \frac{e^{-2 i \vartheta}}{2} \square \biggl[ (\xi - \eta) A(\xi, \eta) \biggr]  \,.
 }
As checks of our Ward identities, in Appendix~\ref{WARDCHECKS} we show that Eqs.~\eqref{WardFinal}--\eqref{WardFinalExtra} are satisfied by the free massless hypermultiplet without a defect and by the stress-tensor superconformal block.

In general, the Laplacian in \eqref{WardFinal} is defined to act on functions of the four coordinates $x = (t, \vec{x})$ or $(t, z, \theta, \phi)$.  In the latter set of coordinates, it is 
\es{LapAdS2S2Gen}{
	\square = z^2 \left( - \partial_t^2 + \partial_z^2 \right) 
	+ \partial_\theta^2 + \cot \theta\, \partial_\theta + \frac{1}{\sin^2 \theta} \partial_\phi^2 \,.
}
For a function $f(\xi, \eta)$, the ${\rm AdS}_2 \times {\rm S}^2$ Laplacian simplifies to
\es{LapA}{
	\square f = (\xi^2 - 1) \partial_\xi^2 f  + 2 \xi \partial_\xi f
	+ (1 - \eta^2) \partial_\eta^2 f - 2 \eta \partial_\eta f \,.
}
The expressions \eqref{WardFinal}--\eqref{LapA} can be trivially continued to Euclidean signature by replacing $t \to -i \tau$ and making no other changes.

The analysis above also shows that $\langle K \rangle = \langle \bar K \rangle = 0$.  Indeed, since $\langle J_{22} \rangle = 0$ due to $\mf{su}(2)_R$ selection rules, we must also have $\langle \delta \delta' J_{22} \rangle = 0$.  From \eqref{ddpJ22} and the fact that the only potentially non-vanishing one-point functions of operators in the flavor current multiplet are those of $K$ and $\bar K$, we find
\es{OnePointWard}{
	\bar \epsilon^{\prime 1}  \epsilon^1\, \langle K \rangle
	+  \bar \epsilon'_{2}    \epsilon_2 \,   \langle \bar K  \rangle = 0 \,.
} 
Since we have enough freedom to choose $\bar \epsilon^{\prime 1}  \epsilon^1$ and $\bar \epsilon'_{2}    \epsilon_2$ independently, we must have $\langle K \rangle = \langle \bar K \rangle = 0$, or equivalently $a_{\mathbb{L}} = \bar{a}_\mathbb{L} = 0$ in \eqref{OnePt} and \eqref{OnePointAdS}.\footnote{When the $\grU(1)$ flavor symmetry is a Cartan subgroup of a simple Lie group $G_F$, the vanishing of one-point functions of the current multiplet also follows from the fact that the superconformal line is $G_F$ invariant.}

\section{Integral constraint on two-point function}
\label{sec:integralconstraint}

In this section we derive one of our main results, namely an integral constraint on the two-point function of the superconformal primary $J_{ij}$ in the current multiplet from mass derivatives of the expectation value of a circular defect with the supersymmetric mass deformation.

We first use the Ward identity \eqref{WardFinal}--\eqref{WardFinalExtra} to simplify the integral constraint \eqref{SecondDerSimp2} that involve two-point functions \eqref{TwoPoint} of the superconformal primary (i.e.,~$\la JJ\ra \propto A$) and those for the scalar descendants (i.e.,~$\la KK\ra\propto B$, $\la \bar K \bar K\ra\propto \bar B$, $\la K\bar K\ra  \propto C$).  The quantity that appears in the latter expression can be written as
\es{BCComb}{
	e^{- 2 i\vartheta} B_{\mathbb{L}, \text{conn}}(\xi, \eta) + 2 C_{\mathbb{L}, \text{conn}}(\xi, \eta) + e^{2 i \vartheta} \bar B_{\mathbb{L}, \text{conn}}(\xi, \eta)  = 
	2 \square \biggl[\xi A_{\mathbb{L}, \text{conn}}(\xi, \eta) \biggr] \,,
}
as per \eqref{WardFinal}--\eqref{WardFinalExtra}.  Thus \eqref{SecondDerSimp} becomes 
\es{SecondDerSimp3}{
	{\cal I}_\mathbb{L}''(0) = -2 \pi  \int_{\HH^2 \times {\rm S}^2}  d^4 x \, \sqrt{g}
	\left[ D_2(\xi) A_{\mathbb{L}, \text{conn}}(\xi, \eta) - D_3(\xi) \square \Bigl( \xi A_{\mathbb{L}, \text{conn}}(\xi, \eta) \Bigr)\right] 
	\,.
}
We can simply integrate by parts twice in this expression to move the Laplacian onto $D_3$.  Using \eqref{DExplicit2} and \eqref{LapA}, one can show that the combination $D_2 - \xi \square D_3$ multiplying $A_{\mathbb{L}, \text{conn}}$ undergoes a remarkable simplification:
\es{D23Rel}{
	D_2(\xi) - \xi \square D_3(\xi)  = 2 \pi \,.
}
When combined with \eqref{SecondDerSimp3} after the double partial integration,\footnote{We can safely integrate by parts because we adopted the functional differentiation definition of correlation functions, which ensures that the supersymmetric Ward identity holds everywhere, including at coincident points.} this yields
\es{SecondDerSimp4}{
	{\cal I}_\mathbb{L}''(0) = -4 \pi^2  \int_{\HH^2 \times {\rm S}^2}  d^4 x \, \sqrt{g}\,  A_{\mathbb{L}, \text{conn}}(\xi, \eta)  \,.
}
The way to interpret this equation is that the point on $\HH^2 \times {\rm S}^2$ that is being integrated over is parameterized by a point on $\HH^2$ with embedding coordinates $X \in \R^{2, 1}$ satisfying $X \cdot X = -1$, and a point on ${\rm S}^2$ described by a unit vector $\hat n \in \R^3$.  The arguments of the function $A$ are $\xi = -X \cdot (0, 0, 1)$ and $\eta = \hat n \cdot (0,0,1)$.

One can simplify this expression further by choosing appropriate coordinates for $\HH^2 \times {\rm S}^2$.   Let us take $X = (\sinh \rho \cos \psi, \sinh \rho \sin \psi, \cosh \rho)$ and $\hat n = (\sin \theta \cos \phi, \sin \theta \sin \phi, \cos \theta)$.  Then $d^4 x \, \sqrt{g}$ becomes $d \rho \, d\psi \, d\theta\, d\phi\, \sinh \rho\, \sin \theta$ and $\xi = \cosh \rho$ and $\eta = \cos \theta$.  One can then see that the integrand in \eqref{SecondDerSimp4} does not depend on the angles $\psi$ and $\phi$, and the integrals over these angles each give a factor of $2 \pi$.  After converting back to $(\xi, \eta)$ from $(\rho, \theta)$, the resulting expression takes the simple form
\es{SecondDerSimp5}{
	{\cal I}_\mathbb{L}''(0) = -16 \pi^4  \int_1^\infty d\xi \int_{-1}^1 d\eta\,   A_{\mathbb{L}, \text{conn}}(\xi, \eta)  \,.
}
This is one of our main results.  In cases where it is possible to calculate ${\cal I}_\mathbb{L}(m)$ using supersymmetric localization, such as ${\cal N} = 4$ SYM or conformal ${\cal N} =2$ SQCD, we can view \eqref{SecondDerSimp4} as an integral constraint that must be obeyed by the connected two-point function $\langle J_{ij} J_{kl} \rangle_{\mathbb{L}, \text{conn}}$ in the flavor current multiplet and all other two-point functions that can be derived from it using Ward identities.

\section{\texorpdfstring{Wilson lines in $\cN=4$ Super-Yang-Mills at strong coupling}{Wilson lines in N = 4 super-Yang-Mills at strong coupling}}\label{sec:sugra}

The half-BPS Wilson line in $\mathcal{N}=4$ SYM provides a canonical example of a line defect that can be studied using the techniques of this paper. In particular, prior results from localization for the Wilson line with a mass deformation \cite{Russo:2013kea,Belitsky:2020hzs,Pufu:2023vwo} and from holography for the Wilson line with two local operators \cite{Barrat:2021yvp} allow us to directly test the integral constraint presented in~\eqref{SecondDerSimp5}. 

For this calculation, we will need the $\mathcal{N} = 4$ vector multiplet, the scalar at the bottom of the $\mathcal{N} = 4$ stress tensor multiplet, and the half-BPS Wilson loop. The $\mathcal{N}=4$ vector multiplet consists of the gauge field $A_\mu$, left-handed and right-handed fermions $\lambda_\alpha,\lambda^\alpha$, $\alpha = 1, \ldots 4$, transforming in the $\bf{4}$ and $\bar{\bf{4}}$  irreps of $\mf{su}(4)_R$, and the six scalars $\phi_m$, $m=1,\ldots,6$, transforming in the $\bf{6}$  irrep of $\mf{so}(6)_R\cong \mf{su}(4)_R$. All fields transform in the adjoint of the gauge group, which we take to be $\SU(N)$\@. We take all these fields to have standard normalizations.  In particular, the Lagrangian is
 \es{KinN4}{
  {\cal L} = \frac{1}{g_\text{YM}^2}  \left[ \frac{1}{4} F_{\mu\nu}^I F^{\mu \nu I}
   + \frac 12 D_\mu \phi_m^I D^\mu \phi_m^I + \bar \lambda^{\alpha I} \slashed{D} \lambda_\alpha^I 
    + \text{(Yukawa terms)} + \text{(potential)} \right] \,,
 }
where $I = 1, \ldots N^2 - 1$ is an adjoint index. In this section we work in Euclidean signature on $\mathbb{R}^4$.

By taking symmetric traceless products of the scalars, one can generate a family of protected chiral primary operators. A special role is played by the dimension $2$ scalar,
\es{20Prime}{
	S(x,u)\equiv u^m u^n S_{mn}(x),\qquad S_{mn}(x)=N_S\left[\phi_m^I(x)\phi_n^I(x)-\frac{1}{6}\delta_{mn} \phi_p^I(x) \phi_p^I(x)\right] \,,
}
which transforms in the $\bf{20}'$ irrep of $\mf{su}(4)_R$ and is the scalar at the bottom of the stress tensor multiplet. For convenience, we introduce a  null polarization vector  $u^m$  to avoid dealing with explicit $\mf{so}(6)_R$ indices.  The chiral primary $S(x,u)$ has a protected two-point (and three-point) function, which can be computed in the free theory, and we fix the normalization $N_S$ so that
\es{S2Pt}{
	\la S(x_1,u_1)S(x_2,u_2) \ra = \frac{(u_1\cdot u_2)^2}{|x_{12}|^4} \,.
}
Using the scalar propagator $\la \phi_m^I(x)\phi_n^J (y)\ra=\frac{g_\text{YM}^2}{4\pi^2}\frac{1}{|x-y|^2}\delta^{IJ}\delta_{mn}$ that follows from \eqref{KinN4}, we find 
\es{SNorm}{
	N_S&=\frac{2\sqrt{2}\pi^2}{g_\text{YM}^2 \sqrt{N^2-1}} \approx \frac{2\sqrt{2}\pi^2}{g_\text{YM}^2 N } \,.
}
In the second step we have taken the large $N$ limit, $N\to \infty$, $g_{\rm YM}\to 0$ with $\lambda\equiv g_{\rm YM}^2N$ fixed.

We also need the half-BPS Wilson line in $\mathcal{N}=4$ SYM, which, when located at $x^1 = x^2 = x^3 = 0$ and oriented along the $\tau = x^4$ direction, takes the following form \cite{Rey:1998ik,Maldacena:1998im,Drukker:1999zq}:
\es{HalfBPSWilsonLine}{
	\mathbb{L}= \tr_{\rm fund} \text{P} \exp \left(\int (i A_4^I(\tau) +\theta^m \phi_m^I(\tau)) T_I \,d\tau \right)\,,
}
where $\theta^m$ normalized by $\theta^m \theta^m=1$ specifies a particular direction on ${\rm S}^5$ (equivalently a polarization  in the $\mf{so}(6)_R$ directions), $T_I$ are the $\SU(N)$ generators, and the trace $\tr_{\rm fund}$ defines the Wilson line in the fundamental representation.

Viewing $\mathcal{N}=4$ SYM as an $\mathcal{N}=2$ theory with flavor symmetry, we can turn on the mass deformation described in Section~\ref{MASSS4}. The $\cN=4$ R-symmetry decomposes as $\mf{su}(4)_R\to \mf{su}(2)_R\times \mf{u}(1)_R \times \mf{su}(2)_F$. The $\mathcal{N}=4$ vector multiplet breaks up into the $\mathcal{N}=2$ vector multiplet---consisting of the gauge field $A_\mu^I$, left-handed and right-handed fermions $\Omega_i^I,\Omega^{iI}$, and complex scalar $X$---and the $\mathcal{N}=2$ hypermultiplet---consisting of complex scalars $Z_A^I$ and left-handed and right-handed fermions $\zeta^{AI}$, $\zeta_A^I$, with $A = 1, 2$ labeling the $\mathfrak{su}(2)_F$ flavor index.  In Appendix~\ref{FREE}, we write down the action for the free ${\cal N} = 4$ Maxwell theory in an ${\cal N} = 2$ notation---see \eqref{N4MaxwellFlat}.  The kinetic terms in the ${\cal N} = 4$ Yang-Mills theory have the same form as in \eqref{N4MaxwellFlat}, the only difference being that each field has an additional adjoint index $I$ that is being summed over.   Thus, to match those conventions, for the adjoint scalars we can choose $Z_1^I=\frac{1}{\sqrt{2} g_\text{YM}}(\phi_1^I+i\phi_4^I)$, $Z_2^I=\frac{1}{\sqrt{2} g_\text{YM}}(\phi_2^I+i\phi_5^I)$, and $X^I=\frac{1}{2}(\phi_3^I+i\phi_6^I)$.  

From \eqref{FlavorOps2} and \eqref{N4Maxwell}, both generalized to the non-Abelian case by introducing an adjoint index for the hypermultiplet fields and summing over it, we read off the moment map operator that appears in the ${\rm S}^4$ mass deformation\footnote{Up to an unimportant overall sign, the same result can be obtained by comparing Eq.~\eqref{SmSphere2} with Eq.~(2.14) of \cite{Bobev:2013cja}, which studied the same mass deformation in $\mathcal{N}=4$ SYM.}
\es{JInTermsOfHyper}{
	J\equiv J_{11}+J_{22}= -\sum_{A=1}^2 (Z^I_A Z^I_A +\bar{Z}^{A,I}\bar{Z}^{A,I}) \,.
}
This operator can also be written in terms of the $\bf{20}'$ operator as
\es{JijInTermsOfS}{
	J&=-\frac{1}{g_\text{YM}^2 N_S}\sum_{A=1}^2 \left[S(x,u_A^+)+S(x,u_A^-)\right],
}
where $u_1^{\pm}=\frac{1}{\sqrt{2}}(1,0,0,\pm i,0,0) $ and $u_2^{\pm}=\frac{1}{\sqrt{2}}(0,1,0,0,\pm i,0)$. These polarizations satisfy $u_A^-\cdot u_B^-=u_A^+\cdot u_B^+=0$ and $u_A^-\cdot u_B^+=\delta_{AB}$.

Finally, we note that after the mass deformation, the supersymmetric Wilson line should couple only to the scalars that remain in the $\mathcal{N}=2$ vector multiplet, so we set
\es{massDeformationWLpolarization}{
\theta^m \phi_m^I=e^{i\vartheta}X^I + e^{-i\vartheta} \bar{X}^I=\phi_3^I \cos\vartheta -\phi_6^I \sin \vartheta \,.
}
Equivalently, $\theta^m$ are the components of the vector $(0, 0, \cos \vartheta, 0, 0, -\sin \vartheta)$.

\subsection{Integrated correlator from localization}

The mass-deformed partition function both with and without the Wilson line can be computed using localization \cite{Pestun:2007rz}. The resulting matrix integral was evaluated at large $N$ and large $\lambda$ in \cite{Russo:2013kea} and was evaluated more systematically in an expansion in $1/N$ and in $1/\lambda$ in \cite{Pufu:2023vwo}. The leading behavior is:
\es{Z''(0)Localization}{
	{\cal I}_\mathbb{L}''(0)=\sqrt{\lambda}+\left(\frac{1}{2}-\frac{\pi^2}{3}\right)+O(1/\sqrt{\lambda},1/N^2)\,. 
}
Next, we will show that the integral constraint in Eq.~\eqref{SecondDerSimp5} combined with existing results from holography reproduces the $\sqrt{\lambda}$ term above.

\subsection{Comparison to supergravity}\label{sec:sugra_check}

Following \cite{Barrat:2021yvp}, we decompose the connected two-point function of two $\mathbf{20}'$ operators in the presence of the Wilson line in \eqref{HalfBPSWilsonLine} into three R-symmetry channels as:
\es{20prime2PtWilsonLine}{
	\la S(x_1,u_1)S(x_2,u_2)\ra_{\mathbb{L}, \text{conn}} &=\frac{(u_1\cdot u_2)^2}{|\vec{x}_1|^2|\vec{x}_2|^2}\left(F_0(\xi,\eta)+\sigma^{-1} F_1(\xi,\eta)+\sigma^{-2} F_2(\xi,\eta)\right) \,,
}
where $\sigma=\frac{(u_1\cdot u_2)}{(u_1\cdot \theta)(u_2\cdot \theta)}$ is an $\mf{so}(6)_R$-invariant cross ratio.

We can relate the two-point function of $J$ in the presence of the Wilson line to \eqref{20prime2PtWilsonLine}. Given that polarization vectors in \eqref{JijInTermsOfS} are orthogonal to the polarization vector of the Wilson line in \eqref{massDeformationWLpolarization} it follows that $\sigma^{-1}=0$, and therefore
\es{J2Pt}{
	\la J(x_1)J(x_2)\ra_{\mathbb{L}, \text{conn}} &=\frac{2\sum_{A=1}^2 (u_A^+\cdot u_A^-)^2}{g_\text{YM}^4 N_S^2}  
	 \frac{F_0(\xi,\eta)}{|\vec{x}_1|^2|\vec{x}_2|^2}
	 =\frac{N^2}{2\pi^4} \frac{F_0(\xi,\eta)}{|\vec{x}_1|^2|\vec{x}_2|^2} \,.
}
Writing this instead as $2\la J_{11}(x_1)J_{22}(x_2)\ra_{\mathbb{L}, \text{conn}}$ and comparing with \eqref{TwoPoint}, we see that 
\es{AWithMinusWithoutDefect}{
	A_{\mathbb{L}, \text{conn}}(\xi,\eta)&=\frac{N^2}{4\pi^4}F_0(\xi,\eta) \,.
}

Finally, the leading contribution to $F_0$ at large $N$ and strong coupling is given in Eq.~(3.33) of \cite{Barrat:2021yvp}, which in our notation becomes\footnote{Note that $\mathcal{O}_{2}^{\rm there}(x,u)=S(x,u)$ and their $z,\bar{z}$ are related to our $\xi,\eta$ by $\xi=\frac{1+z\bar{z}}{2\sqrt{z\bar{z}}}$ and $\eta=\frac{z+\bar{z}}{2\sqrt{z\bar{z}}}$.}$^{,}$\footnote{Note also that close to $\xi = \eta = 1$, we have $A_{\mathbb{L}, \text{conn}}(\xi,\eta) \approx  -\frac{\sqrt{\lambda}}{48 \pi^4}\frac{1}{\xi-\eta} $.  Since for two points with coordinates $x$ and $y$ on $\HH^2 \times {\rm S}^2$, we have $\square \frac{1}{\xi - \eta} = - 8 \pi^2 \delta(x, y)$, the Ward identity \eqref{WardFinal} implies that $\langle K(x) \bar K(y) \rangle$ contains a delta function contribution proportional to $\frac{\sqrt{\lambda}}{6 \pi^2} \delta(x, y)$.  Here $\delta(x, y)$ is the delta function on $\HH^2 \times {\rm S}^2$ normalized so that it integrates to $1$ over the entire space.  This delta function contribution can be interpreted as a non-vanishing one-point function in a scheme where the  $\langle K(x) \bar K(y) \rangle$ correlator does not contain a delta function (see Section~\ref{EXPECTATION} for related discussions).} 
\es{F0LiendoEtAL}{
	F_0(\xi,\eta)&=\frac{\sqrt{\lambda}}{8N^2}\frac{\log(\xi+\sqrt{\xi^2-1})-\xi\sqrt{\xi^2-1}}{(\xi-\eta)(\xi^2-1)^{3/2}}+O(1/N^2,\lambda^{3/2}/N^4) \,.
}
The right hand side of Eq.~\eqref{SecondDerSimp5} at leading order is therefore
\es{IntegralConstraintSupergravity}{
	-16\pi^4 \int_1^\infty d\xi \int_{-1}^1 d\eta \, A_{\mathbb{L}, \text{conn}}&=-\frac{\sqrt{\lambda}}{2}\int_1^\infty d\xi \int_{-1}^1 d\eta \frac{\log(\xi+\sqrt{\xi^2-1})-\xi\sqrt{\xi^2-1}}{(\xi-\eta)(\xi^2-1)^{3/2}}\\&=\sqrt{\lambda} \,,
}
which is in perfect agreement with the localization result in \eqref{Z''(0)Localization}.  This agreement is a consistency check of our result in \eqref{Z''(0)Localization} that can also be viewed as a precision test of the AdS/CFT duality for ${\cal N} = 4$ SYM theory with a defect.

\section{Conclusion}\label{sec:conclusion}

In this work, we have investigated properties of correlation functions of local operators in the flavor current multiplet, in the presence of a half-BPS superconformal line defect, in general 4d $\cN=2$ SCFTs. In particular, we have seen that by working in the conformal frame of ${\rm AdS}_2\times {\rm S}^2$  (and $\mH^2\times {\rm S}^2$ after Wick rotation), which is related to the more familiar setup of a straight line defect in flat spacetime by a Weyl transformation, the analysis of the superconformal Ward identities for the residual $\mf{osp}(4^*|2)$ superconformal symmetry is significantly simplified. The resulting identities among the two-point functions of primary operators in the multiplet are given by familiar differential operators on ${\rm AdS}_2\times {\rm S}^2$. By comparing with the supersymmetric mass deformation of the Euclidean SCFT on ${\rm S}^4$ with a circular defect, where the deformation corresponds to certain integrated insertions of scalar operators in the current multiplet, we derive a simple formula for the two mass derivative of the deformed defect expectation value as an integral of the two-point function of the current multiplet superconformal primary with a trivial integration measure. Our results are further supported by nontrivial consistency checks by analyzing the stress tensor conformal block contribution which applies to all such defect observables, and
also by comparing with known results for the specific case of fundamental Wilson line in the $\cN=4$ super-Yang-Mills theory at large $N$.
Our results pave the way for future studies of such line defects and their generalizations.

A main advantage of the integral constraint on the current multiplet two-point function is that it 
supplies nontrivial coupling-dependent dynamical input to the bootstrap studies and packages  defect OPE data in an integrated correlator that is often accessible by exact methods such as supersymmetric localization. 
In particular, for the half-BPS Wilson-'t Hooft lines in the $\cN=4$ super-Yang-Mills theory, this integrated correlator has been analyzed in detail in \cite{Pufu:2023vwo} in the so-called ``very strong coupling'' large $N$ limit \cite{Binder:2019jwn} where the complexified Yang-Mills coupling $\tau$ is held fixed.  This limit contains much more information about the SCFT (and the defect) than the usual 't Hooft limit at a given order. In particular, one of the main properties of the ${\cal N} = 4$ SYM theory, namely its ${\rm SL}(2,\mZ)$ duality structure, is visible in this very strong coupling limit, thanks to an exact summation over infinitely many instanton contributions \cite{Chester:2019jas,Dorigoni:2021bvj,Dorigoni:2021guq,Dorigoni:2022zcr,Paul:2022piq,Pufu:2023vwo}. 
 In this context, the present work advances the program outlined in \cite{Pufu:2023vwo} by explicitly relating this integrated correlator to the un-integrated two-point function of the superconformal primary in the $\cN=2$ current multiplet, which completes into the $\cN=4$ stress tensor multiplet due to the enhanced superconformal symmetry. The next step, as described in \cite{Pufu:2023vwo}, is to combine this analysis with analytic bootstrap tools, such as the 
defect Mellin amplitudes introduced in
\cite{Goncalves:2018fwx} and recently studied in \cite{Gimenez-Grau:2023fcy}, as well as numerical methods, such as a generalization of the method of determinants in \cite{Gliozzi:2015qsa}. As explained in \cite{Pufu:2023vwo}, the former approach, given the previous success in a similar program carried out for the four-point functions of the stress tensor multiplet (see for instance \cite{Binder:2019jwn,Chester:2019jas}), seems likely to produce the exact stress-tensor two-point function with the half-BPS fundamental Wilson-'t Hooft line defect at least to the first few orders in the $1/N$ expansion at very strong coupling. Via AdS/CFT, such line defects correspond to extended $(p,q)$ strings in the IIB string theory, and the stress tensor two-point function translates into scattering amplitudes of closed string modes reflected off a long string. Consequently, the defect Mellin amplitude that represents the two-point function in the ${\cal N} = 4$ SYM theory determines this reflection amplitude in type IIB string theory on ${\rm AdS}_5\times {\rm S^5}$, with D-instanton effects fully taken into account. Moreover, by taking an appropriate flat space limit, this will produce the first derivation of such an amplitude beyond tree level in the IIB string theory on flat space, and it will shed light on the cancellation of IR divergences between related string theory amplitudes \cite{Periwal:1996pw,Fischler:1996ja}.

Another interesting generalization of this work is to analyze similar integral constraints on correlators with other superconformal defects. Already in $\cN=4$ SYM, there is a large zoo of interesting half-BPS superconformal defects, including the Gukov-Witten surface defects \cite{Gukov:2006jk} as well as interfaces and boundaries \cite{Gaiotto:2008ak}. Such defects also exist in more general $\cN=2$ SCFTs (see, e.g., \cite{Drukker:2010jp}) and we expect much of our analysis will go through for the current multiplet two-point function with the defect. It would be interesting to carry out the determination of relevant localization constraints by analyzing the different 
matrix models that arise from the defect insertions \cite{Gomis:2008qa,Giombi:2009ds,
	Gomis:2009ir,Drukker:2009id,Gomis:2010kv,Giombi:2012ep,Gomis:2014eya,Gomis:2016ljm,Dedushenko:2018tgx,Dedushenko:2020vgd,Dedushenko:2020yzd,Beccaria:2020ykg,Wang:2020seq,Komatsu:2020sup}. In particular, we expect to see  emergent relations between two-point functions of the stress tensor in $\cN=4$ SYM with certain different half-BPS defects 
at large $N$. This is because such defects may arise from the same origin in the bulk IIB string theory. For example, the D5 brane, wrapping submanifolds  ${\rm AdS}_2\times {\rm S}^4$ and ${\rm AdS}_4\times {\rm S}^2$, corresponds to a line defect \cite{Yamaguchi:2006tq} and an interface \cite{Gaiotto:2008ak}, respectively, in the SCFT.

Finally, we emphasize that the integral constraint we derived on the correlators of the current multiplet is applicable to general 4d $\cN=2$ SCFTs with continuous global symmetry and do not require a Lagrangian description. More concretely, for an $\cN=2$ SCFT constructed from twisted  compactification of a 6d $\cN=(2,0)$ SCFT on Riemann surface $\Sigma$ with punctures, known as the Class S construction \cite{Gaiotto:2009we,Gaiotto:2009hg}, half-BPS line defects in the 4d theory descend from surface defects in the 6d theory wrapping a one-cycle on $\Sigma$ \cite{Alday:2009fs,Drukker:2010jp}. The AGT correspondence \cite{Alday:2009aq} states a precise relation between supersymmetric observables of the SCFT on ${\rm S^4}$ decorated by half-BPS line defects and a dual 2d Toda CFT on $\Sigma$ decorated by Verlinde line defects  \cite{Alday:2009fs,Drukker:2010jp}. 
The supersymmetric mass deformation on the SCFT side specifies insertions of local operators at the punctures of $\Sigma$ in the Toda CFT\@. Consequently, the 4d integrated correlator from such a mass deformation translates to certain local OPE data in the 2d CFT, twisted by topological line defects. It would be interesting to explore this connection further (as well as its generalization to other types of half-BPS defects) in order to learn about defects in 4d non-Lagrangian SCFTs.

\section*{Acknowledgements}
This work was supported in part by the US NSF under Grant Nos.~PHY-2111977, PHY-2210420, and PHY-2209997.  RD was also supported in part by an NSF Graduate Research Fellowship. The work of YW was also supported in part by the Simons Junior Faculty Fellows program.

\appendix

\section{\texorpdfstring{Free ${\cal N} = 2$ theories}{Free N = 2 theories}}
\label{FREE}

\subsection{Free massless hypermultiplet}

A hypermultiplet consists of four real scalar fields that can be grouped into the complex combinations $q_{iA}$, with $i =1, 2$, and $A = 1, 2$, obeying the reality condition $(q_{iA})^* = q^{iA} = \varepsilon^{ij} \varepsilon^{AB} q_{jB}$, and left-handed and right-handed fermions $\zeta^A$ and $\zeta_A$, respectively.  The index $A$ is a fundamental index for an $\mathfrak{su}(2)_F$ flavor symmetry.  See Table~\ref{HyperTable}.
\begingroup
\renewcommand{\arraystretch}{1.3}
\begin{table}[htp]
	\begin{center}
		\begin{tabular}{c|c|c|c|c|c}
			field & $\Delta$ & Lorentz rep & $\mf{su}(2)_R$ irrep & $\mf{u}(1)_R$ charge & $\mf{su}(2)_F$ irrep \\
			\hline
			$q_{iA}$ & $1$ & $(0, 0)$ & ${\bf 2}$ & $0$ & ${\bf 2}$ \\
			$\zeta^A$ & $\frac 32$ & $(\frac 12, 0)$ & ${\bf 1}$ & $-\frac 12$ & ${\bf 2}$ \\
			$\zeta_A$ & $\frac 32$ & $(0, \frac 12)$ & ${\bf 1}$ & $\frac 12$  & ${\bf 2}$
		\end{tabular}
	\end{center}
	\caption{Field content of the hypermultiplet.}\label{HyperTable}
\end{table}%
\endgroup

On a conformally flat space (in Lorentzian signature), the action for a free massless hypermultiplet is\footnote{This action can be obtained from (3.163) of \cite{Lauria:2020rhc} after setting all conformal supergravity fields to zero except for the metric and spin connection, and setting all vector multiplet fields to zero as well.  The fields $q^{iA}$ are the same as $A^{iA} = f^{iA}{}_X q^X$ in \cite{Lauria:2020rhc}.  We choose the $f^{iA}{}_X$ such that $q^{iA} = \frac{1}{\sqrt{2}}\begin{pmatrix} 
  i q^3 + q^4 & i q^1 - q^2 \\
  i q^1 + q^2 & - i q^3 + q^4
\end{pmatrix}$.  The dilatation Killing vector is $k_D^X = q^X$, and we take $g_{XY} =\delta_{XY}$ and $d^A{}_B = \delta^A_B$.  See also footnote 25 of \cite{Binder:2021euo}.\label{HyperFootnote1}}
 \es{FreeHyperAction}{
  S  = \int d^4 x\, \sqrt{-g} \left[ - \frac 12 \partial^\mu q^{iA} \partial_\mu q_{iA} - \frac{R}{12} q^{iA} q_{iA}
   -2  \bar \zeta^A \slashed{D} \zeta_A  \right] \,,
 } 
where $R$ is the Ricci scalar.  It is invariant under the supersymmetry transformations
 \es{SUSYFreeHyper}{
  \delta q^{iA} &= - i \bar \epsilon^i \zeta^A + i \bar \epsilon_j \zeta_B \varepsilon^{ji} \varepsilon^{BA} \,, \\
  \delta \zeta^A &= \frac 12 i \slashed{\partial} q^{iA} \epsilon_i - q^{iA} \eta_i \,, \\
  \delta \zeta_A &= -\frac 12 i \slashed{\partial} q_{iA} \epsilon^i - q_{iA} \eta^i \,,
 }
where the transformation parameters must obey the conformal Killing spinor equations \eqref{epsDelta}.

\subsection{\texorpdfstring{Hypermultiplet coupled to a ${\rm U}(1)$ vector multiplet}{Hypermultiplet coupled to a u(1) vector multiplet}}

Since the free hypermultiplet has ${\rm SU}(2)_F$ flavor symmetry, in order to couple it to a ${\rm U}(1)$ vector multiplet we should identify a ${\rm U}(1)$ subgroup of ${\rm SU}(2)_F$.  This can be done by identifying an anti-hermitian generator $T_A{}^B$ of $\mathfrak{su}(2)_F$ (obeying $(T_A{}^B)^* = -T_B{}^A$),\footnote{The $\mathfrak{su}(2)_F$ indices can be raised and lowered with the $\varepsilon$ symbol using the NW-SE convention.  Note that $(T_A{}^B)^* = -T_B{}^A$ and $(T_{AB})^* = T^{AB}$.} such that under infinitesimal ${\rm U}(1)$  gauge transformations, the fields of the vector multiplet and of the hypermultiplet transform as
 \es{U1TransfInfinitesimal}{
  \delta A_\mu &= \partial_\mu \theta \,, \qquad
   \delta \Omega^i = \delta \Omega_i = \delta X = \delta \bar X = \delta Y_{ij} = 0 \,, \\
   \delta q^{iA} &= \theta T_B{}^A q^{iB} \,, \qquad
  \delta q_{iA} = - \theta T_A{}^B q_{iB}  \,, \\
  \delta \zeta^A &= \theta T_B{}^A \zeta^{B} \,, \qquad
   \delta \zeta_A = -\theta T_A{}^B \zeta_{B} \,.
 }
If we want the hypermultiplet to have unit charge under the ${\rm U}(1)$  symmetry, we should take the eigenvalues of $T_A{}^B$ to be $\pm i$, for instance by choosing $T_A{}^B = (i \sigma_2)_A{}^B$.

The gauged hypermultiplet action is\footnote{The coupling to the vector multiplet can also be read off from (3.163) of \cite{Lauria:2020rhc}.  In addition to the conventions in Footnote~\ref{HyperFootnote1}, we take the $\mathfrak{u}(1)$ isometry Killing vector to be $k^X = T_B{}^A q^{iB} f^X{}_{iA}$ and the associated triplet of moment maps $\vec{P}
 = \frac 12 q^{iA} \vec{\tau}_i{}^j T_A{}^B q_{jB}$.  The latter equation can also be written as $P_{ij} =q_{iA} T^{AB} q_{jB}$.  Comparing to \eqref{FlavorOps} below, we see that the moment map $P_{ij}$ is nothing but the $J_{ij}$ operator.}
 \es{GaugedHyper}{
  S  &= \int d^4 x\, \sqrt{-g} \biggl[ - \frac 12 D^\mu q^{iA} D_\mu q_{iA} - \frac{R}{12} q^{iA} q_{iA}
    -2 \bar \zeta^A \slashed{D} \zeta_A   + q_{iA} T^{AB} Y^{ij} q_{jB} \\    
    {}&+  2 X \bar \zeta^A \zeta^B T_{AB} + 2 \bar X \,\bar \zeta_A \zeta_B T^{AB} 
    - 2 i T^{AB} q_{iB} \bar \zeta_A \Omega^i 
    + 2 i T_{AB} q^{iB} \bar \zeta^A \Omega_i  \\
    &{}+ 2 \abs{X}^2 T_B{}^A T_A{}^C q^{iB} q_{iC} \biggr] \,,
 }
where the covariant derivatives are now
 \es{CovDer}{
  D_\mu q^{iA} = \partial_\mu q^{iA} - A_\mu q^{iB} T_B{}^A \,, \qquad
   D_\mu \zeta^A = \partial_\mu \zeta^A + \frac 14 \omega_\mu{}^{ab} \gamma_{ab} \zeta^A 
    - A_\mu \zeta^B T_B{}^A \,.
 } 
In the presence of the vector multiplet, the supersymmetry transformation rules for the hypermultiplet are modified to
 \es{SUSYHyper}{
  \delta q^{iA} &= - i \bar \epsilon^i \zeta^A + i \bar \epsilon_j \zeta_B \varepsilon^{ji} \varepsilon^{BA} \,, \\
  \delta \zeta^A &= \frac 12 i \slashed{D} q^{iA} \epsilon_i
   + i \bar X T_B{}^A q^{iB} \varepsilon_{ij} \epsilon^j - q^{iA} \eta_i \,, \\
  \delta \zeta_A &= -\frac 12 i \slashed{D} q_{iA} \epsilon^i
   - i  X T_A{}^B q_{iB} \varepsilon^{ij} \epsilon_j - q_{iA} \eta^i \,.
 }

\subsection{Massive hypermultiplet in flat space and on the round ${\rm S}^4$}
\label{app:freemassivehyper}

Let us now treat the vector multiplet in the previous subsection as a background.  From comparing the linear terms in the vector multiplet fields with \eqref{SAJ}, one can identify the free field representation for the flavor current multiplet operators\footnote{A minus sign typo in the formula for $\xi_j$ in (3.46) of \cite{Binder:2021euo} is corrected here.}
 \es{FlavorOps}{
    j_\mu &=  \partial_\mu q_{iA} T_B{}^A q^{iB}
     - q_{iA} T_B{}^A \partial_\mu q^{iB} + 
    \bar \zeta_A \gamma_\mu \zeta^B T_B{}^A
     -  \bar \zeta^A \gamma_\mu \zeta_B T_A{}^B \,, \\
  J_{ij} &= q_{iA} T^{AB} q_{jB} \,, \qquad
   K = -2 \bar \zeta^A \zeta^B T_{AB} \,, \qquad
    \bar K = -2 \bar \zeta_A \zeta_B T^{AB} \,, \\
  \xi^i &= -2 i   T_{AB} q^{iB} \zeta^A \,, \qquad
   \xi_i = 2 i T^{AB} q_{iB} \zeta_B \,.
 }
Note that these expressions should be used for computing correlation functions using Wick contractions only at separated points.  See Section~\ref{EXPECTATION}.

As discussed above \eqref{FlatSpaceMass}, on $\R^{1, 3}$ one can give the hypermultiplet a supersymmetry preserving mass by setting $X = \mathfrak{m}/2$ and $\bar X = \bar{\mathfrak{m}}/2$, in which case \eqref{GaugedHyper} becomes
 \es{MassiveHyper}{
  S_\text{flat}  = \int d^4 x\, &\biggl[ - \frac 12 \partial^\mu q^{iA} \partial_\mu q_{iA} 
    - 2 \bar \zeta^A \slashed{D} \zeta_A +  \mathfrak{m} \bar \zeta^A \zeta^B T_{AB} + \bar{\mathfrak{m}} \bar \zeta_A \zeta_B T^{AB} \\
     {}&+ \frac{\abs{\mathfrak{m}}^2}2  T_B{}^A T_A{}^C q^{iB} q_{iC} \biggr] \,,
 }
where the linear terms in $(\mathfrak{m}, \bar{\mathfrak{m}})$ reproduce \eqref{FlatSpaceMass}.  (In order to avoid clutter, we suppressed the subscripts ``flat'' when writing \eqref{MassiveHyper}.)

On a round ${\rm S}^4$, the vector multiplet fields should take the expectation values in \eqref{ExpValues} and \eqref{mdeltaChoice} in order to produce the massive supersymmetric hypermultiplet.  After going to Euclidean signature and using $R = 12/r^2$ for a round ${\rm S}^4$ of radius $r$, \eqref{GaugedHyper} becomes
 \es{MassiveHyperS4}{
  S_\text{sphere}  &= \int d^4 x\, \sqrt{g} \biggl[  \frac 12 \partial^\mu q^{iA} \partial_\mu q_{iA} + \frac{1}{r^2}  q^{iA} q_{iA}
    + 2 \bar \zeta^A \slashed{D} \zeta_A   \mp  i (\tau_2)^{ij}  \frac{m}{2r} q_{iA} T^{AB} q_{jB} \\    
    {}&-  m e^{-i \vartheta} \bar \zeta^A \zeta^B T_{AB} -  m e^{i \vartheta} \bar \zeta_A \zeta_B T^{AB}  - \frac{ m^2}{2} T_B{}^A T_A{}^C q^{iB} q_{iC} \biggr] \,,
 }
Again, the linear terms in $(\mathfrak{m}, \bar{\mathfrak{m}})$ reproduce \eqref{SmSphere}.  (As above, we suppressed the subscripts ``sphere'' when writing \eqref{MassiveHyperS4}.)

These formulas become more transparent if we use the notation
 \es{ZDefs}{
  Z_A \equiv q_{1A} \,, \qquad (Z_A)^* = \bar Z^A = \varepsilon^{AB} q_{2B} \,,
 }
and we choose, for instance, $T_A{}^B = (i \sigma_2)_A{}^B$, so that $T^{AB} =T_{AB} = -\delta_{AB}$.  With these choices, the scalar operators in the flavor current multiplet become
 \es{FlavorOps2}{
  J_{11} &= - \sum_{A=1}^2 Z_A Z_A  \,, \qquad  J_{22} = -\sum_{A=1}^2 \bar Z^A \bar Z^A  \,, \qquad
    J_{12} = \sum_{A=1}^2 (Z_1 \bar Z^2 - Z_2 \bar Z^1)  \,, \\
   K &= 2 \sum_{A=1}^2 \bar \zeta^A \zeta^A  \,, \qquad
    \bar K = 2 \sum_{A=1}^2 \bar \zeta_A \zeta_A \,.
 }
The massive hypermultiplet action in flat space with this notation is
  \es{MassiveHyper2}{
  S_\text{flat}  = \int d^4 x\, \sum_{A=1}^2 &\biggl[ -\abs{\partial_\mu Z_A}^2 
    - 2 \bar \zeta^A \slashed{D} \zeta_A -  \mathfrak{m} \bar \zeta^A \zeta^A  - \bar{\mathfrak{m}} \bar \zeta_A \zeta_A  - \abs{\mathfrak{m}}^2  \abs{Z_A}^2 \biggr] \,,
 }
and on ${\rm S}^4$ it is
 \es{MassiveHyperS42}{
  S_\text{sphere}  &= \int d^4 x\, \sqrt{g} \sum_{A=1}^2 \biggl[  \partial^\mu \bar Z^A \partial_\mu  Z_A +  \frac{2}{r^2} \bar Z^A Z_A 
    + 2 \bar \zeta^A \slashed{D} \zeta_A   \\    
    {}&+  m e^{-i \vartheta} \bar \zeta^A \zeta^A +  m e^{i \vartheta} \bar \zeta_A \zeta_A
    \mp  i  \frac{m}{2r} \left( Z_A Z_A + \bar Z^A \bar Z^A \right)  +  m^2 \bar Z^A Z_A \biggr] \,.
 }

\subsection{Free ${\cal N} = 2$ Abelian vector multiplet}

For a dynamical Abelian vector multiplet, the Maxwell action on a conformally flat space is\footnote{This action can be read off from (20.89) of \cite{Freedman:2012zz} with the prepotential $F(X) = i X^2 / (2 g_\text{YM}^2) $ with all supergravity fields set to zero except for the metric and spin connection.  Then $N = 2 \abs{X}^2 / g_\text{YM}^2$.}
 \es{Maxwell}{
  S = \frac{1}{g_\text{YM}^2} \int d^4x \, \sqrt{-g} 
   \biggl[ 
     -\frac 14 F_{\mu\nu} F^{\mu\nu} - 2 \partial^\mu \bar X \partial_\mu X - \frac{R}{3} \abs{X}^2 + Y_{ij} Y^{ij}   -  \bar \Omega^i \slashed{D} \Omega_i 
   \biggr]  \,.
 }
This action is invariant under the superconformal transformation rules \eqref{SUSYVarsVector} provided that the conformal Killing spinor equations \eqref{epsDelta} are satisfied.  We can of course rescale the scalar and fermionic fields in order to obtain an action with canonically normalized kinetic terms, but in that case $g_\text{YM}$ would appear in the transformation rules, and we prefer to not do that.  From \eqref{Maxwell}, one can immediately obtain the action in flat space by setting $R =0$ or on ${\rm S}^4$ by setting $R = 12/r^2$ and flipping the signs of all the terms.

\subsection{${\cal N} = 4$ Maxwell theory}

Lastly, we can put together the ingredients mentioned above and write down the ${\cal N} = 4$ Maxwell theory.  The conformal ${\cal N} =4$ theory on $\R^{1, 3}$ can be obtained by adding together \eqref{MassiveHyper2} with $\mathfrak{m} = \bar{\mathfrak{m}} = 0$ and \eqref{Maxwell}:
 \es{N4MaxwellFlat}{
  S^{{\cal N} = 4}_\text{flat}
   &=  \int d^4x \,
    \frac{1}{g_\text{YM}^2} \biggl[ 
     -\frac 1{4} F_{\mu\nu} F^{\mu\nu} - 2 \partial^\mu \bar X \partial_\mu X  + Y_{ij} Y^{ij}   - \bar \Omega^i \slashed{D} \Omega_i 
   \biggr] \\
   &{}+\sum_{A=1}^2 \biggl[  -\partial^\mu \bar Z^A \partial_\mu  Z_A 
    - 2 \bar \zeta^A \slashed{D} \zeta_A    \biggr] \,.
 }
The supersymmetric mass-deformed ${\cal N} = 4$ theory on ${\rm S}^4$ is obtained by adding together \eqref{Maxwell} and \eqref{MassiveHyperS42}:
 \es{N4Maxwell}{
  S^{{\cal N} = 4}_\text{sphere}
   &=  \int d^4x \, \sqrt{g} 
    \frac{1}{g_\text{YM}^2} \biggl[ 
     \frac 1{4} F_{\mu\nu} F^{\mu\nu} + 2 \partial^\mu \bar X \partial_\mu X + \frac{4}{r^2} \abs{X}^2 - Y_{ij} Y^{ij}   + \bar \Omega^i \slashed{D} \Omega_i 
   \biggr] \\
   &{}+\sum_{A=1}^2 \biggl[  \partial^\mu \bar Z^A \partial_\mu  Z_A +  \frac{2}{r^2} \bar Z^A Z_A 
    + 2 \bar \zeta^A \slashed{D} \zeta_A   +  m e^{-i \vartheta} \bar \zeta^A \zeta^A +  m e^{i \vartheta} \bar \zeta_A \zeta_A \\
    &{}\mp  i  \frac{m}{2r} \left( Z_A Z_A + \bar Z^A \bar Z^A \right)  +  m^2 \bar Z^A Z_A \biggr] \,.
 }

\section{\texorpdfstring{Evaluation of $D_\Delta$}{Evaluation of DΔ}}
\label{DDELTA}

To evaluate \eqref{DFunc}, let us make the choice $X = (0 , \frac{1 - \zeta^2}{2\zeta}, \frac{1 + \zeta^2}{2\zeta})$ and parameterize $X_2$ as $X_2 = (\frac{\tau_2}{z_2},  \frac{1 - \tau_2^2 - z_2^2}{2z_2}, \frac{1 + \tau_2^2 + z_2^2}{2z_2})$.  Thus,
 \es{DDelta}{
  D_\Delta \left( \frac{1 + \zeta^2}{2 \zeta} \right)  = \int \frac{d\tau_2\, dz_2}{z_2^2}
   \left( \frac{2 z_2}{1 + z_2^2 + \tau_2^2} \right)^{4 - \Delta} 
   \left( \frac{2 \zeta z_2}{\zeta^2 + z_2^2 + \tau_2^2} \right)^{4 - \Delta} 
    \,.
 }
For the cases of interest $\Delta = 2, 3$ the integrals can be done in closed form,  producing
 \es{DExplicit}{
   D_2 \left( \frac{1 + \zeta^2}{2 \zeta} \right) &= -8 \pi \zeta^2 \frac{1 - \zeta^2 + (1 + \zeta^2) \log \zeta}{(1 - \zeta^2)^3} \,, \\
   D_3  \left( \frac{1 + \zeta^2}{2 \zeta} \right) &= -4 \pi \zeta \frac{ \log \zeta}{1 - \zeta^2} \,.
 } 
Note the general relation $-X \cdot X_* = \xi =  \frac{1 + \zeta^2}{2 \zeta}$. Then \eqref{DExplicit2} follows by a change of variable from $\zeta$ to $\xi$.

\section{Additional steps in the Ward identity computation}
\label{WARDDETAILS}

Plugging \eqref{epsAdS} into \eqref{jaVanish}, we obtain
 \es{jaVanish2}{
  (\bar \alpha^{\prime 1} -\bar \beta^{\prime 1} \gamma_3)    \gamma^a (\alpha_2 + \gamma_3 \beta_2) 
   =(\bar \alpha'_{2} - \bar \beta'_2 \gamma_3) \gamma^a  (\alpha^1 + \gamma_3 \beta^1)  \,.
 }
Using \eqref{OSpParams} to solve $\alpha^1 = e^{-i \vartheta} \gamma^0 \alpha_2$, $\beta^1 = e^{-i \vartheta} \gamma^0 \beta_2$, and similarly for the primed quantities, \eqref{jaVanish2} reduces to
 \es{jaVanish2prime}{
  \bar \alpha_2' \{\gamma^0,   \gamma^a \} \alpha_2+ \bar \alpha_2' \{ \gamma^0 ,  \gamma^a \gamma^3 \} \beta_2 -\bar \beta_2' \{ \gamma^0,  \gamma^3 \gamma^a \} \alpha_2
  -\bar \beta_2' \{ \gamma^0,  \gamma^3  \gamma^a \gamma^3 \} \beta_2  
   =0 \,.
 }
Giving values to $a = 0, 1, 2, 3$, one obtains the equation \eqref{jaVanish3} quoted in the main text.

The simplification of \eqref{WardIdExplicit} proceeds as follows.   Using \eqref{epsAdS}, this expression becomes
  \es{WardIdExplicit2}{
    & (\bar \alpha^{1} -\bar \beta^1 \gamma_3)  ( \alpha^{\prime 1}
      + \gamma_3 \beta^{\prime 1} ) \, B(x)
    + (\bar \alpha_{2} - \bar \beta_2 \gamma_3)   (  \alpha'_2 + \gamma_3 \beta'_2)  \,   C(x)  \\
     &{}= - \square \Bigl[\frac{ \bar \alpha_2 \alpha_2' +  x^a ( \bar \alpha_2 \gamma_a \beta_2'  
    -  \bar \beta_2 \gamma_a \alpha_2') -  \bar \beta_2 \beta_2' (-t^2 + \abs{\vec{x}}^2) }{\abs{\vec{x}}}   A(x) \Bigr] \,,
 }
which,  after applying \eqref{OSpParams} to the LHS, can be written as 
 \es{WardIdExplicit2prime}{
    &{}-e^{-2 i\vartheta} (-\bar \alpha_{2}  \alpha'_2 +\bar \beta_2   \beta'_2
    + \bar \alpha_{2}   \gamma^3 \beta'_2
       -\bar \beta_2  \gamma^3  \alpha'_2
       )   \, B(x) \\
    &{}+ (\bar \alpha_{2} \alpha'_2 - \bar \beta_2  \beta'_2 + \bar \alpha_{2} \gamma^3 \beta'_2 
     - \bar \beta_2 \gamma^3 \alpha'_2 )   \,   C(x)  \\
     &{}= - \square \Bigl[\frac{ \bar \alpha_2 \alpha_2' +  x^a ( \bar \alpha_2 \gamma_a \beta_2'  
    -  \bar \beta_2 \gamma_a \alpha_2') -  \bar \beta_2 \beta_2' (-t^2 + \abs{\vec{x}}^2) }{\abs{\vec{x}}}   A(x) \Bigr] \,.
 }
By combing the above with \eqref{jaVanish3}, we then obtain \eqref{WardIdExplicit3}.

\section{Ward identity checks}
\label{WARDCHECKS}

\subsection{Free massless hypermultiplet check}
\label{FREECHECK}

The first check of the Ward identities \eqref{WardFinal} we provide is for a free massless hypermultiplet in the absence of the defect.  Let us start with the flat space action in \eqref{MassiveHyper2} with $\mathfrak{m} = \bar{\mathfrak{m}} = 0$.  The propagators for the scalars $(Z_A, \bar Z^A)$ and fermions $(\zeta_A, \zeta^A)$ are
 \es{Propagators}{
  \langle Z_A(x) \bar Z^B(y) \rangle &= \frac{1}{4 \pi^2 (x - y)^2} \delta_A^B \,, \\
  \langle \zeta_A(x) \bar \zeta^B(y) \rangle &= -\frac 12 \slashed{\partial} \frac{1}{4 \pi^2 (x - y)^2} P_R \delta_A^B = \frac{\gamma_a (x - y)^a P_R}{4 \pi^2 (x - y)^2} \delta_A^B \,, \\
 \langle \zeta^A(x) \bar \zeta_B(y) \rangle &= -\frac 12 \slashed{\partial} \frac{1}{4 \pi^2 (x - y)^2} P_L \delta_A^B = \frac{\gamma_a (x - y)^a P_L}{4 \pi^2 (x - y)^2} \delta_A^B \,.
 }
To check that these equations are consistent with supersymmetry, we can start with the supersymmetry variations \eqref{SUSYFreeHyper}, which in this notation read
 \es{SUSYFreeHyperZZb}{
  \delta Z_A &= i \bar \epsilon_1 \zeta_A + i \bar \epsilon^2 \zeta^B \varepsilon_{BA} \,, \\
  \delta \bar Z^A &= - i \bar \epsilon^1 \zeta^A - i \bar \epsilon_2 \zeta_B  \varepsilon^{BA} \,, \\
  \delta \zeta^A &= \frac 12 i (  \slashed{\partial}  \bar Z^A \epsilon_1 - \slashed{\partial}  \varepsilon^{AB} Z_B \epsilon_2 ) - ( \bar Z^A \eta_1 -\varepsilon^{AB} Z_B \eta_2 ) \,, \\
  \delta \zeta_A &= -\frac 12 i  \left(  \slashed{\partial}  Z_A \epsilon^1 +  \slashed{\partial}  \bar Z^B \varepsilon_{BA} \epsilon^2 \right)  - (Z_A \eta^1  + \bar Z^B \varepsilon_{BA} \eta^2 ) \,.
 }
Then, taking $(\epsilon_i, \epsilon^i)$ to be constant and $\eta_i = \eta^i = 0$, we find
 \es{deltazetaZ}{
  \delta \langle \zeta_A(x) \bar Z^B(y) \rangle 
   = - \frac{i}{2} \langle  \slashed{\partial}Z_A(x) \epsilon^1 \bar Z^B(y) \rangle 
    - i \langle \zeta_A(x) \bar \zeta^B(y) \epsilon^1 \rangle \,.
 }
This equation indeed vanishes upon using \eqref{Propagators}, thus verifying the relative normalization of the scalar and fermion propagators in \eqref{Propagators}.

Using \eqref{Propagators} and the expressions for $J_{ij}$, $K$, $\bar K$ in \eqref{FlavorOps}, we then perform the required Wick contractions to determine the two-point functions
 \es{JJKK}{
  \langle J_{11}(x) J_{22}(y) \rangle 
   &= \langle Z_A(x) Z_A(x) \bar Z^B(y) \bar Z^B(y) \rangle 
     = \frac{1}{4 \pi^4 (x - y)^4} \,, \\
  \langle K(x) \bar K(y) \rangle  
   &= 4 \langle \bar \zeta^A(x) \zeta^A(x) \bar \zeta_B(y) \zeta_B(y) \rangle 
    = \frac{2}{\pi^4 (x-y)^6} \,, \\
    \langle K(x) K(y) \rangle  &= \langle \bar K(x) \bar K(y) \rangle = 0 \,.
 } 
From these expressions, the definitions of the functions $A$, $B$, $\bar B$, and $C$ in \eqref{TwoPoint}, and the definitions of the $\xi$ and $\eta$ invariants in \eqref{xietaDef}, we conclude that, at separated points,\footnote{As per the comment below \eqref{FlavorOps}, we expect a match only at separated points.}
 \es{ABCFree}{
  A = \frac{1}{16 \pi^4 (\xi - \eta)^2} \,, \qquad
   B = \bar B =  0 \,, \qquad C = \frac{1}{4 \pi^4 (\xi - \eta)^3} \,.
 }
These expressions are indeed consistent with the Ward identities \eqref{WardFinal}--\eqref{WardFinalExtra}.  Therefore, this calculation provides a consistency check on \eqref{WardFinal}--\eqref{WardFinalExtra}.

\subsection{Stress tensor superconformal block check}
\label{STRESSCHECK}

Let us now perform another check of the superconformal Ward identities in \eqref{WardFinal}--\eqref{WardFinalExtra} by focusing on the contribution of the stress tensor multiplet to the two-point functions involving $J_{ij}$, $K$, and $\bar{K}$. The point is that a pair of correlators related by a superconformal Ward identity will satisfy the same identity after being projected onto a particular superconformal multiplet because the projection commutes with the supercharges.

The contribution of the stress tensor multiplet to the two-point function of two scalars in the current multiplet can be packaged into a superconformal block, which can be written as a finite sum of conformal blocks. The only conformal primaries in the stress tensor multiplet that can contribute are the scalar at the bottom of the multiplet and the stress tensor, whose quantum numbers are summarized in Table~\ref{StessTensorTable}. This is because only operators with spin $(\frac{\ell}{2},\frac{\ell}{2})$, appear in the OPE of two scalars, and only operators with even $\ell$ can have non-zero one-point functions in the presence of a line defect. 

\begingroup
\renewcommand{\arraystretch}{1.3}
\begin{table}[htp]
	\begin{center}
		\begin{tabular}{c|c|c|c|c}
			operator & $\Delta$ & Lorentz rep & $\mf{su}(2)_R$ irrep & $\mf{u}(1)_R$ charge \\
			\hline
			$M$ & $2$ & $(0, 0)$ & ${\bf 1}$ & 0 \\
			$T_{\mu\nu}$ & $4$ & $(1, 1)$ & ${\bf 1}$ & $0$ \\
		\end{tabular}
	\end{center}
	\caption{For the check of the Ward identities, the relevant conformal primaries in the stress tensor multiplet are the scalar $M$ at the bottom of the multiplet and the stress tensor $T_{\m\n}$. }\label{StessTensorTable}
\end{table}%
\endgroup

The two-point function of scalars of equal dimension $\Delta_\phi$ in the presence of a line defect oriented along the $\tau$ direction in $\mathbb{R}^4$ can be decomposed into a sum of conformal blocks as follows:
\es{scalar2ptExpandedInBlocks}{
    \la \phi_1(\tau_1,\vec{x}_1)\phi_2(\tau_2,\vec{x}_2) \ra &=\frac{1}{|x_{12}|^{2\Delta_\phi}}\sum_{\mathcal{O}}c_{\mathcal{O}}f_{\Delta,\ell}(\xi,\eta)\\&=\frac{1}{|\vec{x}_1|^{\Delta_\phi} |\vec{x}_2|^{\Delta_\phi}}\sum_{\mathcal{O}}\frac{c_{\mathcal{O}}}{2^{\Delta_\phi} (\xi-\eta)^{\Delta_\phi}} f_{\Delta,\ell}(\xi,\eta) \,.
}
Here, $\mathcal{O}$ is a primary with dimension $\Delta$ and spin $(\frac{\ell}{2},\frac{\ell}{2})$ that appears in the OPE of $\phi_1$ and $\phi_2$,  $c_{\mathcal{O}}$ is a product of the $\phi_1\phi_2\mathcal{O}$ OPE coefficient and the one-point function coefficient of $\mathcal{O}$, and $f_{\Delta,\ell}$ is the conformal block. The cross ratios $\xi$ and $\eta$ are given in \eqref{xietaEucDef}. 

Applying Eq.~\eqref{scalar2ptExpandedInBlocks} to Eq.~\eqref{TwoPoint}, the contribution of the stress tensor multiplet to the conformally invariant functions $A$, $B$, $\bar B$, $C$ appearing in the $J_{ij}$, $K$, and $\bar{K}$ two-point functions takes the following form 
\es{ABCStressTensor}{
    A^{\rm ST}(\xi,\eta)&=(\xi-\eta)^{-2}\left[a_1 f_{2,0}(\xi,\eta)+a_2 f_{4,2}(\xi,\eta)\right] \,, \\
    B^{\rm ST}(\xi,\eta)&=(\xi-\eta)^{-3}\left[b_1 f_{2,0}(\xi,\eta)+b_2 f_{4,2}(\xi,\eta)\right] \,, \\
    \bar B^{\rm ST}(\xi,\eta)&=(\xi-\eta)^{-3}\left[\bar b_1 f_{2,0}(\xi,\eta)+\bar b_2 f_{4,2}(\xi,\eta)\right] \,, \\    
    C^{\rm ST}(\xi,\eta)&=(\xi-\eta)^{-3}\left[c_1 f_{2,0}(\xi,\eta)+c_2 f_{4,2}(\xi,\eta)\right] \,,
}
for some coefficients $a_i$, $b_i$, $\bar b_i$, and $c_i$. We note that $T_{\mu\nu}$ and $M$ do not appear in the $KK$ and $\bar K \,\bar K$ OPEs because of the $\mf{u}(1)_R$ symmetry, which means $b_1=b_2 = \bar b_1 = \bar b_2 = 0$. By contrast, the stress tensor should appear in the $K\bar{K}$ OPE and have a non-zero one-point function with the line defect (if not topological), so at a minimum we expect $a_2\neq 0$. Consequently, the Ward identities in \eqref{WardFinal}--\eqref{WardFinalExtra} projected onto the stress tensor multiplet imply
\es{BlockWardId}{
    \frac{1}{2}\square((\xi-\eta)A^{\rm ST}(\xi,\eta))&=0,\quad \frac{1}{2}\square((\xi+\eta)A^{\rm ST}(\xi,\eta))=C^{\rm ST}(\xi,\eta) \,.
}
Checking whether these equations have a consistent solution with $a_2\neq 0$ provides a fairly stringent check of the Ward identities we have derived in the main text.

The last ingredient we need to perform this check are explicit expressions for the conformal blocks. The conformal blocks of two local scalars with a line defect were studied in \cite{Billo:2016cpy,Lauria:2017wav} and an explicit formula for the blocks as series expansions in the distance between the two local operators was derived in \cite{Isachenkov:2018pef,Liendo:2019jpu} (see also appendix A of \cite{Barrat:2021yvp}). We find it convenient to expand the blocks in powers of $\bar{\xi}=\xi-1$ and $\bar{\eta}=1-\eta$, which are (one half of the square of) the chordal distances on $\mathbb{H}^2$ and ${\rm S}^2$ respectively. Up to an unimportant normalization, the first six terms of the two conformal blocks we need are:\footnote{One can get Eq.~\eqref{f20f42BlocksExplicit} from (A.1) of \cite{Barrat:2021yvp} if we map our conformal cross ratios $\xi,\eta$ to their cross ratios $z,\bar{z}$ using $\xi=\frac{1+z\bar{z}}{2\sqrt{z\bar{z}}}$, $\eta=\frac{z+\bar{z}}{2\sqrt{z\bar{z}}}$. We can also expand the conformal blocks to higher orders if desired.}
\es{f20f42BlocksExplicit}{
    f_{2,0}(\bar{\xi},\bar{\eta})&=(\bar{\xi}+\bar{\eta})+\frac{1}{6}(\bar{\eta}^2-\bar{\xi}^2)+\frac{1}{60}(\bar{\xi}+\bar{\eta})(3\bar{\xi}^2-2\bar{\xi}\bar{\eta}+3\bar{\eta}^2)+\frac{1}{280}(\bar{\eta}^2-\bar{\xi}^2)(5\bar{\eta}^2+2\bar{\eta}\bar{\xi}+5\bar{\xi}^2)\\&+\frac{1}{5040}(\bar{\xi}+\bar{\eta})(35\bar{\xi}^4-20\bar{\xi}^3\bar{\eta}+18\bar{\xi}^2\bar{\eta}^2-20\bar{\xi}\bar{\eta}^3+35\bar{\eta}^4)\\&+\frac{1}{22176}(\bar{\eta}^2-\bar{\xi}^2)(63\bar{\xi}^4+28\bar{\xi}^3\bar{\eta}+58\bar{\xi}^2\bar{\eta}^2+28\bar{\xi}\bar{\eta}^3+63\bar{\eta}^4)+\ldots\\
    f_{4,2}(\bar{\xi},\bar{\eta})&=\bar{\xi}^2-\bar{\eta}^2-\frac{1}{7}(\bar{\xi}+\bar{\eta})(3\bar{\xi}^2-5\bar{\xi}\bar{\eta}+3\bar{\eta}^2)+\frac{1}{28}(5\bar{\xi}^4-4\bar{\xi}^3\bar{\eta}+4\bar{\eta}^3\bar{\xi}-5\bar{\eta}^4)\\&-\frac{1}{462}(\bar{\xi}+\bar{\eta})(35\bar{\xi}^4-65\bar{\xi}^3\bar{\eta}+72\bar{\xi}^2\bar{\eta}^2-65\bar{\xi}\bar{\eta}^3+35\bar{\eta}^4)\\&+\frac{25}{48048}\left(63\bar{\xi}^6-56\bar{\xi}^5\bar{\eta}+19\bar{\xi}^4\bar{\eta}^2-19\bar{\xi}^2\bar{\eta}^4+56\bar{\xi}\bar{\eta}^5-63\bar{\eta}^6\right)\\&-\frac{1}{16016}(\bar{\xi}+\bar{\eta})(231\bar{\xi}^6-441\bar{\xi}^5\bar{\eta}+525\bar{\xi}^4\bar{\eta}^2-550\bar{\xi}^3\bar{\eta}^3+525\bar{\xi}^2\bar{\eta}^4-441\bar{\xi}\bar{\eta}^5+231\bar{\eta}^6)+\ldots
}
One can readily check that these obey the Casimir equation
\es{CasimirEqn}{
    \mathcal{C}f_{\Delta,\ell}(\xi,\eta)=c_{\Delta,\ell}f_{\Delta,\ell}(\xi,\eta) \,,
}
where $c_{\Delta,\ell}=\Delta(\Delta-4)+\ell(\ell+2)$ and the Casimir operator in differential form is\footnote{See Eq.~(4.12) of \cite{Billo:2016cpy}, with $\xi_{\rm there}=2(\xi-\eta)$, $\cos{\phi_{\rm there}}=\eta$.}
\es{CasimirDifferentialOp}{
    \mathcal{C}&=2(1-\xi\eta)(1-\xi^2)\frac{\partial^2}{\partial \xi^2}+2(1-\xi\eta)(1-\eta^2)\frac{\partial^2}{\partial\eta^2}+4(1-\eta^2)(1-\xi^2)\frac{\partial^2}{\partial \xi \partial \eta}\\&+2(\eta-2\xi+\eta\xi^2)\frac{\partial}{\partial \xi}+2(\xi-2\eta+\eta^2\xi)\frac{\partial}{\partial \eta} \,.
}

Given \eqref{f20f42BlocksExplicit}, we can check that
\es{BlockSatisfyId1}{
    \square\left((\xi-\eta)^{-1}f_{2,0}(\xi,\eta)\right)=\square\left((\xi-\eta)^{-1}f_{4,2}(\xi,\eta)\right)=0 \,,
}
up to the order that we have expanded the conformal blocks, which is consistent with the first identity in \eqref{BlockWardId}. Furthermore, substituting \eqref{f20f42BlocksExplicit} into the second equation in \eqref{BlockWardId}, we see that the second identity is satisfied up to the first six orders in the series expansion if we set $a_1=-\frac{15}{2}a_2$, $c_1=0$, $c_2=6a_2$.

\bibliographystyle{ssg}
\bibliography{line_defect}

\begingroup\raggedright\begin{thebibliography}{10}

\bibitem{Kondo:1964nea}
J.~Kondo, ``{Resistance Minimum in Dilute Magnetic Alloys},'' {\em Prog. Theor.
  Phys.} {\bf 32} (1964), no.~1 37--49.

\bibitem{Affleck:1995ge}
I.~Affleck, ``{Conformal field theory approach to the Kondo effect},'' {\em
  Acta Phys. Polon. B} {\bf 26} (1995) 1869--1932,
  \href{https://arxiv.org/abs/cond-mat/9512099}{{\tt cond-mat/9512099}}.

\bibitem{Parisen_Toldin_2017}
F.~Parisen~Toldin, F.~F. Assaad, and S.~Wessel, ``Critical behavior in the
  presence of an order-parameter pinning field,'' {\em Physical Review B} {\bf
  95} (Jan., 2017).

\bibitem{Allais:2014fqa}
A.~Allais and S.~Sachdev, ``{Spectral function of a localized fermion coupled
  to the Wilson-Fisher conformal field theory},'' {\em Phys. Rev. B} {\bf 90}
  (2014), no.~3 035131, \href{https://arxiv.org/abs/1406.3022}{{\tt
  1406.3022}}.

\bibitem{Cuomo:2021kfm}
G.~Cuomo, Z.~Komargodski, and M.~Mezei, ``{Localized magnetic field in the O(N)
  model},'' {\em JHEP} {\bf 02} (2022) 134,
  \href{https://arxiv.org/abs/2112.10634}{{\tt 2112.10634}}.

\bibitem{Popov:2022nfq}
F.~K. Popov and Y.~Wang, ``{Non-perturbative defects in tensor models from
  melonic trees},'' {\em JHEP} {\bf 11} (2022) 057,
  \href{https://arxiv.org/abs/2206.14206}{{\tt 2206.14206}}.

\bibitem{wilson1974confinement}
K.~G. Wilson, ``Confinement of quarks,'' {\em Physical review D} {\bf 10}
  (1974), no.~8 2445.

\bibitem{t1978phase}
G.~t~Hooft, ``On the phase transition towards permanent quark confinement,''
  {\em Nuclear Physics B} {\bf 138} (1978), no.~1 1--25.

\bibitem{Rey:1998ik}
S.-J. Rey and J.-T. Yee, ``{Macroscopic strings as heavy quarks in large N
  gauge theory and anti-de Sitter supergravity},'' {\em Eur. Phys. J. C} {\bf
  22} (2001) 379--394, \href{https://arxiv.org/abs/hep-th/9803001}{{\tt
  hep-th/9803001}}.

\bibitem{Maldacena:1998im}
J.~M. Maldacena, ``{Wilson loops in large N field theories},'' {\em Phys. Rev.
  Lett.} {\bf 80} (1998) 4859--4862,
  \href{https://arxiv.org/abs/hep-th/9803002}{{\tt hep-th/9803002}}.

\bibitem{Kapustin:2005py}
A.~Kapustin, ``{Wilson-'t Hooft operators in four-dimensional gauge theories
  and S-duality},'' {\em Phys. Rev. D} {\bf 74} (2006) 025005,
  \href{https://arxiv.org/abs/hep-th/0501015}{{\tt hep-th/0501015}}.

\bibitem{Billo:2016cpy}
M.~Bill\`o, V.~Gon\c{c}alves, E.~Lauria, and M.~Meineri, ``{Defects in
  conformal field theory},'' {\em JHEP} {\bf 04} (2016) 091,
  \href{https://arxiv.org/abs/1601.02883}{{\tt 1601.02883}}.

\bibitem{Lauria:2018klo}
E.~Lauria, M.~Meineri, and E.~Trevisani, ``{Spinning operators and defects in
  conformal field theory},'' {\em JHEP} {\bf 08} (2019) 066,
  \href{https://arxiv.org/abs/1807.02522}{{\tt 1807.02522}}.

\bibitem{Kobayashi:2018okw}
N.~Kobayashi and T.~Nishioka, ``{Spinning conformal defects},'' {\em JHEP} {\bf
  09} (2018) 134, \href{https://arxiv.org/abs/1805.05967}{{\tt 1805.05967}}.

\bibitem{Poland:2018epd}
D.~Poland, S.~Rychkov, and A.~Vichi, ``{The Conformal Bootstrap: Theory,
  Numerical Techniques, and Applications},'' {\em Rev. Mod. Phys.} {\bf 91}
  (2019) 015002, \href{https://arxiv.org/abs/1805.04405}{{\tt 1805.04405}}.

\bibitem{Chester:2019wfx}
S.~M. Chester, ``{Weizmann lectures on the numerical conformal bootstrap},''
  {\em Phys. Rept.} {\bf 1045} (2023) 1--44,
  \href{https://arxiv.org/abs/1907.05147}{{\tt 1907.05147}}.

\bibitem{Poland:2022qrs}
D.~Poland and D.~Simmons-Duffin, ``{Snowmass White Paper: The Numerical
  Conformal Bootstrap},'' in {\em {2022 Snowmass Summer Study}}, 3, 2022.
\newblock \href{https://arxiv.org/abs/2203.08117}{{\tt 2203.08117}}.

\bibitem{Pestun:2016zxk}
V.~Pestun {\em et.~al.}, ``{Localization techniques in quantum field
  theories},'' {\em J. Phys.} {\bf A50} (2017), no.~44 440301,
  \href{https://arxiv.org/abs/1608.02952}{{\tt 1608.02952}}.

\bibitem{Binder:2018yvd}
D.~J. Binder, S.~M. Chester, and S.~S. Pufu, ``{Absence of $D^4 R^4$ in
  M-Theory From ABJM},'' {\em JHEP} {\bf 04} (2020) 052,
  \href{https://arxiv.org/abs/1808.10554}{{\tt 1808.10554}}.

\bibitem{Binder:2019jwn}
D.~J. Binder, S.~M. Chester, S.~S. Pufu, and Y.~Wang, ``{$ \mathcal{N} $ = 4
  Super-Yang-Mills correlators at strong coupling from string theory and
  localization},'' {\em JHEP} {\bf 12} (2019) 119,
  \href{https://arxiv.org/abs/1902.06263}{{\tt 1902.06263}}.

\bibitem{Chester:2019jas}
S.~M. Chester, M.~B. Green, S.~S. Pufu, Y.~Wang, and C.~Wen, ``{Modular
  invariance in superstring theory from $ \mathcal{N} $ = 4
  super-Yang-Mills},'' {\em JHEP} {\bf 11} (2020) 016,
  \href{https://arxiv.org/abs/1912.13365}{{\tt 1912.13365}}.

\bibitem{Chester:2020dja}
S.~M. Chester and S.~S. Pufu, ``{Far beyond the planar limit in
  strongly-coupled $ \mathcal{N} $ = 4 SYM},'' {\em JHEP} {\bf 01} (2021) 103,
  \href{https://arxiv.org/abs/2003.08412}{{\tt 2003.08412}}.

\bibitem{Chester:2020vyz}
S.~M. Chester, M.~B. Green, S.~S. Pufu, Y.~Wang, and C.~Wen, ``{New modular
  invariants in $ \mathcal{N} $ = 4 Super-Yang-Mills theory},'' {\em JHEP} {\bf
  04} (2021) 212, \href{https://arxiv.org/abs/2008.02713}{{\tt 2008.02713}}.

\bibitem{Binder:2019mpb}
D.~J. Binder, S.~M. Chester, and S.~S. Pufu, ``{AdS$_{4}$/CFT$_{3}$ from weak
  to strong string coupling},'' {\em JHEP} {\bf 01} (2020) 034,
  \href{https://arxiv.org/abs/1906.07195}{{\tt 1906.07195}}.

\bibitem{Binder:2020ckj}
D.~J. Binder, S.~M. Chester, M.~Jerdee, and S.~S. Pufu, ``{The 3d $ \mathcal{N}
  $ = 6 bootstrap: from higher spins to strings to membranes},'' {\em JHEP}
  {\bf 05} (2021) 083, \href{https://arxiv.org/abs/2011.05728}{{\tt
  2011.05728}}.

\bibitem{Chester:2021gdw}
S.~M. Chester, R.~R. Kalloor, and A.~Sharon, ``{Squashing, Mass, and Holography
  for 3d Sphere Free Energy},'' {\em JHEP} {\bf 04} (2021) 244,
  \href{https://arxiv.org/abs/2102.05643}{{\tt 2102.05643}}.

\bibitem{Binder:2021cif}
D.~J. Binder, S.~M. Chester, and M.~Jerdee, ``{ABJ Correlators with Weakly
  Broken Higher Spin Symmetry},'' {\em JHEP} {\bf 04} (2021) 242,
  \href{https://arxiv.org/abs/2103.01969}{{\tt 2103.01969}}.

\bibitem{Alday:2021ymb}
L.~F. Alday, S.~M. Chester, and H.~Raj, ``{ABJM at strong coupling from
  M-theory, localization, and Lorentzian inversion},'' {\em JHEP} {\bf 02}
  (2022) 005, \href{https://arxiv.org/abs/2107.10274}{{\tt 2107.10274}}.

\bibitem{Chester:2021aun}
S.~M. Chester, R.~Dempsey, and S.~S. Pufu, ``{Bootstrapping $ \mathcal{N} $ = 4
  super-Yang-Mills on the conformal manifold},'' {\em JHEP} {\bf 01} (2023)
  038, \href{https://arxiv.org/abs/2111.07989}{{\tt 2111.07989}}.

\bibitem{Chester:2022sqb}
S.~M. Chester, ``{Bootstrapping 4d $ \mathcal{N} $ = 2 gauge theories: the case
  of SQCD},'' {\em JHEP} {\bf 01} (2023) 107,
  \href{https://arxiv.org/abs/2205.12978}{{\tt 2205.12978}}.

\bibitem{Behan:2023fqq}
C.~Behan, S.~M. Chester, and P.~Ferrero, ``{Gluon scattering in AdS at finite
  string coupling from localization},'' {\em JHEP} {\bf 02} (2024) 042,
  \href{https://arxiv.org/abs/2305.01016}{{\tt 2305.01016}}.

\bibitem{Chester:2023ehi}
S.~M. Chester, R.~Dempsey, and S.~S. Pufu, ``{Level Repulsion in $\mathcal{N} =
  4$ super-Yang-Mills via Integrability, Holography, and the Bootstrap},''
  \href{https://arxiv.org/abs/2312.12576}{{\tt 2312.12576}}.

\bibitem{Chester:2023qwo}
S.~M. Chester, S.~S. Pufu, Y.~Wang, and X.~Yin, ``{Bootstrapping M-theory
  Orbifolds},'' \href{https://arxiv.org/abs/2312.13112}{{\tt 2312.13112}}.

\bibitem{Behan:2024vwg}
C.~Behan, S.~M. Chester, and P.~Ferrero, ``{Towards Bootstrapping F-theory},''
  \href{https://arxiv.org/abs/2403.17049}{{\tt 2403.17049}}.

\bibitem{Pufu:2023vwo}
S.~S. Pufu, V.~A. Rodriguez, and Y.~Wang, ``{Scattering From $(p,q)$-Strings in
  $\text{AdS}_5 \times \text{S}^5$},''
  \href{https://arxiv.org/abs/2305.08297}{{\tt 2305.08297}}.

\bibitem{Herzog:2020bqw}
C.~P. Herzog and A.~Shrestha, ``{Two point functions in defect CFTs},'' {\em
  JHEP} {\bf 04} (2021) 226, \href{https://arxiv.org/abs/2010.04995}{{\tt
  2010.04995}}.

\bibitem{Agmon:2020pde}
N.~B. Agmon and Y.~Wang, ``{Classifying Superconformal Defects in Diverse
  Dimensions Part I: Superconformal Lines},''
  \href{https://arxiv.org/abs/2009.06650}{{\tt 2009.06650}}.

\bibitem{Alday:2009fs}
L.~F. Alday, D.~Gaiotto, S.~Gukov, Y.~Tachikawa, and H.~Verlinde, ``{Loop and
  surface operators in N=2 gauge theory and Liouville modular geometry},'' {\em
  JHEP} {\bf 01} (2010) 113, \href{https://arxiv.org/abs/0909.0945}{{\tt
  0909.0945}}.

\bibitem{Drukker:2009id}
N.~Drukker, J.~Gomis, T.~Okuda, and J.~Teschner, ``{Gauge Theory Loop Operators
  and Liouville Theory},'' {\em JHEP} {\bf 02} (2010) 057,
  \href{https://arxiv.org/abs/0909.1105}{{\tt 0909.1105}}.

\bibitem{Drukker:2010jp}
N.~Drukker, D.~Gaiotto, and J.~Gomis, ``{The Virtue of Defects in 4D Gauge
  Theories and 2D CFTs},'' {\em JHEP} {\bf 06} (2011) 025,
  \href{https://arxiv.org/abs/1003.1112}{{\tt 1003.1112}}.

\bibitem{Dolan:2002zh}
F.~A. Dolan and H.~Osborn, ``{On short and semi-short representations for
  four-dimensional superconformal symmetry},'' {\em Annals Phys.} {\bf 307}
  (2003) 41--89, \href{https://arxiv.org/abs/hep-th/0209056}{{\tt
  hep-th/0209056}}.

\bibitem{Barrat:2021yvp}
J.~Barrat, A.~Gimenez-Grau, and P.~Liendo, ``{Bootstrapping holographic defect
  correlators in $\mathcal{N}=4$ super Yang-Mills},''
  \href{https://arxiv.org/abs/2108.13432}{{\tt 2108.13432}}.

\bibitem{Billo:2023ncz}
M.~Bill\`o, F.~Galvagno, M.~Frau, and A.~Lerda, ``{Integrated correlators with
  a Wilson line in $ \mathcal{N} $ = 4 SYM},'' {\em JHEP} {\bf 12} (2023) 047,
  \href{https://arxiv.org/abs/2308.16575}{{\tt 2308.16575}}.

\bibitem{BilloToAppear}
M. Bill\`o, F. Galvagno, M. Frau, and A. Lerda, to appear.

\bibitem{Freedman:2012zz}
D.~Z. Freedman and A.~Van~Proeyen, {\em {Supergravity}}.
\newblock Cambridge Univ. Press, Cambridge, UK, 5, 2012.

\bibitem{Lauria:2020rhc}
E.~Lauria and A.~Van~Proeyen, {\em {${\cal N}=2$ Supergravity in $D=4,5,6$
  Dimensions}}, vol.~966.
\newblock 3, 2020.

\bibitem{Cordova:2016xhm}
C.~Cordova, T.~T. Dumitrescu, and K.~Intriligator, ``{Deformations of
  Superconformal Theories},'' {\em JHEP} {\bf 11} (2016) 135,
  \href{https://arxiv.org/abs/1602.01217}{{\tt 1602.01217}}.

\bibitem{Binder:2021euo}
D.~J. Binder, D.~Z. Freedman, S.~S. Pufu, and B.~Zan, ``{The holographic
  contributions to the sphere free energy},'' {\em JHEP} {\bf 01} (2022) 171,
  \href{https://arxiv.org/abs/2107.12382}{{\tt 2107.12382}}.

\bibitem{Pestun:2007rz}
V.~Pestun, ``{Localization of gauge theory on a four-sphere and supersymmetric
  Wilson loops},'' {\em Commun. Math. Phys.} {\bf 313} (2012) 71--129,
  \href{https://arxiv.org/abs/0712.2824}{{\tt 0712.2824}}.

\bibitem{Osborn:1993cr}
H.~Osborn and A.~C. Petkou, ``{Implications of conformal invariance in field
  theories for general dimensions},'' {\em Annals Phys.} {\bf 231} (1994)
  311--362, \href{https://arxiv.org/abs/hep-th/9307010}{{\tt hep-th/9307010}}.

\bibitem{Erdmenger:1996yc}
J.~Erdmenger and H.~Osborn, ``{Conserved currents and the energy momentum
  tensor in conformally invariant theories for general dimensions},'' {\em
  Nucl. Phys. B} {\bf 483} (1997) 431--474,
  \href{https://arxiv.org/abs/hep-th/9605009}{{\tt hep-th/9605009}}.

\bibitem{Closset:2012vg}
C.~Closset, T.~T. Dumitrescu, G.~Festuccia, Z.~Komargodski, and N.~Seiberg,
  ``{Contact Terms, Unitarity, and F-Maximization in Three-Dimensional
  Superconformal Theories},'' {\em JHEP} {\bf 10} (2012) 053,
  \href{https://arxiv.org/abs/1205.4142}{{\tt 1205.4142}}.

\bibitem{Closset:2012vp}
C.~Closset, T.~T. Dumitrescu, G.~Festuccia, Z.~Komargodski, and N.~Seiberg,
  ``{Comments on Chern-Simons Contact Terms in Three Dimensions},'' {\em JHEP}
  {\bf 09} (2012) 091, \href{https://arxiv.org/abs/1206.5218}{{\tt 1206.5218}}.

\bibitem{Gomis:2015yaa}
J.~Gomis, P.-S. Hsin, Z.~Komargodski, A.~Schwimmer, N.~Seiberg, and S.~Theisen,
  ``{Anomalies, Conformal Manifolds, and Spheres},'' {\em JHEP} {\bf 03} (2016)
  022, \href{https://arxiv.org/abs/1509.08511}{{\tt 1509.08511}}.

\bibitem{Gomis:2016sab}
J.~Gomis, Z.~Komargodski, H.~Ooguri, N.~Seiberg, and Y.~Wang, ``{Shortening
  Anomalies in Supersymmetric Theories},'' {\em JHEP} {\bf 01} (2017) 067,
  \href{https://arxiv.org/abs/1611.03101}{{\tt 1611.03101}}.

\bibitem{Papadimitriou:2017kzw}
I.~Papadimitriou, ``{Supercurrent anomalies in 4d SCFTs},'' {\em JHEP} {\bf 07}
  (2017) 038, \href{https://arxiv.org/abs/1703.04299}{{\tt 1703.04299}}.

\bibitem{Schwimmer:2018hdl}
A.~Schwimmer and S.~Theisen, ``{Moduli Anomalies and Local Terms in the
  Operator Product Expansion},'' {\em JHEP} {\bf 07} (2018) 110,
  \href{https://arxiv.org/abs/1805.04202}{{\tt 1805.04202}}.

\bibitem{Papadimitriou:2019gel}
I.~Papadimitriou, ``{Supersymmetry anomalies in $\mathcal{N}=1$ conformal
  supergravity},'' {\em JHEP} {\bf 04} (2019) 040,
  \href{https://arxiv.org/abs/1902.06717}{{\tt 1902.06717}}.

\bibitem{Katsianis:2019hhg}
G.~Katsianis, I.~Papadimitriou, K.~Skenderis, and M.~Taylor, ``{Anomalous
  Supersymmetry},'' {\em Phys. Rev. Lett.} {\bf 122} (2019), no.~23 231602,
  \href{https://arxiv.org/abs/1902.06715}{{\tt 1902.06715}}.

\bibitem{Katsianis:2020hzd}
G.~Katsianis, I.~Papadimitriou, K.~Skenderis, and M.~Taylor, ``{Supersymmetry
  anomaly in the superconformal Wess-Zumino model},'' {\em JHEP} {\bf 21}
  (2020) 209, \href{https://arxiv.org/abs/2011.09506}{{\tt 2011.09506}}.

\bibitem{Bzowski:2020tue}
A.~Bzowski, G.~Festuccia, and V.~Proch\'azka, ``{Consistency of supersymmetric
  \textquoteright{}t Hooft anomalies},'' {\em JHEP} {\bf 02} (2021) 225,
  \href{https://arxiv.org/abs/2011.09978}{{\tt 2011.09978}}.

\bibitem{Russo:2013kea}
J.~G. Russo and K.~Zarembo, ``{Massive ${\cal N}=2$ Gauge Theories at Large
  $N$},'' {\em JHEP} {\bf 11} (2013) 130,
  \href{https://arxiv.org/abs/1309.1004}{{\tt 1309.1004}}.

\bibitem{Belitsky:2020hzs}
A.~V. Belitsky and G.~P. Korchemsky, ``{Circular Wilson loop in N=2* super
  Yang-Mills theory at two loops and localization},'' {\em JHEP} {\bf 04}
  (2021) 089, \href{https://arxiv.org/abs/2003.10448}{{\tt 2003.10448}}.

\bibitem{Drukker:1999zq}
N.~Drukker, D.~J. Gross, and H.~Ooguri, ``{Wilson loops and minimal
  surfaces},'' {\em Phys. Rev. D} {\bf 60} (1999) 125006,
  \href{https://arxiv.org/abs/hep-th/9904191}{{\tt hep-th/9904191}}.

\bibitem{Bobev:2013cja}
N.~Bobev, H.~Elvang, D.~Z. Freedman, and S.~S. Pufu, ``{Holography for $N =
  2^*$ on $S^4$},'' {\em JHEP} {\bf 07} (2014) 001,
  \href{https://arxiv.org/abs/1311.1508}{{\tt 1311.1508}}.

\bibitem{Dorigoni:2021bvj}
D.~Dorigoni, M.~B. Green, and C.~Wen, ``{Novel Representation of an Integrated
  Correlator in $\mathcal N$ = 4 Supersymmetric Yang-Mills Theory},'' {\em
  Phys. Rev. Lett.} {\bf 126} (2021), no.~16 161601,
  \href{https://arxiv.org/abs/2102.08305}{{\tt 2102.08305}}.

\bibitem{Dorigoni:2021guq}
D.~Dorigoni, M.~B. Green, and C.~Wen, ``{Exact properties of an integrated
  correlator in $ \mathcal{N} $ = 4 SU(N) SYM},'' {\em JHEP} {\bf 05} (2021)
  089, \href{https://arxiv.org/abs/2102.09537}{{\tt 2102.09537}}.

\bibitem{Dorigoni:2022zcr}
D.~Dorigoni, M.~B. Green, and C.~Wen, ``{Exact results for duality-covariant
  integrated correlators in $\mathcal{N}=4$ SYM with general classical gauge
  groups},'' \href{https://arxiv.org/abs/2202.05784}{{\tt 2202.05784}}.

\bibitem{Paul:2022piq}
H.~Paul, E.~Perlmutter, and H.~Raj, ``{Integrated correlators in $ \mathcal{N}
  $ = 4 SYM via SL(2, \ensuremath{\mathbb{Z}}) spectral theory},'' {\em JHEP}
  {\bf 01} (2023) 149, \href{https://arxiv.org/abs/2209.06639}{{\tt
  2209.06639}}.

\bibitem{Goncalves:2018fwx}
V.~Goncalves and G.~Itsios, ``{A note on defect Mellin amplitudes},''
  \href{https://arxiv.org/abs/1803.06721}{{\tt 1803.06721}}.

\bibitem{Gimenez-Grau:2023fcy}
A.~Gimenez-Grau, ``{The Witten Diagram Bootstrap for Holographic Defects},''
  \href{https://arxiv.org/abs/2306.11896}{{\tt 2306.11896}}.

\bibitem{Gliozzi:2015qsa}
F.~Gliozzi, P.~Liendo, M.~Meineri, and A.~Rago, ``{Boundary and Interface CFTs
  from the Conformal Bootstrap},'' {\em JHEP} {\bf 05} (2015) 036,
  \href{https://arxiv.org/abs/1502.07217}{{\tt 1502.07217}}. [Erratum: JHEP 12,
  093 (2021)].

\bibitem{Periwal:1996pw}
V.~Periwal and O.~Tafjord, ``{D-brane recoil},'' {\em Phys. Rev. D} {\bf 54}
  (1996) R3690--R3692, \href{https://arxiv.org/abs/hep-th/9603156}{{\tt
  hep-th/9603156}}.

\bibitem{Fischler:1996ja}
W.~Fischler, S.~Paban, and M.~Rozali, ``{Collective coordinates for
  D-branes},'' {\em Phys. Lett. B} {\bf 381} (1996) 62--67,
  \href{https://arxiv.org/abs/hep-th/9604014}{{\tt hep-th/9604014}}.

\bibitem{Gukov:2006jk}
S.~Gukov and E.~Witten, ``{Gauge Theory, Ramification, And The Geometric
  Langlands Program},'' \href{https://arxiv.org/abs/hep-th/0612073}{{\tt
  hep-th/0612073}}.

\bibitem{Gaiotto:2008ak}
D.~Gaiotto and E.~Witten, ``{S-Duality of Boundary Conditions In N=4 Super
  Yang-Mills Theory},'' {\em Adv. Theor. Math. Phys.} {\bf 13} (2009), no.~3
  721--896, \href{https://arxiv.org/abs/0807.3720}{{\tt 0807.3720}}.

\bibitem{Gomis:2008qa}
J.~Gomis, S.~Matsuura, T.~Okuda, and D.~Trancanelli, ``{Wilson loop correlators
  at strong coupling: From matrices to bubbling geometries},'' {\em JHEP} {\bf
  08} (2008) 068, \href{https://arxiv.org/abs/0807.3330}{{\tt 0807.3330}}.

\bibitem{Giombi:2009ds}
S.~Giombi and V.~Pestun, ``{Correlators of local operators and 1/8 BPS Wilson
  loops on S**2 from 2d YM and matrix models},'' {\em JHEP} {\bf 10} (2010)
  033, \href{https://arxiv.org/abs/0906.1572}{{\tt 0906.1572}}.

\bibitem{Gomis:2009ir}
J.~Gomis, T.~Okuda, and D.~Trancanelli, ``{Quantum 't Hooft operators and
  S-duality in N=4 super Yang-Mills},'' {\em Adv. Theor. Math. Phys.} {\bf 13}
  (2009), no.~6 1941--1981, \href{https://arxiv.org/abs/0904.4486}{{\tt
  0904.4486}}.

\bibitem{Gomis:2010kv}
J.~Gomis and B.~Le~Floch, ``{'t Hooft Operators in Gauge Theory from Toda
  CFT},'' {\em JHEP} {\bf 11} (2011) 114,
  \href{https://arxiv.org/abs/1008.4139}{{\tt 1008.4139}}.

\bibitem{Giombi:2012ep}
S.~Giombi and V.~Pestun, ``{Correlators of Wilson Loops and Local Operators
  from Multi-Matrix Models and Strings in AdS},'' {\em JHEP} {\bf 01} (2013)
  101, \href{https://arxiv.org/abs/1207.7083}{{\tt 1207.7083}}.

\bibitem{Gomis:2014eya}
J.~Gomis and B.~Le~Floch, ``{M2-brane surface operators and gauge theory
  dualities in Toda},'' {\em JHEP} {\bf 04} (2016) 183,
  \href{https://arxiv.org/abs/1407.1852}{{\tt 1407.1852}}.

\bibitem{Gomis:2016ljm}
J.~Gomis, B.~Le~Floch, Y.~Pan, and W.~Peelaers, ``{Intersecting Surface Defects
  and Two-Dimensional CFT},'' {\em Phys. Rev. D} {\bf 96} (2017), no.~4 045003,
  \href{https://arxiv.org/abs/1610.03501}{{\tt 1610.03501}}.

\bibitem{Dedushenko:2018tgx}
M.~Dedushenko, ``{Gluing II: boundary localization and gluing formulas},'' {\em
  Lett. Math. Phys.} {\bf 111} (2021), no.~1 18,
  \href{https://arxiv.org/abs/1807.04278}{{\tt 1807.04278}}.

\bibitem{Dedushenko:2020vgd}
M.~Dedushenko and D.~Gaiotto, ``{Algebras, traces, and boundary correlators in
  $ \mathcal{N} $ = 4 SYM},'' {\em JHEP} {\bf 12} (2021) 050,
  \href{https://arxiv.org/abs/2009.11197}{{\tt 2009.11197}}.

\bibitem{Dedushenko:2020yzd}
M.~Dedushenko and D.~Gaiotto, ``{Correlators on the wall and sln spin chain},''
  {\em J. Math. Phys.} {\bf 63} (2022), no.~9 092301,
  \href{https://arxiv.org/abs/2009.11198}{{\tt 2009.11198}}.

\bibitem{Beccaria:2020ykg}
M.~Beccaria and A.~A. Tseytlin, ``{On the structure of non-planar strong
  coupling corrections to correlators of BPS Wilson loops and chiral primary
  operators},'' {\em JHEP} {\bf 01} (2021) 149,
  \href{https://arxiv.org/abs/2011.02885}{{\tt 2011.02885}}.

\bibitem{Wang:2020seq}
Y.~Wang, ``{Taming defects in $ \mathcal{N} $ = 4 super-Yang-Mills},'' {\em
  JHEP} {\bf 08} (2020), no.~08 021,
  \href{https://arxiv.org/abs/2003.11016}{{\tt 2003.11016}}.

\bibitem{Komatsu:2020sup}
S.~Komatsu and Y.~Wang, ``{Non-perturbative defect one-point functions in
  planar $\mathcal{N}=4$ super-Yang-Mills},'' {\em Nucl. Phys. B} {\bf 958}
  (2020) 115120, \href{https://arxiv.org/abs/2004.09514}{{\tt 2004.09514}}.

\bibitem{Yamaguchi:2006tq}
S.~Yamaguchi, ``{Wilson loops of anti-symmetric representation and
  D5-branes},'' {\em JHEP} {\bf 05} (2006) 037,
  \href{https://arxiv.org/abs/hep-th/0603208}{{\tt hep-th/0603208}}.

\bibitem{Gaiotto:2009we}
D.~Gaiotto, ``{N=2 dualities},'' {\em JHEP} {\bf 08} (2012) 034,
  \href{https://arxiv.org/abs/0904.2715}{{\tt 0904.2715}}.

\bibitem{Gaiotto:2009hg}
D.~Gaiotto, G.~W. Moore, and A.~Neitzke, ``{Wall-crossing, Hitchin systems, and
  the WKB approximation},'' {\em Adv. Math.} {\bf 234} (2013) 239--403,
  \href{https://arxiv.org/abs/0907.3987}{{\tt 0907.3987}}.

\bibitem{Alday:2009aq}
L.~F. Alday, D.~Gaiotto, and Y.~Tachikawa, ``{Liouville Correlation Functions
  from Four-dimensional Gauge Theories},'' {\em Lett. Math. Phys.} {\bf 91}
  (2010) 167--197, \href{https://arxiv.org/abs/0906.3219}{{\tt 0906.3219}}.

\bibitem{Lauria:2017wav}
E.~Lauria, M.~Meineri, and E.~Trevisani, ``{Radial coordinates for defect
  CFTs},'' {\em JHEP} {\bf 11} (2018) 148,
  \href{https://arxiv.org/abs/1712.07668}{{\tt 1712.07668}}.

\bibitem{Isachenkov:2018pef}
M.~Isachenkov, P.~Liendo, Y.~Linke, and V.~Schomerus, ``{Calogero-Sutherland
  Approach to Defect Blocks},'' {\em JHEP} {\bf 10} (2018) 204,
  \href{https://arxiv.org/abs/1806.09703}{{\tt 1806.09703}}.

\bibitem{Liendo:2019jpu}
P.~Liendo, Y.~Linke, and V.~Schomerus, ``{A Lorentzian inversion formula for
  defect CFT},'' {\em JHEP} {\bf 08} (2020) 163,
  \href{https://arxiv.org/abs/1903.05222}{{\tt 1903.05222}}.

\end{thebibliography}\endgroup

\end{document}